
\magnification = 1200
\def\lapp{\hbox{$ {	\lower.40ex\hbox{$<$}
		   \atop \raise.20ex\hbox{$\sim$}
		   }	 $}  }
\def\rapp{\hbox{$ {	\lower.40ex\hbox{$>$}
		   \atop \raise.20ex\hbox{$\sim$}
		   }	 $}  }
\def\barre#1{{\not\mathrel #1}}
\def\krig#1{\vbox{\ialign{\hfil##\hfil\crcr
	   $\raise0.3pt\hbox{$\scriptstyle \circ$}$\crcr\noalign
	   {\kern-0.02pt\nointerlineskip}
$\displaystyle{#1}$\crcr}}}
\def\upar#1{\vbox{\ialign{\hfil##\hfil\crcr
	   $\raise0.3pt\hbox{$\scriptstyle \leftrightarrow$}$\crcr\noalign
	   {\kern-0.02pt\nointerlineskip}
$\displaystyle{#1}$\crcr}}}
\def\ular#1{\vbox{\ialign{\hfil##\hfil\crcr
	   $\raise0.3pt\hbox{$\scriptstyle \leftarrow$}$\crcr\noalign
	   {\kern-0.02pt\nointerlineskip}
$\displaystyle{#1}$\crcr}}}

\def\svec#1{\skew{-2}\vec#1}

\def\Tr{\,{\rm Tr}\,}

\def\g5{\gamma_5}

\def\leff{{\cal L}_{{\rm eff}}}
\def\lqcd{{\cal L}_{{\rm QCD}}}
\def\m{{\cal M}}
\def\su3{SU(3)}
\def\l{\Lambda}
\def\hen{{\rm high \, \, energy}}
\def\len{{\rm low \, \, energy}}
\def\zva{{\cal Z}[v,a,s,p]}
\topskip=0.60truein
\leftskip=0.18truein
\vsize=8.8truein
\hsize=6.5truein
\tolerance 10000
\hfuzz=20pt

\baselineskip 12pt plus 1pt minus 1pt
\pageno=0
\centerline{{\bf RECENT DEVELOPMENTS IN CHIRAL PERTURBATION THEORY
}\footnote{*}{Work supported in part by Deutsche
Forschungsgemeinschaft and by Schweizerischer Nationalfonds.}}
\vskip 40pt
\centerline{Ulf-G. Mei{\ss}ner\footnote{$^\dagger$}{Heisenberg
Fellow.}}
\vskip	4pt
\centerline{\it Universit\"at Bern,
Institut f\"ur Theoretische Physik}
\centerline{{\it Sidlerstr. 5,
CH--3012 Bern,\ \ Switzerland}}
\vskip 3truecm
\baselineskip 12pt plus 1pt minus 1pt
{\bf Abstract:} I review recent developments in chiral perturbation theory
(CHPT) which is the effective field theory of the standard model below the
chiral symmetry breaking scale. The effective chiral Lagrangian formulated in
terms of the pseudoscalar
Goldstone bosons ($\pi, \, K, \, \eta$) is briefly discussed. It
is shown how one can gain insight into the ratios of the light quark masses
and to what extent these statements are model--independent. A few selected
topics concerning the dynamics and interactions of the Goldstone bosons are
considered. These are $\pi \pi$ and $\pi K$ scattering, some non--leptonic
kaon decays and the problem of strong pionic final state interactions. CHPT
also allows to make precise statements about the temperature dependence of
QCD Green functions and the finite size effects related to the propagation of
the (almost) massless pseudoscalar mesons. A central topic is the inclusion
of matter fields, baryon CHPT. The relativistic and the heavy fermion
formulation of coupling the baryons to the Goldstone fields are discussed. As
applications, photo--nucleon processes, the $\pi N$ $\Sigma$--term and
non--leptonic hyperon decays are presented. Implications of the spontaneously
broken chiral symmetry on the nuclear forces and meson exchange currents are
also described. Finally, the use of effective field theory methods in the
strongly coupled Higgs sector and in the calculation of oblique electroweak
corrections is touched upon.
\vfill
\centerline{Commissioned article for J. Phys. G: Nucl. Part. Phys.}
\vskip 20pt
\noindent BUTP--93/01 \hfill January 1993
\eject
\baselineskip 12pt plus 1pt minus 1pt
\noindent CONTENTS
\bigskip
\item{1.} Introduction \hfill 2
\medskip
\item{2.} Chiral effective Lagrangian
\smallskip
2.1. \quad QCD in the presence of external sources \hfill 7

2.2. \quad Effective theory to lowest order \hfill 8

2.3. \quad Chiral counting scheme \hfill 10

2.4. \quad Effective theory at next--to--leading
order \hfill 12

2.5. \quad The low--energy constants \hfill 16

2.6. \quad Extensions  \hfill 19
\medskip
\item{3.} Light quark mass ratios
\smallskip
3.1. \quad Lowest order estimates \hfill 21

3.2. \quad Next--to--leading order estimates \hfill 22
\medskip
\item{4.} The meson sector -- selected topics
\smallskip
4.1. \quad $\pi \pi$ scattering \hfill 29

4.2. \quad Testing the mode of quark condensation   \hfill 34

4.3. \quad $\pi K$ scattering \hfill 36

4.4. \quad Beyond one loop \hfill 38

4.5. \quad The decays $K \to 2 \pi$ and $K \to 3 \pi$  \hfill 43
\medskip
\item{5.} Finite temperatures and sizes
\smallskip
5.1. \quad Effective theory at finite temperature   \hfill 45

5.2. \quad Melting condensates	 \hfill 47

5.3. \quad Effective theory in a box   \hfill 50

5.4. \quad Finite size effects: CHPT versus Monte Carlo  \hfill 53

5.5. \quad An application to high--$T_c$ superconductivity  \hfill 54
\medskip
\item{6.} Baryons
\smallskip
6.1. \quad Relativistic formalism \hfill 57

6.2. \quad Non--relativistic formalism \hfill 63

6.3. \quad Photo--nucleon processes \hfill 68

6.4. \quad Baryon masses and the $\sigma$--term   \hfill 74

6.5. \quad Non--leptonic hyperon decays \hfill 77

6.6. \quad Nuclear forces and exchange currents \hfill 79
\medskip
\item{7.} Strongly coupled Higgs sector
\smallskip
7.1. \quad Basic ideas \hfill 83

7.2. \quad Effective Lagrangian at next--to--leading order \hfill 86

7.3. \quad Longitudinal vector boson scattering   \hfill 88

7.4. \quad Electroweak radiative corrections \hfill 92
\medskip
\item{8.} Miscelleaneous omissions  \hfill 95
\medskip
\noindent \quad Appendix A: The case of
$SU(2) \times SU(2)$  \hfill 97
\smallskip
\noindent \quad Appendix B: $SU(3)$ meson--baryon
lagrangian  \hfill 98
\medskip
\noindent \quad References
\hfill 99
\vfill
\eject
\noindent{\bf I. \quad INTRODUCTION}
\medskip
Effective field theories (EFTs) have become a popular tool in particle
and nuclear physics. An effective field theory differs from a conventional
renormalizable ("fundamental") quantum field theory in the following
respect. In EFT, one only works at low energies (where "low" is defined
with respect to some scale specified later) and expands the theory in
powers of the energy/characteristic scale. In that case,
renormalizability at all scales
is not an issue and one has to handle strings of
non--renormalizable interactions. Therefore, at a given order
in the energy expansion, the theory is specified by a finite
number of coupling (low--energy) constants (this allows {\it e.g.}
for an order--by--order renormalization). All observables are
parametrized in terms of these few constants and thus there is a host
of predictions for many different processes. Obviously, at some high
energy this effective theory fails and one has to go over to a better
$\hen$ theory (which again might be an EFT of some fundamental
theory). The trace of this underlying $\hen$ theory are the particular
values of the $\len$ constants. The most complete and most worked  with
EFT is chiral perturbation theory (CHPT) which will be the central topic
of this review. Before elaborating on the particular aspects of CHPT,
let me make some general comments concerning the applications of
EFTs.

\midinsert
\vskip 7.0truecm
{\noindent\narrower \it Fig.~1:\quad
Scattering of light by light. At very low photon energies,
the electrons (solid lines) can be integrated out and one is left
with the EFT discussed in the text.
\smallskip}
\endinsert
EFTs come into play when the underlying fundamental theory contains
massless (or very light) particles. These induce poles and cuts
and conventional Taylor expansions in powers of momenta fail. A
typical example is QED where gauge invariance protects the photon
from acquiring a mass. One photon exchange involves a propagator
$\sim 1/t$, with $t$ the invariant four--momentum transfer squared.
Such a potential can not be Taylor expanded. Here, EFT comes into the
game. Euler and Heisenberg [1] considered the scattering of light
by light at very low energies, $\omega \ll m_e$, with $\omega$
the photon energy and $m_e$ the electron mass (cf. Fig.1). To
calculate the scattering amplitude, one does not need full QED but
rather integrates out the electron from the theory. This leads to an
effective Lagrangian of the form
$$\leff = {1 \over 2} ({\vec E}^2 - {\vec B}^2)
 +{e^4 \over 360 \pi^2 m_e^4} \biggl[({\vec E}^2 - {\vec B}^2)^2
 + 7 ({\vec E} \cdot {\vec B})^2 \biggr] + \ldots
 \eqno(1.1) $$
 which is nothing but a derivative expansion since ${\vec E}$ and
 ${\vec B}$ contain derivatives of the gauge potential. Stated
 differently, since the photon energy is small, the electromagnetic
 fields are slowly varying.
 From eq.(1.1)
 one reads off that corrections to the leading term are suppressed
by powers of $(\omega/m_e)^4$. Straightforward calculation leads
to the cross section $\sigma  (\omega ) \sim \omega^6 / m_e^8$. This
can, of course, also be done using full QED, but the EFT caculation
is much simpler.
A more detailed discussion of effective field theory methods in
QED can be found in the monograph by Dittrich and Reuter [2].

A similar situation arises in QCD which is a non--abelian gauge
theory of colored quarks and gluons,
$$ \eqalign{
\lqcd &= -{1 \over 4 g^2} G_{\mu \nu}^a G^{\mu \nu,a}
  + \bar q i \gamma^\mu D_\mu q  - \bar q \m q \cr
&=
\,\,\,\,\,\,\,\,\,\,\,\,\,\,\,\,\,\,\,\,\,\,\,\,\,\,
\lqcd^0
\,\,\,\,\,\,\,\,\,\,\,\,\,\,\,\,\,\,\,\,\,\,\,\,\,\,\,\,\,
+  \lqcd^{\rm I} \cr}
\eqno(1.2)$$
with $\m = {\rm diag} (m_u,m_d,m_s, \ldots)$ the quark mass matrix.
For the full theory, there is a conserved charge for every quark
flavor separately since the quark masses are all different. However,
for the first three flavors ($u,d,s)$ it is legitimate to set the
quark masses to zero since they are small on a typical hadronic scale
like {\it e.g.} the $\rho$--meson mass. The absolute values of the
running quark masses at 1 GeV are $m_u \simeq 5$ MeV,
$m_d \simeq 9$ MeV,
$m_s \simeq 175$ MeV,
{\it i.e.} $m_u/M_\rho \simeq 0.006$, $m_d/M_\rho \simeq 0.012$ and
$m_s/M_\rho \simeq 0.23$ [3]. If one sets the quark masses to zero, the
left-- and right--handed quarks defined by
$$
q_L = {1 \over 2} ( 1 - \gamma_5) \,q, \,\,\,\,\,
q_R = {1 \over 2} ( 1 + \gamma_5)\, q
\eqno(1.3) $$
do not interact with each other and the whole theory admits an $U(3)
\times U(3)$ symmetry. This is further reduced by the axial anomaly,
so that the actual symmetry group of three flavor massless QCD is
$$ G= \su3_L \times \su3_R \times U(1)_{L+R} \eqno(1.4) $$
The U(1) symmetry related to baryon number conservation will not be
discussed in any further detail. The conserved charges which come along
with the chiral $\su3 \times \su3$ symmetry generate the corresponding
Lie algebra. In the sixties and seventies, manipulations of the
commutation relations between the conserved vector (L+R) and
axial--vector (L-R) charges were called "PCAC relations" or
"current algebra calculations" and lead to a host of $\len$ theorems
and predictions [4]. These rather tedious manipulations have
nowadays been replaced by EFT methods, in particular by CHPT (as will
be discussed later on). Let me come back to QCD. One quickly realizes
that the ground state does not have the full symmetry $G$, eq.(1.4). If
that were the case, every known hadron would have a partner of the
same mass but with opposite parity. Clearly, this is in contradiction
with the observed particle spectrum. Another argument involves the
two--point correlation functions of the vector ($V_\mu^a$) and
axial--vector ($A_\mu^a)$ currents. These can be determined from the
semi--leptonic $\tau$ decays $\tau \to n \pi + \nu_\tau$ [5]
and they show a rather different behaviour. While the VV--correlator
is peaked around $M_\rho \simeq 770$ MeV and smooth otherwise, the
AA--correlator shows a broad enhancement around $M_{A_1} \simeq 1200$
MeV and is smooth elsewhere. The physical ground state must therefore
be asymmetric under the chiral $\su3_L \times \su3_R$ [6].
In fact, the chiral symmetry is spontaneously broken down (hidden) to
the vectorial subgroup of isospin and hypercharge, generated by the
vector currents,
$$ H = \su3_{L+R} \times U(1)_{L+R} \eqno(1.5) $$
As mandated by Goldstone's theorem [7], the spectrum of massless QCD
must therefore contain $N_f^2 - 1 = 9-1 = 8$ massless bosons
with quantum numbers $J^P = 0^-$ (pseudoscalars) since the axial
charges
do not annihilate the vacuum. Reality is a bit more complex. The
quark masses are not exactly zero which gives rise to an explicit
chiral symmetry breaking (as indicated by the term $\lqcd^I$ in
eq.(1.2)). This is in agreement with the observed particle spectrum --
there are no massless strongly interacting particles. However, the
eight lightest hadrons are indeed pseudoscalar mesons. These are the
pions ($\pi^\pm\, , \,\, \pi^0$), the kaons ($K^\pm\,,\,\, {\bar K}^0\, ,
\,\, K^0)$ and the eta ($\eta$). One observes that $M_\pi \ll M_K
\approx M_\eta$ which indicates that the masses of the quarks in  the
$SU(2)$ subgroup (of isospin) should be considerably smaller than the
strange quark mass. This expectation is borne out by actual calculation
of quark mass ratios as discussed later on. Also, from the relative
size of the quark masses $m_{u,d} \ll m_s$ one expects the chiral
expansion to converge much more rapidly in the two--flavor case than
for $\su3_f$. These basic features of QCD can now be explored
in a similar fashion as outlined before for the case of QED. The
technical details will be given in later sections, here I just wish to
outline the main ideas beyond CHPT.

As already noted, the use of EFTs in the context of strong interactions
preceeds QCD. The Ward identities related to the spontaneously broken
chiral symmetry were explored in great detail in the sixties in the
context of current algebra and pion pole dominance [4,8].  The work
of Dashen and Weinstein [9], Weinberg [10] and Callan, Coleman, Wess
and Zumino [11] clarified the relation between current algebra
calculations and the use of effective Lagrangians (at tree level).
However, only with Weinberg's [12] seminal paper in 1979 it became
clear how one could systematically generate loop corrections to the
tree level (current algebra) results. In fact, he showed that these
loop corrections are suppressed by powers of ($E/\l)^2$, with $E$
a typical energy (four--momentum) and $\l$ the scale below which
the EFT can be applied (typically the mass of the first non--Goldstone
resonance, in QCD $\l \simeq M_\rho$). The method was systematized
by Gasser and Leutwyler for $SU(2)_f$ in Ref.[13] and for $\su3_f$
in Ref.[14] and has become increasingly popular ever since. The basic
idea of using an effective Lagrangian instead of the full theory is
based on a universality theorem for $\len$ properties of field
theories containing massless (or very light) particles. Consider
a theory (like QCD) at low energies. It exhibits the following properties:
\medskip
\item{$\bullet$}${\cal L}$ is symmetric under some Lie group $G$ (in QCD:
$G = \su3_L \times \su3_R$).
\medskip
\item{$\bullet$}The ground state $|0>$ is symmetric under
$ H \subset G$
(in QCD: $H = SU(3)_V$). To any broken generator of $G$ there appears
a massless Goldstone boson (called "pion") with the corresponding
quantum numbers ($J^P = 0^-$ in QCD).
\medskip
\item{$\bullet$}The Goldstone bosons have a finite transition amplitude
to decay into the vacuum (via the current associated with the broken
generators). This matrix element carries a scale $F$, which is of
fundamental importance for the $\len$ sector of the theory (in QCD:
$<0| A_\mu^a | \pi^b> = i p_\mu \delta^{ab} F$, with $F$ the pion decay
constant in the chiral limit).
\medskip
\item{$\bullet$}There exists no other massless (strongly interacting)
particles.
\medskip
\item{$\bullet$}Explicit symmetry breaking (like the quark mass term
in QCD) can be treated in a perturbative fashion.
\medskip
Now any theory with these properties looks the same (in more than
two space-time dimensions). This means that to leading order the solution
to the Ward identities connected to the broken symmetry is unique
and only contains the scale $F$. Thus, the EFT to lowest order is
uniquely fixed and it is most economical to formulate it in terms
of the Goldstone fields.
In fact, one collects the pions in a matrix--valued function (generally
denoted '$U$') which transforms linearly under the full action of $G$.
In QCD, a popular choice is $U(x)
= \exp [i  \lambda^a  \pi^a (x)
/F]$  with $\lambda^a\, ( \,\, a = 1, \ldots, 8)$ the Gell--Mann
matrices and $U'(x) = R U(x) L^\dagger$
under chiral $\su3_L \times \su3_R$
(with $L,R$ an element of $\su3_{L,R}$). Accordingly, the pion fields
transform in a highly non--linear fashion [10,11]. This is a
characteristic feature of EFTs.

While the discussion so far was mostly centered around QCD, the
universality theorem immediately allows one to relate QCD to the Higgs
model of standard electroweak symmetry breaking. The Higgs model has
$G =	   O (4)$ broken down to $H =	    O (3)$, which is a
structure isomorphic to two--flavour massless QCD. The only difference
is that while in QCD one has $F \simeq 93$ MeV, in the Higgs model
$F \simeq 250$ GeV. The longitudinal degrees of freedom of the
$W^{\pm}$ and $Z^0$ bosons play the role of the pions as discussed in
more detail later on. Where these two theories really differ is at
next--to--leading order. While to leading order all Green functions
are given in terms of the scale $F$ (and by some parameters related
to explicit symmetry breaking), the solutions to the Ward identities
at next--to--leading order can only be given modulo some unknown
coefficients, the low--energy constants. These have two purposes:
First, they allow to absorb the divergences inflicted by the loop
contributions which have to be accounted for and, second, their
finite values reflect the underlying $\hen$ theory. In QCD, these
$\len$ constants are functions of the scale $\l_{{\rm QCD}}$ and
the masses of the heavy quarks and are in principle calculable.
However, at present  one has to determine them phenomenologically
or use some models to estimate them. This will be discussed in the
following section. It is important to notice that in this way one can
set up a consistent order-by-order renormalization scheme. This allows
one to make predictions once one has used a certain set of processes
to pin down the low--energy constants. In what follows, I will always
consider EFTs beyond tree level, i.e. taking into account loop
corrections. An important argument for doing this is unitarity. Tree
graphs are always real and thus unitarity is violated. Evidently, loop
graphs have imaginary parts and one can restore unitarity in a
perturbative fashion. It was already noted long ago that there are
large unitarity corrections to some processes [15], but it should be
stressed that there are also other corrections. CHPT accounts for all
of them in a {\bf systematic} manner.
The general framework of EFTs is therefore based on some general
principles like Lorentz invariance, analyticity, unitarity and
cluster decomposition. If one has a finite number of massless or light
particle types, one simply writes down an effective non--linear
Lagrangian of these particles in accordance with the pertinent
symmetry requirements. Corrections to the leading order predictions
can be worked out in a systematic fashion. For doing that, one invokes
higher dimensional operators whose contributions are suppressed by
powers of $ 1 / M_{\rm new}$, with $M_{\rm new}$
a typical scale of new physics
(like the $\rho$ in QCD or the techni-rho in technicolor). As we will
see in what follows, there exists special cases were the one loop
corrections are already large at low energies. This forces one to go
beyond the generally well working one loop approximation (as will be
discussed in some detail later on).

We are now in the position to
consider more specific examples build on these general ideas.
The material is organized as follows. In section II, I review the
basic construction of the next--to--leading order chiral Lagrangian
of the strong interactions. It also contains the pertinent chiral
counting rules which are at the heart of the systematic expansion in
small momenta and quark masses. Extensions to incorporate the
energy--momentum tensor and the $U(1)_A$ anomaly are briefly touched
upon. The physics related to the anomalous sector (Wess-Zumino-Witten
term) is not discussed in any detail. In section III, it is shown
how chiral
perturbation theory can be used to gain insight about the ratios of
the light quark masses. Section IV is concerned with some applications
in the meson sector like the classical field of pion--pion scattering.
I also elucidate some problems due to strong pionic final state
interactions. The extension to non-leptonic weak decays is briefly
discussed in the context of the decays $K \to 2 \pi, \, 3 \pi$. In
section V effects of finite temperatures and volumes are considered.
In particular, the melting of the quark and gluon condensates and
finite size effects in relation to lattice gauge theories are
discussed. The following section VI is devoted to the field of baryon
CHPT, with special emphasis on the newly developed methods from heavy
quark EFTs and applications to photo-nucleon processes. In section VII,
it is shown how CHPT methods can be used in the strongly interacting
Higgs sector. The topics of $W_L W_L$ scattering and oblique
corrections to the $Z$ propagator are discussed. In section VIII,
miscellaneous applications and developments are touched upon.
\bigskip
\noindent{\bf II. \quad CHIRAL EFFECTIVE LAGRANGIAN}
\medskip
In this section, I briefly review how to construct the effective
chiral Lagrangian of the strong interactions at next--to--leading order,
following closely the work of Gasser and Leutwyler [14]. It is most
economical to use the external field technique since it avoids
any complication related to the non--linear transformation properties of
the pions. The basic objects to consider are currents and densities
with external fields coupled to them [16] in accordance with the
symmetry requirements. The associated Green functions automatically obey
the pertinent Ward identities and higher derivative terms can be
constructed systematically. The S--matrix elements for processes
involving physical mesons follow then via standard LSZ reduction.
\bigskip
\noindent{\bf 2.1. QCD in the presence of external sources}
\medskip
Consider the vacuum--to--vacuum transition amplitude in the presence
of external fields
$$
{\rm e}^{i\zva} = <0 \, \, {\rm out} | 0 \, \, {\rm in}>_{v,a,s,p}
\eqno(2.1)$$
based on the QCD Lagrangian
$$\qquad \eqalign{ &
{\cal L} = \lqcd^0 + \bar q (\gamma^\mu v_\mu (x) + \gamma^5 \gamma^\mu
a_\mu (x))q - \bar q (s(x) - ip(x)) q
- { \theta (x) \over 16 \pi^2}
G_{\mu \nu}^a \tilde G^{\mu \nu,a}
\cr
& \lqcd^0 = -{1 \over 2 g^2} G_{\mu \nu}^a G^{\mu \nu, a} + \bar q
i \gamma^\mu (\partial_\mu - i A_\mu) q
\cr}
\eqno(2.2) $$
with $A_\mu$ the gluon field, $G_{\mu \nu}$ the corresponding field
strength tensor and $\tilde G_{\mu \nu} = {1 \over 2} \epsilon_{
\mu \nu \alpha \beta} G^{\alpha \beta}$ its dual. $\lqcd^0$ is
massless QCD with vanishing vacuum angle. The external vector
$(v_\mu)$, axial--vector($a_\mu)$, pseudoscalar $(p)$ and scalar
$(s)$ fields are hermitean $3 \times 3$ matrices in flavor space.
The quark mass matrix $\m$,
$$ \m = {\rm diag} (m_u\, , \, m_d \, , \, m_s)  \eqno(2.3) $$
is contained in the scalar field $s(x)$. The Green functions of
massless QCD are obtained by expanding the generating functional
around $v_\mu = a_\mu = s= p =0$ and $\theta (x) = \theta_0$.
For the real world, one has to expand around
$v_\mu = a_\mu = p =0\, , \, \, s(x) = \m$ and $\theta (x) = \theta_0$.
The Lagrangian ${\cal L}$ is invariant under local $\su3 \times \su3$
chiral transformations if the quark and external fields transform as
follows:
$$\eqalign{
q'_R &= Rq \, \, \, \, ; \, \, \, \, q'_L = Lq \cr
v'_\mu + a'_\mu & = R(v_\mu + a_\mu) R^\dagger
+ i R \partial_\mu R^\dagger \cr
v'_\mu - a'_\mu &= L(v_\mu - a_\mu) L^\dagger
+ i L \partial_\mu L^\dagger \cr
s' + i  p' &= R(s + i p) L^\dagger \cr}
\eqno(2.4) $$
with $L,R$ elements of $\su3_{L,R}$ (in general, these are elements
of $U(3)_{L,R}$, but we already account for the axial anomaly to be
discussed later). The path integral representation of ${\cal Z}$
reads:
$$ {\rm e}^{i \zva} = \int [D A_\mu] [Dq] [D \bar q] {\rm e}^{
\int i d^4x {\cal L}(q,\bar q, G_{\mu \nu}\, ; \, v,a,s,p)}
\eqno(2.5) $$
It allows one to make contact to the effective meson theory. Since
we are interested in processes were the momenta are small (the low
energy sector of the theory), we can expand the Green functions
in powers of the external momenta.
This amounts to an expansion in derivatives of the external fields.
As already pointed out, the low energy expansion is not a simple
Taylor expansion since the Goldstone bosons generate poles at $q^2 =0$
(in the chiral limit) or $q^2 = M_\pi^2$ (for finite quark masses).
The $\len$ expansion involves two small parameters, the external
momenta $q$ and the quark masses $\m$. One expands in powers of these
with the ratio $\m /q^2$ fixed. The effective meson Lagrangian	to carry
out this procedure follows from the $\len$ representation of the
generating functional
$$ {\rm e}^{i \zva} = \int [DU]
{\rm e}^{
\int i d^4x {\cal L}_{{\rm eff}}(U \, ; \, v,a,s,p)}
\eqno(2.6) $$
where the matrix $U$ collects the meson fields. The low energy
expansion is now obtained from a perturbative expansion of the meson
EFT,
$$ \leff = {\cal L}_2 + {\cal L}_4 +  \ldots
\eqno(2.7)$$
where the subscript ($n = 2$, 4, $\ldots$) denotes the low energy dimension
(number of derivatives and/or quark mass terms). The various terms
in this expansion will now be discussed.
\medskip
\noindent{\bf 2.2. Effective theory to leading order}
\medskip
Let us now construct the leading term (called ${\cal L}_2$) in the
$\len$ expansion (2.7). The mesons are described by a unitary
$3 \times 3$ matrix in flavor space,
$$ U^\dagger U = 1 \, \,  , \, \, \, \, {\rm det} \, U = 1 \eqno(2.8) $$
The matrix $U$ transforms linearly under chiral symmetry, $U' =
R U L^\dagger$. The lowest order Lagrangian consistent with Lorentz
invariance, chiral symmetry, parity, G--parity and charge conjugation
reads [14]
$$
{\cal L} = {1 \over 4} F^2 \biggl\lbrace {\rm Tr} [ \nabla_\mu U^\dagger
\nabla^\mu U + \chi^\dagger U + \chi U^\dagger] \biggr\rbrace
+ {1 \over 12} H_0 \nabla_\mu \theta \nabla^\mu \theta
\eqno(2.9) $$
The covariant derivative $\nabla_\mu U$ transforms linearly  under
chiral $\su3 \times \su3$ and contains the couplings to the external
vector and axial fields,
$$ \nabla_\mu U = \partial_\mu U - i (v_\mu + a_\mu) U + iU(v_\mu - a_\mu)
\eqno(2.10) $$
The field $\chi$ embodies the scalar and pseudoscalar externals,
$$\chi = 2 B (s + ip)
\eqno(2.11) $$
The last term in (2.9) is related to CP violation in the strong interaction.
For the moment, we will set this term to zero (which agrees with the
empirical observation that the electric dipole moment of the neutron
is tiny). Let us now discuss the various constants appearing in
eqs.(2.9,2.11). The scale $F$ is related to the axial vector currents,
$$A_\mu^a = -F \partial_\mu \pi^a + \ldots
\eqno(2.12)$$
and thus can be identified with the pion decay constant in the chiral
limit, $ F= F_\pi \lbrace 1 + {\cal O}(\m) \rbrace$, by direct
comparison with the matrix--element $<0|A_\mu^a|\pi^b> = i p_\mu
\delta^{ab} F$. The constant $B$, which appears in the field $\chi$,
is related to the explicit chiral symmetry breaking. Consider the
symmetry breaking part of the Lagrangian and expand it in powers of
the pion fields (with $p=0$, $s = \m$ so that
$\chi = 2 B \m$)\footnote{*}{remember that $\pi$ stands as a generic
symbol for the pions, kaons and the $\eta$.}
$${\cal L}_2^{SB} = {1 \over 2} F^2 B {\rm Tr}[\m(U + U^\dagger)]=
(m_u + m_d + m_s) B [F^2 - {\pi^2 \over 2} + {\pi^4 \over 24 F^2}
+ \ldots ]
\eqno(2.13) $$
with $\pi = \lambda^a \pi^a$. The first term on the right hand side
of eq.(2.13) is obviously related to the vacuum energy, while the
second and third are meson mass and interaction terms, respectively.
Since $\partial H_{{\rm QCD}} / \partial m_q = \bar q q$ it follows
from (2.13) that
$$
<0| \bar u u|0> =
<0| \bar d d|0> =
<0| \bar s s|0> = - F^2 B \lbrace 1 + {\cal O}(\m) \rbrace
\eqno(2.14) $$
This shows that the constant $B$ is related to the vev's of the scalar
quark densities $<0| \bar q q| 0>$, the order parameter of the
spontaneous chiral symmetry breaking. The relation (2.14) is only
correct modulo higher order corrections in the quark masses as
indicated by the term ${\cal O}(\m)$. One can furthermore read off the
pseudoscalar mass terms from (2.13). In the case of isospin symmetry
($m_u = m_d = \hat m)$, one finds
$$\eqalign{
M_\pi^2&= 2 \hat m B \lbrace 1 + {\cal O}(\m)\rbrace
= \krig M_\pi^2  \lbrace 1 + {\cal O}(\m)\rbrace \cr
M_K^2&= ( \hat m + m_s) B \lbrace 1 + {\cal O}(\m)\rbrace
= \krig M_K^2	 \lbrace 1 + {\cal O}(\m)\rbrace \cr
M_\eta^2&= {2 \over 3}( \hat m +2 m_s) B \lbrace 1 + {\cal O}(\m)\rbrace
= \krig M_\eta^2    \lbrace 1 + {\cal O}(\m)\rbrace \cr}
\eqno(2.15) $$
with $\krig M_P$ denoting the leading term in the quark mass expansion
of the pseudoscalar meson masses. For the $\krig M_P$, the Gell--Mann--Okubo
relation is exact, $4 \krig M_K^2 = \krig M_\pi^2 + 3 \krig M_\eta^2$
[17]. In the case of isospin breaking, which leads to $\pi^0 - \eta$
mixing, these mass formulae are somewhat more complicated (see section 3
and ref.[14]). Eq.(1.15) exhibits nicely the Goldstone character of the
pions -- when the quark masses are set to zero, the pseudoscalars
are massless and $\su3 \times \su3$ is an exact symmetry. For small
symmetry breaking, the mass of the pions is proportional to the square
root of the symmetry breaking parameter, {\it i.e.} the quark masses
(an alternative scenario is discussed in section 4.2). From eqs.(2.13)
and (2.15) one can eliminate the constant $B$ and gets the celebrated
Gell--Mann--Oakes--Renner [18] relations
$$\eqalign{
F_\pi^2 M_\pi^2&= - 2 \hat m <0| \bar u u |0> + {\cal O}(\m^2) \cr
F_K^2 M_K^2&= -(\hat m + m_s) <0| \bar u u |0> + {\cal O}(\m^2) \cr
F_\eta^2 M_\eta^2&=
-{2 \over 3} (\hat m + m_s) <0| \bar u u |0> + {\cal O}(\m^2) \cr}
\eqno(2.16) $$
where we have used $F_P = F \lbrace 1 + {\cal O}(\m) \rbrace$ ($P=
\pi$, $K$, $\eta$), {\it i.e.} the differences in the physical
decay constants $F_\pi \ne F_K \ne F_\eta$ appear in the terms
of order ${\cal M}^2$.

{}From this discussion we realize that to leading order the strong
interactions are characterized by two scales, namely $F$ and $B$.
Numerically, using the sum rule value $<0| \bar u u |0> = (-225$ MeV)$^3$
[19], one has
$$\eqalign{
&F \simeq F_\pi \simeq 93 \, \,  {\rm MeV}\cr
& B \simeq 1300 \, \, {\rm MeV} \cr}
\eqno(2.17) $$
The large value of the ratio $B/F \simeq 14$ has triggered some
investigations of alternative scenarios concerning the mode of
quark condensation as will be discussed later.

One can now calculate tree diagrams using the effective Lagrangian
${\cal L}_2$ and derive with ease all so--called current algebra
predictions ($\len$ theorems). Current algebra is, as should have
become evident by now, only the first term in a systematic $\len$
expansion. Working out tree graphs using ${\cal L}_2$ can not be
the whole story -- tree diagrams are always real and thus unitarity
is violated. One has to include higher order corrections to cure this.
To do this in a consistent fashion, one needs a counting scheme to
be discussed next.
\bigskip
\noindent{\bf 2.3. Chiral counting scheme}
\medskip
The leading term in the $\len$ expansion of $\leff$ (2.7) was denoted
${\cal L}_2$ because it has dimension (chiral power) two. It contains
two derivatives or one power of the quark mass matrix. If one assumes
the matrix $U$ to be order one, $U = {\cal O}(1)$, a consistent
power counting scheme for local terms containing $U$, $\partial_\mu  U$,
$v_\mu$, $a_\mu$, $s$, $p$, $\ldots$ goes as follows. Denote by $q$
a generic small momentum (for an exact definition of 'small', see
eq.(2.27)). Derivatives count as order $q$ and so do the external fields
which occur linearly in the covariant derivative $\nabla_\mu U$. For
the scalar and pseudoscalar fields, it is most convenient to book them
as order $q^2$. This can be traced back to the fact
that the scalar field $s(x)$ contains the quark mass matrix, thus
$s(x) \sim \m \sim M_\pi^2 \sim q^2$. With these rules, all terms
appearing in (2.9) are of order $q^2$, thus the notation ${\cal L}_2$
(notice that a term of order one is a constant since $U^\dagger U
= 1$ and can therefore be disregarded. Odd powers of $q$ clash with
parity requirements). To summarize, the building blocks of all terms
containing derivatives and/or quark masses have the following
dimension:
$$\eqalign{
& \partial_\mu U(x) \, , \,\, v_\mu (x) \, , \, \, a_\mu(x) =
{\cal O}(q) \cr
& s(x) \, , \, \, p(x) \, , \, \, F_{\mu \nu}^{L,R} (x) =
{\cal O}(q^2) \cr}
\eqno(2.18) $$
where we have introduced the field strength $F_{\mu \nu}^{L,R}$ for
later use. They are defined via
$$\eqalign{
F_{\mu \nu}^I&= \partial_\mu F_\nu^I - \partial_\nu F_\mu^I
- i [F_\mu^I, F_\nu^I] \, , \, \, \, I= L,R \cr
F_{\mu \nu}^R&= v_\mu + a_\mu \, \, ; \, \, \, \,
F_{\mu \nu}^L= v_\mu - a_\mu \cr}
\eqno(2.19) $$
As already mentioned, unitarity calls for pion loop graphs. Weinberg
[12] made the important observation that diagrams with $n$ ($n = 1$,
$2$, $\ldots$) meson loops are suppressed by powers of ${(q^2)}^n$
with respect to the leading term. His rather elegant argument
goes as follows. Consider the S--matrix for a reaction involving
$N_e$ external pions
$$ S = \delta(p_1 + p_2 + \ldots + p_{N_e}) M
\eqno(2.20)$$
with $M$ the transition amplitude. The dimension of $M$ is $[M] = 4-N_e$
since any external pion wave function scales like [mass]$^{-1}$
according to the usual PCAC relation. Now $M$ depends on the total
momentum flowing through the amplitude, on the pertinent coupling
constants $g$ and the renormalization  scale $\mu$ (the loop
diagrams are in general divergent and need to be regularized),
$$\eqalign{
M&= M (q \, , \, \, g \, , \, \, \mu) = q^D f(q/\mu \, , \, \, g) \cr
D&= 4 - N_e - \Delta \cr}
\eqno(2.21) $$
Here, $\Delta$ is the dimension of the different couplings and propagators
entering the transition amplitude. For the effective meson theories under
consideration, the coupling constants associated with the pionic
interactions scale as $4 - d$, with $d$ the number of derivatives as
mandated by Goldstone's theorem (in the chiral limit). If $M$ contains
$N_d$ such vertices and $N_i$ internal propagators, we have
$$
\Delta = \sum_d N_d (d-4) - 2 N_i -N_e
\eqno(2.22)$$
The total number of loops $(N_l)$ follows from simple topological
arguments to be
$$ N_e = N_i - \sum_d N_d - 1 \eqno(2.23) $$
so that the total scaling dimension $D$ of $M$ is
$$ D = 2 + \sum_d N_d (d-2) + 2 N_l
\eqno(2.24) $$
The dominant graphs at low energy carry the smallest value of $D$. The
leading terms with $d=2$ scale like $q^2$ at tree level $(N_l =0)$,
like $q^4$ at one loop level ($N_l = 1$) and so on. Higher derivative
terms with $d=4$ scale as $q^4$ at tree level, as $q^6$ at one--loop
order {\it etc}. This power suppression of loop diagrams is at
the heart of the $\len$ expansion in EFTs like {\it e.g.} CHPT.

Up to now, I have been rather casual with the meaning of the word
"small". By small momentum or small quark mass I mean this with
respect to some typical hadronic scale, also called the scale of
chiral symmetry breaking (denoted by $\l_\chi$). Georgi and Manohar
[20] have argued that a consistent chiral expansion is possible if
$$ \l_\chi \le 4 \pi F_\pi \simeq 1 \, \, \,  {\rm GeV}
\eqno(2.25) $$
Their argument is based on the observation that under a change of
the renormalization scale of order one typical loop contributions
(say to the $\pi \pi$ scattering amplitude) will correspond to
changes in the effective couplings of the order $F_\pi^2 / \l_\chi^2
\simeq 1/(4 \pi)^2$. Setting $\l_\chi = 4 \pi F_\pi$ and cutting
the logarithmically divergent loop integrals at this scale, quantum
corrections are of the same order of magnitude as changes in the
renormalized interaction terms. The factor $(4 \pi)^2$ is related to
the dimensionality of the loop integrals (for a more detailed
outline of this argument see the monograph by Georgi [21]).
Another type of argument was already mentioned in section 1. Consider
$\pi \pi$ scattering in the $I=J=1$ channel. There, at $\sqrt s =
770$ MeV, one hits the $\rho$--resonance. This is a natural barrier
to the derivative expansion of the Goldstone mesons and therefore
serves as a cut off. The appearance of the $\rho$ signals the
regime of the non--Goldstone particles and describes new physics.
It is therefore appropriate to choose
$$\l_\chi \simeq M_\rho \simeq 770 \, \, \, {\rm MeV}
\eqno(2.26) $$
which is not terribly different from the estimate (2.25). In summary,
small external momenta $q$ and small quark masses $\m$ means
$$q/M_\rho \ll 1 \, \, ; \, \, \, \, \m/M_\rho \ll 1 \, \,.
\eqno(2.27) $$
\bigskip
\noindent{\bf 2.4. Effective theory at next--to--leading order}
\medskip
We have now assembled all tools to discuss the generating functional
${\cal Z}$ at next--to--leading order, {\it i.e.} at ${\cal O}(q^4)$.
It consists of three different contributions:
\medskip
\item{1)}The anomaly functional is of order $q^4$ (it contains four
derivatives). We denote the corresponding functional by ${\cal Z}_A$.
The explicit construction was given by Wess and Zumino [22] and can
also be found in ref.[14]. A geometric interpretation is provided by
Witten [23].
\medskip
\item{2)}The most general effective Lagrangian of order $q^4$ which
is gauge invariant. It leads to the action ${\cal Z}_2 + {\cal Z}_4
= \int d^4x {\cal L}_2
+ \int d^4x {\cal L}_4$.
\medskip
\item{3)}One loop graphs associated with the lowest order term,
${\cal L}_2$. These also scale as terms of order $q^4$.
\medskip
Let me first discuss the anomaly functional ${\cal Z}_A$. It subsumes
all interactions which break the intrinsic parity and is responsible
{\it e.g.} for the decay $\pi^0 \to 2 \gamma$. It also generates
interactions between five or more Goldstone bosons [23]. In what
follows, we will not consider this sector in any detail (for a review,
see ref.[24] and recent work in the context of CHPT is quoted in
section 8).

What is now the most general Lagrangian at order $q^4$? The building
blocks and their $\len$ dimensions were already discussed -- we can
have terms with four derivatives and one quark mass or with two
quark masses (and, correspondingly, the other external fields).
In $\su3$, the only invariant tensors are $g_{\mu \nu}$ and
$\epsilon_{\mu \nu \alpha \beta}$, so one is left with (imposing
also P, G and gauge invariance)
$$ {\cal L} = \sum_{i=1}^{10} L_i P_i + \sum_{j=1}^2 H_j \tilde P_j
\eqno(2.28) $$
with
$$\eqalign{
P_1 &= \Tr (\nabla^\mu U^\dagger \nabla_\mu U)^2  \cr
P_2 &= \Tr (\nabla_\mu U^\dagger \nabla_\nu U)
       \Tr (\nabla^\mu U^\dagger \nabla^\nu U) \cr
P_3 &= \Tr (\nabla^\mu U^\dagger \nabla_\mu U
	    \nabla^\nu U^\dagger \nabla_\nu U) \cr
P_4 &= \Tr (\nabla^\mu U^\dagger \nabla_\mu U)
       \Tr (\chi^\dagger U + \chi U^\dagger) \cr
P_5 &= \Tr (\nabla^\mu U^\dagger \nabla_\mu U)
	   (\chi^\dagger U + \chi U^\dagger) \cr
P_6 &= \bigl[\Tr (\chi^\dagger U + \chi U^\dagger)\bigr]^2 \cr
P_7 &= \bigl[\Tr (\chi^\dagger U - \chi U^\dagger)\bigr]^2 \cr
P_8 &= \Tr (\chi^\dagger U
	    \chi^\dagger U +
	    \chi U^\dagger
	    \chi U^\dagger) \cr
P_9 &=
    - i \Tr (F_{\mu \nu}^R \nabla^\mu U \nabla^\nu U^\dagger )
    \Tr (F_{\mu \nu}^L \nabla^\mu U^\dagger \nabla^\nu U) \cr
P_{10} &= \Tr(U^\dagger F^R_{\mu \nu} U F^{L, \mu \nu}) \cr
\tilde P_1 &=
	\Tr(F^R_{\mu \nu} U F^{R, \mu \nu}
	 +  F^L_{\mu \nu} U F^{L, \mu \nu} ) \cr
\tilde P_2 &= \Tr(\chi^\dagger \chi) \cr}
\eqno(2.29) $$
For the two flavor case, not all of these terms are independent. The
pertinent $q^4$ effective Lagrangian is discussed in Appendix A. The
first ten terms of (2.28) are of physical relevance for the $\len$
sector, the last two are only necessary for the consistent renormalization
procedure discussed below. These terms proportional to $\tilde P_j
(j = 1$, $2)$ do not contain the Goldstone fields and are therefore
not directly measurable at low energies.
The constants $L_i \, \, (i=1 , \ldots , 10)$ appearing in (2.28) are
the so--called low--energy constants. They are not fixed by the
symmetry and have the generic structure
$$ L_i = L_i^r \, + \, L_i^{\rm inf}   \eqno(2.30)$$
These constants serve to renormalize the infinities of the pion loops
$(L_i^{{\rm inf}})$
and the remaining finite pieces $(L_i^r)$ have to be
fixed phenomenologically or to be estimated by some model (see below).
At next--to--leading order, the strong interactions dynamics is
therefore determined in terms of twelve parameters -- $B$, $F$, $L_1,$
$\ldots$, $L_{10}$ (remember that we have disregarded the singlet
vector and axial currents). In the absence of external fields, only
the first three terms in (2.28) have to be retained.

Finally, we have to consider the loops generated by the lowest order
effective Lagrangian. These are of dimension $q^4$ (one loop
approximation) as mandated by Weinberg's scaling rule. To evaluate
these loop graphs one considers the neighbourhood of the solution $\bar
U (x)$ to the classical equations of motion. In terms of the
generating functional, this reads
$$ {\rm e}^{i {\cal Z}} =
{\rm e}^{i \int d^4 x [{\cal L}_2(\bar U)
+ {\cal L}_4 (\bar U)]} \int [DU]
{\rm e}^{i \int d^4 x [{\cal L}_2( U)
- {\cal L}_2 (\bar U)]}
\eqno(2.31) $$
The bar indicates that the Lagrangian is evaluated at the classical
solution. According to the chiral counting, in the second factor of
(2.31) only the term ${\cal L}_2$ is kept. This leads to
$${\cal Z} = \int d^4x (
\bar {\cal L}_2 +
\bar {\cal L}_4) + {i \over 2} {\rm ln} \, {\rm det} D
\eqno(2.32) $$
The operator $D$ is singular at short distances. The ultraviolet
divergences contained in ${\rm ln} \, {\rm det} D$
can be determined via the
heat kernel expansion [25]. Using dimensional regularization, the
UV divergences in four dimensions take the form
$$ -{1 \over (4 \pi)^2} {1 \over d-4} {\rm Sp}({1 \over 2} \hat \sigma^2
+{1 \over 12} \hat \Gamma_{\mu \nu} \hat \Gamma^{\mu \nu})
\eqno(2.33) $$
The explicit form of the operators $\hat \sigma$ and $\hat \Gamma_{\mu
\nu}$ can be found in ref.[14]. Using their explicit expressions, the
poles in ${\rm ln} \, {\rm det} D$ can be absorbed by the following
renormalization of the $\len$ constants:
$$ \eqalign{
L_i&= L_i^r + \Gamma_i \lambda , \, \, \, \, i = 1, \ldots, 10 \cr
H_j&= H_j^r + \tilde
\Gamma_j \lambda , \, \, \, \, j = 1, 2 	 \cr}
\eqno(2.34)$$
with
$$\eqalign{
\lambda & = {1 \over 16 \pi^2} \mu^{d-4} \biggl\lbrace {1 \over d-4}
- {1 \over 2} [ {\rm ln}(4 \pi) + \Gamma'(1) + 1 ] \biggr\rbrace \cr
\Gamma_1 & = {3 \over 32} \, , \, \, \, \,
\Gamma_2   = {3 \over 16} \, , \, \, \, \,
\Gamma_3   = 0		  \, , \, \, \, \,
\Gamma_4   = {1 \over 8} \, , \, \, \, \, \cr
\Gamma_5 & = {3 \over 8} \, , \, \, \, \,
\Gamma_6   = {11 \over 144} \, , \, \, \, \,
\Gamma_7   = 0		  \, , \, \, \, \,
\Gamma_8   = {5 \over 48} \, , \, \, \, \, \cr
\Gamma_9 & = {1 \over 4} \, , \, \, \, \,
\Gamma_{10}   = -{1 \over 4} \, , \, \, \, \,
\tilde \Gamma_{1}   = -{1 \over 8} \, , \, \, \, \,
\tilde \Gamma_{2}  = {5 \over 24} \, , \, \, \, \, \cr}
\eqno(2.35) $$
and $\mu$ is the scale of dimensional regularization. The $q^4$
contribution ${\cal Z}_4 + {\cal Z}_{{\rm 1-loop}}$ is finite at
$d = 4$ when expressed in terms of the renormalized coupling constants
$L_i^r$ and $H_i^r$. The next step consists in the expansion
of the differential operator $D$ in powers of the external fields. This
gives the explicit contributions of the one--loop graphs to a given
Green function. The full machinery is spelled out in Gasser and
Leutwyler [14]. In general, one groups the loop contributions into
tadpole and unitarity corrections. While the tadpoles contain one
vertex and one loop, the unitarity corrections contain one loop
and two vertices. The tadpole contributions renormalize the
couplings of the effective Lagrangian. Both of these loop contributions
also depend on the scale of dimensional regularization. In contrast,
physical observables are $\mu$--independent. For actual calculations,
however, it is sometimes convenient to choose a particular value
of $\mu$, say, $\mu = M_\eta$ or $\mu = M_\rho$.

\midinsert
\vskip 7.0truecm
{\noindent\narrower \it Fig.~2:\quad
Goldstone boson scattering in the one--loop approximation.
At lowest order only tree diagrams (a) contribute. At next--to--leading
order, one has tadpole (b), unitarity (c) and higher derivative (as
denoted by the black box in (d)) corrections.
\smallskip}
\endinsert
To one--loop order, the generating functional therefore takes the form
$${\cal Z} =
{\cal Z}_2 +
{\cal Z}_4 +
{\cal Z}_{{\rm one-loop}} +
{\cal Z}_{{\rm anom}}
\eqno(2.36) $$
and what remains to be done is to determine the values of the
renormalized $\len$ constants, $L_i^r (i=1$, $\ldots$, $10)$.
Before doing this, let me graphically show the one--loop approximation
to the $\pi \pi$--scattering amplitude (fig.2). The first term
on the right hand side is the famous tree level term which leads to
Weinberg's current algebra prediction for the scattering amplitude.
The next two terms depict a tadpole and a unitarity correction. Finally,
the term with the black box is a higher derivative term accompanied
by a $\len$ constant. The analytical form of this amplitude will be
discussed in section 4.
\bigskip
\noindent{\bf 2.5. The low--energy constants}
\medskip
The $\len$ constants are in principle calculable from QCD, they depend
on $\Lambda_{{\rm QCD}}$ and the heavy quark masses
$$ L_i^r = L_i^r( \Lambda_{{\rm QCD}} \, ; \, \, m_c \, , \,  m_b\, ,
\, m_t)    \eqno(2.37)$$
In practice, such a calculation is not feasible. One therefore resorts
to phenomenology and determines the $L_i^r$ from data. However, some
of these constants are not easily extracted from empirical information.
Therefore, one uses constraints from the large $N_c$ world [16,27].
In the limit of
$N_c$ (with
$N_c$ the number of colors) going to infinity and keeping $g^2 N_c$
fixed, the Green functions are proportional to some power of
$N_c$ [26, 27, 28]. Furthermore, in this limit the $\eta '$ becomes
massless, $M_{\eta '} \sim 1/N_c$. This leads to an enhancement of the
coupling $L_7$,
$$ L_7^{\eta '} = - {\gamma ^2 \over 48} {F^2 \over M_\eta^2}
\eqno(2.38) $$
with $\gamma$ measuring the strength of $\eta \eta '$ mixing. The
large--$N_c$ counting rules for all $L_i^r$ have been worked out
by Gasser and Leutwyler [14]:
$$ \eqalign{
{\cal O} (N_c^2) \, &: \, \, \, \, L_7 \cr
   {\cal O} (N_c) \, &: \, \, \, \, L_1 \, ,  \, \, L_2 \, , \, \,
    L_3 \, ,  \, \, L_4 \, , \, \,
    L_8 \, ,  \, \, L_{10}  \cr
   {\cal O} (1)  \, &: \, \, \, \, 2 L_1 - L_2
\, ,  \, \, L_4 \, , \, \, L_6 \cr}
\eqno(2.39) $$
Using this and experimental information from $\pi \pi$ scattering,
$F_K / F_\pi$, the electromagnetic radius of the pion and so on, one
ends up with the values for the $L_i^r (\mu = M_\eta )$ given in table 1
(large $N_c$ arguments are used to estimate $L_1 \, , \, L_4$ and $L_6$
). For comparison, we also give the values at $\mu = M_\rho$. It should
be noted that the large $N_c$ suppression  of $ 2 L_1 - L_2$ has
recently been tested using data from $K_{\ell 4}$ decay. Riggenbach et
al. [29] find $( 2 L_1 - L_2 ) / L_3 = -0.19^{+0.55}_{-0.80}$, i.e. the
large $N_c$ prediction works within one standard deviation. More
accurate data will allow to further pin down this quantity. In the case
of $SU(2)$, one can define scale--independent couplings $\bar{\ell}_i$
 ($i = 1, \ldots , 7$). These are discussed in appendix A.

Can one now understand the values of the $L_i^r$ from some first
principles? Already in their 1984 paper, Gasser and Leutwyler [13]
made the following observation. They considered an effective theory
of $\rho$ mesons coupled to the pseudoscalars. Eliminating the heavy
field by use of the equations of motion in the region of momenta
much smaller than the $\rho$ mass, one ends up with terms of order
$q^4$. The values of the corresponding $\len$ constants are given
in terms of $M_\rho$ and the $\rho$--meson coupling strengths to
photons and pions. This leads to a fair description of the $\len$
constants. To make this statement more transparent, consider the
vector form factor of the pion
$$ <\pi^+ (p') | J_\mu^{\rm em} | \pi^+ (p) > = F_V (t) ( p+p')_\mu
\eqno(2.40) $$
with $t = (p'-p)^2$. For small $t$, the form factor can be Taylor
expanded,
$$F_V(t) = 1 + {1 \over 6} <r^2>_V^\pi	     t + {\cal O}(t^2)
\eqno(2.41) $$
Calculating $F_V(t)$ with ${\cal L}_2$ alone, one gets $F_V (t) =1$.
The one--loop result has the form
$$\eqalign{
F_V (t) = 1 & + \biggl[ {2 L_9 \over F_\pi^2} - {1 \over 96 \pi^2
F_\pi^2} \bigl(\ln \bigl({M_\pi^2 \over \mu^2} \bigr) -{1 \over 3}
\bigr)\biggr] t   \cr
& + { t - 4 M_\pi^2 \over 96 \pi^2 F_\pi^2} \biggl[ 2 + \sigma \ln
\bigl| {\sigma - 1 \over \sigma + 1} \bigr| + i \pi \Theta  (t -
4 M_\pi^2) \biggr] \cr}
\eqno(2.42) $$
with $ \sigma = \sqrt {1 - 4 M_\pi^2/t}$. The last term contains
the imaginary part required by unitarity and some rescattering
corrections. Clearly, one has to pin down $L_9^r$ to have a prediction
for the form factor. As it is well known, $F_V (t)$ is well described
within the vector meson dominance (VMD) picture,
$$ F_V (t) = {M_\rho^2 \over M_\rho^2 - t} = 1 + {t \over M_\rho^2}
+{t^2 \over M_\rho^4} + \ldots	  \eqno(2.43)  $$
where I have neglected imaginary parts and alike (a better form is
given in Ref.[30]). Of course, the expansion in $t/ M_\rho^2$ is
only useful as long as $t/ M_\rho^2 \ll 1$. Comparing the terms linear
in $t$, we find from (2.42) and (2.43) that $L_9^r = F_\pi^2 / 2 M_\rho
^2 \simeq 7.3 \cdot 10^{-3} $, which is close to the phenomenological
value of
$L_9^r (M_\rho)$. The small discrepancy is due to the fact that
VMD does not accurately give the vector radius, $<r^2>_{\pi\, ,\,
{\rm VMD}}^V = 6 / M_\rho^2 = 0.40$ fm$^2$  while experimentally
$<r^2>_\pi^V = 0.439 \pm 0.008$ fm$^2$ [31].

This method has been generalized by Ecker {\it et al}. [32] and by
Donoghue {\it et al}. [33]. They consider the lowest order effective
theory of Goldstone bosons coupled to resonance fields (R). These
resonances are of vector (V), axial--vector (A), scalar (S) and
non--Goldstone pseudoscalar (P) type. For the latter category, only the
$\eta '$ is of practical importance (cf. eq(2.38)). The form of the
pertinent couplings is dictated by chiral symmetry in terms of a
few coupling constants which can be determined from data (from
meson--meson and meson--photon decays). At low momenta, one
integrates out the resonance fields. Since their couplings to the
Goldstone bosons are of order $q^2$, resonance exchange produces terms
of order $q^4$ and higher. Symbolically, this reads
$$ \int [dR] {\rm exp}\biggl( i \int d^4 x \tilde \leff [U,R]\biggr) =
	     {\rm exp}\biggl( i \int d^4 x \leff [U]\biggr)
\eqno(2.44) $$
So to leading order $(q^4)$, one only sees the momentum--independent
part of the resonance propagators,
$$ {1 \over M_R^2 - t} = {1 \over M_R^2} \biggl(1 + {t \over M_R^2 - t}
\biggr)      \eqno(2.45)    $$
and thus the $L_i^r (\mu \simeq M_R)$  can be expanded in terms of the
resonance coupling constants and their masses. This leads to
$$L_i^r (\mu) = \sum_{R=V,A,S,P} L_i^{\rm Res} + \hat L_i (\mu)
\eqno(2.46)   $$
with $\hat L_i (\mu)$ a remainder. For this scenario to make sense,
one has to choose $\mu$ somewhere in the resonance region. A preferred
choice is $\mu = M_\rho$ (as shown in Ref.[32], any value of $\mu$
between 500 MeV and 1 GeV does the job). As an example, let me show the
result for the $\len$ constant $L_3^r (M_\rho)$
$$ \eqalign{
L_3^r = L_3^V + L_3^S & = -{3 \over 4} {G_V^2 \over M_V^2} + {1 \over 2
} {c_d^2 \over M_S^2}	  \cr
&= (- 3.55 + 0.53 ) \cdot 10^{-3}  \cr}
\eqno(2.47) $$
As advocated, only the resonance masses $(M_V = 770$ MeV, $M_S = 983$
MeV) and couplings appear ($|G_V| = 53$ MeV, $|c_d| = 32$ MeV). In
table 1, we show the corresponding values for all $\len$ constants.
It is apparent that the resonances almost completely saturate
the $L_i^r$, with no need for additional contributions. This method
of estimating $L_i^r$ is sometimes called QCD duality or the QCD version
of VMD. In fact, it is rather natural that the higher lying hadronic
states leave their imprints in the sector of the light pseudoscalars
-- as already stated, the typical resonance mass is the scale of new
physics not described by the Godstone bosons.

Finally, I mention some other attempts to estimate these $\len$
constants. These are based on Nambu--Jona-Lasinio models [34],
extended Nambu--Jona-Lasinio approaches [35], QCD coupled to constituent
quarks [36] or solution to Coulomb gauge QCD in the ladder approximation
[37]. All of this give results similar to the procedure described above.
It is important to have such a tool since in many circumstances
($q^6$ corrections, non--leptonic weak interactions) the number of
coupling constants is too large to allow for a systematic
phenomenological determination.
\bigskip
$$\hbox{\vbox{\offinterlineskip
\def\strut{\hbox{\vrule height	8pt depth  8pt width 0pt}}
\hrule
\halign{
\strut\vrule# \tabskip 0.1in &
\hfil#\hfil  &
\vrule# &
\hfil#\hfil &
\hfil#\hfil &
\hfil#\hfil &
\vrule# \tabskip 0.0in
\cr
\noalign{\hrule}
&  $i$
&& $L_i^r (M_\eta)$  & $L_i^r (M_\rho)$   & $L_i^{\rm RES}$
& \cr
\noalign{\hrule}
& 1   && $0.9 \pm 0.3$	& $0.7 \pm 0.3$ & 0.6 & \cr
& 2   && $1.7 \pm 0.7$	& $1.3 \pm 0.7$ & 1.2 & \cr
& $3^*$   && $-4.4 \pm 2.5$  & $-4.4 \pm 2.5$ & $-3.0$ & \cr
& 4   && $0.0 \pm 0.5$	& $-0.3 \pm 0.5$ & 0.0 & \cr
& 5   && $2.2 \pm 0.5$	& $ 1.4 \pm 0.5$ & 1.4 & \cr
& 6   && $0.0 \pm 0.3$	& $ -0.2 \pm 0.3$ & 0.0 & \cr
& $7^*$   && $-0.4 \pm 0.15$  & $-0.4 \pm 0.15$ & $-0.3$ & \cr
& 8   && $1.1 \pm 0.3$	& $  0.9 \pm 0.3$ & 0.9 & \cr
& 9   && $7.4 \pm 0.2$	& $  6.9 \pm 0.2$ & 6.9 & \cr
& 10  && $-5.7 \pm 0.3$  & $ -5.2 \pm 0.3$ & $-6.0$ & \cr
\noalign{\hrule}}}}$$
\smallskip
{\noindent\narrower Table 1:\quad Low--energy constants for $\su3_L \times
\su3_R$. The first two columns give the phenomenologically determined
values at $\mu = M_\eta$ and $\mu = M_\rho$. The $L_i^r (i=1$, $\ldots$,
8) are from ref.[14], the $L_{9,10}$ from ref.[43]. The '$*$' denotes
the constants which are not renormalized. The third column shows the
estimate based on resonance exchange [32]. The constant $L_{10}$ has
recently been reexamined by Donoghue and Holstein [261].
\smallskip}
\goodbreak
\noindent{\bf 2.6. Extensions}
\medskip
Here I will briefly discuss some extensions of the chiral effective
Lagrangian at next--to--leading order. I will be sketchy on these
topics and the interested reader should consult the pertinent
literature.

The first extension concerns the energy--momentum tensor in chiral
theories. Let $\Theta_{\mu \nu}$ denote the energy--momentum tensor.
Matrix elements of the type $<\pi \pi | \Theta^\mu_\mu | 0>$
appear in the decay of a light Higgs [40,49]
(which is by now excluded
experimentally) or in the $\pi \pi$ spectrum of the transitions
$\psi ' \to \psi \pi \pi$ and $\Upsilon' \to \Upsilon \pi \pi$
(when one makes use of the multipole expansion [38]). Donoghue and
Leutwyler have enumerated the new terms and performed the pertinent
renormalization procedure [39].
There are three new terms which couple the
meson fields to the Ricci tensor and the curvature scalar plus three
contact terms involving squares of the curvature. These have no
meson matrix elements. The resulting energy--momentum tensor to
order $q^4$ is given in Ref.[39],
$$\Theta^{\mu \nu} =
  \Theta^{\mu \nu}_2 +
  \Theta^{\mu \nu}_{4,g} +
  \Theta^{\mu \nu}_{4,R} + {\cal O} (q^6)
\eqno(2.48) $$
where
 $\Theta^{\mu \nu}_{4,g}$ are the conventional terms arising from (2.29)
 but with Lorentz indices raised and lowered with $g_{\mu \nu}$
and  $\Theta^{\mu \nu}_{4,R}$ is generated by the new couplings.
The one particle matrix elements $<p' | \Theta_{\mu \nu} | p>$
(for $\pi$, $K$, $\eta$) are analysed. They contain two form factors,
a scalar and a tensor since $\Theta_{\mu \nu}$ contains a spin--zero
and a spin--two part. Comparison with a dispersive analysis of the
scalar form factors [40] allows to pin down $L_{11}^r$ and $L_{13}^r$,
$L_{12}^r$ is estimated via tensor $f_2$--exchange. The physical
significance of these new couplings is the following. While
$L_{12}^r$ determines the slope of the tensor form factor, the slope
of the scalar form factor is given by $4 L_{11}^r + L_{12}^r$. The
combination $L_{11}^r - L_{13}^r$ measures the flavour asymmetries
generated by the quark masses.
In this context, one can also make contact to the dilaton model of
the conformal anomaly [41]. In that model,
the breaking of the conformal symmetry is given in terms of an
effective scalar field. The pertinent low--energy constants are
saturated by this scalar and a comparison to the empirical
values of $L_{11}^r$ and $L_{13}^r$ shows that the model can be
considered semi--quantitative at best. Translating the
$L_{11}^r$ into the mass $M_\sigma$, one finds
560 MeV $\le M_\sigma
\le$ 720 MeV,
whereas the empirical value for $L_{13}^r$ leads to a
larger scalar mass,
1000 MeV $\le M_\sigma
\le$ 1100 MeV.

The second extension I want to address concerns
the axial $U(1)$ transformations.
Even in the limit of vanishing quark masses, the $U(1)_A$ Noether
current is not conserved
$$ i \partial^\mu J_{5 \mu}^{(0)} = { 3 \alpha_s \over 8 \pi}
G \tilde G + 2 \sum_{q=u,d,s} m_q \bar q \gamma_5 q
\eqno(2.49)$$
Since $G \tilde G$ is a total divergence, one can define a gauge
invariant current which generates $U(1)_A$ symmetry transformations
in the chiral limit. A chiral rotation with its associated charge $\tilde Q_5$
shifts the $\theta$ vacuum of QCD, $\exp \lbrace i \alpha \tilde
Q_5 \rbrace | 0> = | \theta - 6 \alpha >$ with $\alpha$ the angle of the
chiral rotation. One therefore adds a $\theta$ source term to the
QCD Lagrangian (cf. eq.(2.2)). In the basis where the quark mass
matrix is diagonal QCD without sources is characterized by a vacuum
angle $\theta(x) = \bar \theta$. $\theta$ can be added to the
chiral Lagrangian in a straightforward manner [14,42]. The lowest
order term is already given in eq.(2.9) provided one substitutes
the external field $\chi$ by $\tilde \chi = \chi \exp (i \theta/3)$.
The $\theta$--dependence is then entirely contained in this new
source (apart from a contact term). Clearly, $\tilde \chi$ is invariant
under global $U(1)_A$ rotations. At next--to--leading order, one
has the same terms as before (with $\chi \to \tilde \chi$) and
five additional operators involving derivatives of $\theta$ plus higher
order contact terms with no meson matrix elements. These terms are
enumerated in ref.[42], for the later applications we only need
one of them
$${\cal L}_4^{D \theta} = i L_{14} D_\mu \theta D^\mu \theta
\Tr (\tilde \chi^\dagger U - U^\dagger \tilde \chi) + \ldots
\eqno(2.50) $$
Finally one can also couple a pseudoscalar $SU(3)$--singlet
field like {\it e.g.} the $\eta '$. This is spelled out in detail
in the original work of Gasser and Leutwyler [14] and will not
be needed in what follows.
\bigskip
\noindent{\bf III. LIGHT QUARK MASS RATIOS}
\medskip
The current quark masses characterize the strength of the flavour symmetry
breaking. They are small on typical hadronic scales as indicated by the fact
that the $\su3$ predictions work fairly well. Expanding physical quantities in
powers of the quark masses, the deviations from the symmetry limit are probing
the quark masses. To lowest order, this is a clean method. The perturbation is
of the form $\bar q \m q$ and forming ratios, one can eliminate the dependence
on the unknown operator $\bar q q$. I will review how CHPT allows to fix the
ratios of the light quark masses at leading and next--to--leading order. In the
latter case, some model dependence enters as will be discussed later on. For a
fairly comprehensive review on the history of the subject and the
determinations of the absolute values of the quark masses, the reader is
referred to Gasser and Leutwyler [3]. Also, as a cautionary remark, let me
mention that the mass parameters appearing in the Lagrangian should not be
confused with the so--called constituent masses. These represent the energy of
quarks confined inside hadrons. They do not follow directly from QCD but rather
from some effective model.
\bigskip
\noindent{\bf 3.1. Lowest order estimates}
\medskip
The best place to investigate the light quark mass ratios is the spectrum of
the pseudoscalar Goldstone bosons. As already discussed in section 2, their
masses are directly proportional to the quark masses. Neglecting any
electromagnetic and higher order corrections, the standard GMOR scenario [18]
(see also ref.[44]) gives
$$\eqalign{
M_{\pi^+}^2 &= (m_u + m_d) \, B \, \, \, , \cr
M_{K^+}^2 &= (m_u + m_s) \, B \, \, \, , \cr
M_{K^0}^2 &= (m_d + m_s) \, B \, \, \, , \cr} \eqno(3.1) $$
which allows to extract
$$ {m_u \over m_d} = 0.66 \, \, \, , \, \, \,
{m_s \over m_d} = 20.1 \, \, \, , \, \, \,
{\hat m \over m_s} = {1 \over 24.2} \eqno(3.2) $$
with $\hat m = (m_u + m_d) / 2$ the average light quark mass. This estimate is,
however, somewhat too naive. The charged pions and kaons are surrounded by a
photon cloud which leads to a mass shift of order $e^2$ (isospin breaking). In
the chiral limit, one can resort to Dashen's theorem [45] which states that the
electromagnetic self-energy of a pseudoscalar meson $P$ is proportional to its
charge $Q_P$,
$$ M_P^2 = (M_P^2)_{\rm strong} + e^2 Q_P C + {\cal O}(e^2 \m )
\eqno(3.3) $$
As shown by Das et al. [46], the constant $C$ can be expressed as an integral
over the vector and axial--vector spectral functions. The latter have been
determined from the semi--leptonic $\tau \to \nu_\tau + n \pi$
decays [5].
Another way of fixing this constant is based on the empirical fact that for the
pion the mass difference is almost entirely of electromagnetic origin,
$(M_{\pi^+} - M_{\pi^0})_{\rm strong} \sim (m_u - m_d)^2$. This strong
splitting is due to the
$\pi^0 \eta$ mixing which causes the neutral pion to become lighter than its
charged partner. Numerically, this effect is tiny, $\sim 0.2$ MeV [3]. Thus,
the pion mass difference fixes the value of $C$, which for the kaons results in
the following leading--order strong splitting:
$$ (M_{K^0}^2 - M_{K^+}^2)_{\rm strong} = M_{K^0}^2 - M_{K^+}^2 + M_{\pi^+}^2 -
M_{\pi^0}^2	\eqno(3.4) $$
which amounts to $(M_{K^0} - M_{K^+})_{\rm strong} = (5.3 - 1.3)$ MeV = 4 MeV
[3] and modifies the quark mass ratios accordingly
$$ {m_u \over m_d} = 0.55 \, \, \, , \, \, \,
{m_s \over m_d} = 20.1 \, \, \, , \, \, \,
{\hat m \over m_s} = {1 \over 25.9} \eqno(3.5) $$
This is the celebrated result of Weinberg [47]. Before discussing the ${\cal O}
(\m^2 )$ corrections to these numbers, let us pause for a moment. The ratio
$m_u / m_d$ is rather different from one, so why does one not see large isospin
violations in physical processes? The answer lies in the fact that although
$m_d \simeq 2 m_u$, both masses are so small compared to the typical hadronic
 scale
that this effect is almost perfectly screened. One exception to be discussed
 later
is the decay $\eta \to 3 \pi$, whose amplitude is proportional to $m_d - m_u$.
Weinberg [47] has also proposed to look into isospin breaking effects in the
pion--nucleon scattering lengths, but considering the present status of the
empirical information, this does not seem to be a realistic proposal. However,
before elaborating further on these topics, let us consider the corrections of
order $\m^2$.
\bigskip
\noindent{\bf 3.2. Next--to--leading order estimates}
\medskip
Before discussing various attempts to pin down the order $\m^2$ corrections to
the quark mass ratios (3.5), we have to discuss one subtlety which arises at
next--to--leading order. As noted by Kaplan and Manohar [48], the chiral
symmetry does not differentiate between the conventional mass matrix $\m =
{\rm diag} (m_u , m_d , m_s)$ (with $m_{u,d,s}$ real) or $\m$' defined by the
eigenvalues
$$\eqalign{
m_u' = \alpha_1 m_u \, + \, \alpha_2 m_d m_s \, \, \, ,  \cr
m_d' = \alpha_1 m_d \, + \, \alpha_2 m_u m_s \, \, \, ,  \cr
m_s' = \alpha_1 m_s \, + \, \alpha_2 m_u m_d \, \, \, ,  \cr}
\eqno(3.6) $$
with $\alpha_1 , \alpha_2$ two arbitrary constants. As already noted, only the
product $B \m$ (via the scalar field $s(x)$) enters the chiral Lagrangian and
thus $\alpha_1$ merely renormalizes the value of $B$. The constant $\alpha_2$
enters at order $\m^2$ and contributes to ${\cal L}_4$. To be more specific, it
contributes to the terms with two powers of $\chi$. One can therefore
completely hide this symmetry by redefining
$$ B' = B / \alpha_1 \, \, , \, \, L_6 ' = L_6 - c \, \, ,
\, \, L_7 ' = L_7 - c \, \, , \, \, L_8 ' = L_8 + 2 c \, \, .
\eqno(3.7) $$
with $c = (\alpha_2 / \alpha_1) \, (F^2 / 32 B)$. This means that ${\cal
L}_2 + {\cal L}_4$ is invariant under the symmetry (3.7). Notice that the term
proportional to $L_5^r$ is not affected since it only contains one power of the
quark mass matrix. Furthermore, it should be stressed that this is not a
symmtery of QCD but rather an artefact of the truncated chiral expansion (for a
different view, see Ref.[42]).	In any case, some theoretical arguments are
necessary to overcome this ambiguity if one wants to extract the corrections of
order $\m^2$ to the quark mass ratios.

Let us consider first the classical analysis of Gasser and Leutwyler [14]. The
chiral expansion of the pseudoscalar masses at next--to--leading order reads
$$\eqalign{
M_\pi^2 &= \krig M_\pi^2
\bigl\lbrace
1 + \mu_\pi - {1 \over 3} \mu_\eta + 2 \hat m
K_3 + K_4 \bigr\rbrace \, \, , \cr
M_K^2 &= \krig M_K^2 \bigl\lbrace 1 + {2 \over 3} \mu_\eta + ( \hat m + m_s )
K_3 + K_4 \bigr\rbrace \, \, , \cr
M_\eta^2 &= \krig M_\eta^2 \bigl\lbrace 1 + 2 \mu_K - {4 \over 3} \mu_\eta +
{2 \over 3} (\hat m + m_s) K_3 + K_4 \bigr\rbrace + K_5 + \krig M_\pi^2
\bigl\lbrace -\mu_\pi + {2 \over 3} \mu_K + {1 \over 3} \mu_\eta
\bigr\rbrace \, \, .
\cr} \eqno(3.8) $$
with
$$\eqalign{
&\mu_P = {M_P^2 \over 16 \pi^2 F^2} \ln ( M_P / \mu ) \, \, , \cr
&K_3 = 2 \kappa ( 2 L_8^r - L_5^r) \, \, , \, \, K_4 = 4 ( 2 \hat m + m_s )
\kappa ( 2 L_6^r - L_4^r) \, \, , \cr
&K_5 = {8 \over 9} \kappa^2 ( 3 L_7 + L_8^r) \, \, , \, \, K_6 =
\kappa	 L_5^r / F	\, \, , \cr
&K_7 = 2 \kappa ( 2 \hat m + m_s ) L_4^r / F \, \, , \, \, \kappa = 4 B / F
\, \, . \cr} \eqno(3.9) $$
neglecting terms of the order $m_u - m_d$. Inspired by the discussion of the
lowest order mass ratios, one forms the dimensionless quantities
$$\eqalign{
Q_1 &= {M_K^2 \over M_\pi^2} = {\hat m + m_s \over 2 \hat m} ( 1 + \Delta ) \,
\, \cr
Q_2 &= {(M_{K^0}^2 - M_{K^+}^2 )_{\rm strong} \over M_K^2 - M_\pi^2} =
{m_d - m_u \over m_s - \hat m} ( 1 + \Delta ) \, \, . \cr}
\eqno(3.10) $$
Remarkably, in both cases the correction $\Delta$ is the same ( for an explicit
expression, see ref.[14]). Therefore, up to corrections of order $\m^2$ one can
form a ratio independent of the $\len$ constants (which are hidden in the
explicit value of $\Delta$):
$$ \bigl({m_u \over m_d}\bigr)^2
+ {Q_2 \over Q_1} \bigl({m_s \over m_d}\bigr)^2 = 1
\eqno(3.11) $$
which defines an ellipsis with semi--axis 1 and $\sqrt {Q_2 / Q_1}$,
respectively. Making use of Dashen's theorem, this leads to $\sqrt {Q_2 / Q_1}
= 23.6$. Therefore, eq.(3.11) constitutes a low--energy theorem -- any value of
the mass ratios $m_u / m_d$ and $m_s / m_d$ must fulfill this elliptic
constraint (modulo corrections to Dashen's theorem, see below).  However, the
actual values of these ratios depend on the value of the correction $\Delta$.
This depends on the low--energy constants $L_5^r$ and $L_8^r$. Redefinitions of
$L_{5,8}^r$ allow one to move along the ellipse. In particular, one can choose
$m_u = 0$ which is favored in connection with the strong CP-problem [50]
(see also the instanton gas calculation reported in ref.[51]).
However, there are strong phenomenological constraints on the low--energy
constants.
First,
the value of $L_5^r$ is rather narrowly pinned down by the
coupling constant ratio $F_K / F_\pi =1.22 \pm 0.01$ [52] leading to the value
given in table 1  (it is also consistent with scalar resonance exchange for
$M_S
\simeq 1$ GeV). Concerning $L_8^r$, the situation is less favorable. Making use
of the deviation from the Gell-Mann--Okubo relation (cf. eq.(3.8)), which
involves the couplings $L_5^r, L_7$ and $L_8^r$ and is well known, Leutwyler
[53]
has proposed to estimate $L_7 $ via $\eta '$ exchange as given in eq(2.38).
$L_7$ also enters the determination of the $\eta \eta '$ mixing angle, which is
empirically constrained to $\Theta_{\eta \eta '} = (22 \pm 4)^\circ$
[54]. This allows
to fix $L_8^r$. The elliptic constraint together with the one from the $\eta
\eta '$ mixing leads to
$$ {m_u \over m_d} = 0.55 \pm 0.12 \, \, , \, \,
{m_s \over m_d} = 20   \pm 2.2	\, \, , \eqno(3.12) $$
rather consistent with Weinberg's estimate (3.5). A further check comes from
the baryon mass spectrum. Gasser [55] and Gasser and Leutwyler [3] have
analyzed the next--to--leadig order quark mass corrections which scale as
$\m^{3/2}$. Combining the isospin breaking mass splittings $m_p - m_n$,
$m_{\Sigma^+} - m_{\Sigma^-}$ and $m_{\Xi^0} - m_{\Xi^-}$ one arrives at
$$ R = {m_s - \hat m \over m_d - m_u}  = 43.5 \pm 3.2  \eqno(3.13) $$
This result for $R$ is also supported by examining the $\rho \omega$ splitting
at next--to-leading order [3]. Imposing this constraint on the results (3.12),
one ends up with
$$ {m_u \over m_d} = 0.56  \pm 0.06 \, \, , \,
\, \,  {m_s \over m_d} = 20  \pm 2 \, \, , \, \,
{\hat m \over m_s} = {1 \over 25.6 \pm 2.0} \eqno(3.14) $$
which shows that the meson spectrum (consistently with the baryon mass
splittings) determines the quark mass ratios rather accurately. This is a
consistent picture. However, what is definitively missing is a better treatment
of the corrections to Dashen's theorem.

Donoghue and Wyler [42] have taken a somewhat different path to get a handle on
the second
order corrections to the quark mass ratios. They were inspired by the
instanton gas calculation of Choi et al. [51]. In that calculation, in	which an
originally massless up quark travels through the instanton gas, it acquires an
effective mass,
$$ m_u^{\rm eff} \sim m_d m_s {\rm e}^{i \theta}     \eqno(3.15) $$
due to quantum effects	('t Hooft's six--fermion term which has been shown to
be of utmost importance in understanding the problem of flavor mixing in
theories with quark degrees of freedom [56]). Notice that the vacuum angle
$\theta$ enters the game. The $\theta$-dependence is the main ingredient in the
calculation of ref.[42] to overcome the Kaplan-Manohar ambiguity. The basic
observation is that the operator $G \tilde G$, which follows from $\lqcd$ after
variation by $\theta$, is immune to the reparametrization invariance of the
quark masses. However, since
one can not directly measure the $\theta$-dependence
from strong CP-violating effects, one has to resort to the multipole expansion
for heavy quark systems [38]. The decays $V \to V + M$ with $V = \Psi$ or
$\Upsilon$ and $M = \pi^0 , \eta, 3 \pi$ are determined by the operator $G
\tilde G$ to leading order in the heavy quark expansion. Taking ratios, one can
eliminate almost all model dependence. The strength of $G \tilde G$ can be
parametrized in terms of the number $r_{G \tilde G}$,
$$ \qquad
\eqalign{ &  r_{G \tilde G}  = { <0 | G \tilde G | \pi^0 > \over
<0 | G \tilde G | \eta >} = {3 \sqrt 3 \over 4} ( {m_d - m_u \over m_s - \hat m
}) {F_\eta \over F_\pi}   \cr
&\times \biggl\lbrace
 1 - {32 B \over F_\pi^2} (m_s - \hat m ) (L_7 + L_8^r ) +
{4 L_{14}^r \over F_\pi^2} ( M_\pi^2 - M_\eta^2 )  - ( 3 \mu_\pi - 2 \mu_K
- \mu_\eta ) \biggr\rbrace  \cr}		\eqno(3.16) $$
To arrive at this result, one uses $\partial \leff / \partial \theta \sim G
\tilde G$ and sandwiches the operator between the pertinent states. Clearly,
all the  terms in $\leff$ which contain $\tilde \chi = \chi \exp (i \theta / 3
)$ can contribute and there is also the term proportional to $L_{14}^r$ (cf.
eq.(2.50)) which is linear in $\theta$. The first term on the right hand side
of (3.16) agrees with the result of Novikov et al. [57]. One can furthermore
form one particular combination of the mass ratios which is free of chiral
logarithms (contained in the $\mu_P$) and only depends on one low--energy
constant [42]
$$ {m_d - m_u \over m_d + m_u} {m_s + \hat m \over m_s - \hat m} = { 4
( F_K^2 M_K^2 - F_\pi^2 M_\pi^2 ) \over 3 \sqrt{3} F_\pi^2 M_\pi^2} {F_\pi
\over F_\eta} r_{G \tilde G} [ 1 - \Delta_{GMO} ]
\biggl[1 + { 4 L_{14}^r \over
F\pi^2} ( M_\eta^2 - M_\pi^2 ) \biggr]
\eqno(3.17) $$
so that one has to determine $r_{G \tilde G}$ and $L_{14}^r$ to get this mass
ratio. The first quantity can be determined from
$$ {\Gamma (V' \to V \pi^0 ) \over \Gamma (V' \to V \eta )} = r_{G \tilde G}^2
\bigl( {p_\pi \over p_\eta} \bigr)^3	      \eqno(3.18) $$
with $p_\pi ( p_\eta )$ the momentum of the $\pi ( \eta )$. The data on the
decays $\Psi ' \to J/\Psi + \pi^0$ and $\Psi ' \to J/\Psi + \eta$ lead to
$r_{G \tilde G} = 0.043 \pm 0.0055$. Furthermore, one can use the principle of
resonance saturation to estimate $L_{14}^r$. The authors of ref.[42] give the
following limits: $0 \le 4 L_{14}^r / F_\pi^2 \le M_{\eta '}^2$. Combining
pieces and adding the errors (theoretical and empirical) in quadrature, this
leads to
$$ {m_d - m_u \over m_d + m_u} {m_s + \hat m \over m_s - \hat m} =
0.59 \pm 0.11	    \eqno(3.19)$$
or equivalently (since $m_s \gg \hat m$):
$$ {m_u \over m_d} = 0.30 \pm 0.07 \, \, \, \, .    \eqno(3.20) $$
This result gives further credit to the notion that $m_u$ is indeed unequal
zero, but its rather large deviation to the value given in (3.14) should be
noted. Such an effect can only occur if there are substantial corrections to
Dashen's theorem, much bigger than assumed in the calculation of Leutwyler
[53]. This will be discussed below. The estimate of Donoghue and Wyler can be
criticied on two grounds.
First, it corresponds to a value of $L_7$ which is not
in particular good agreement with phenomenology. Second, it is not clear that
the multipole expansion is justified in the $\Psi ' ( \Psi )$ system. A
measurement for the upsilon system would be very much needed to calrify the
uncertainties inflicted by the heavy quark expansion on the result (3.20).

Donoghue, Holstein and Wyler [58] have extended this latter analysis by
considering also the $\eta \to 3 \pi$ decay and a model to estimate the
corrections to Dashen's theorem (see also ref.[59] for an earlier attempt).
They consider the particular mass ratio
$$ R_1 = {m_d - m_u \over m_s - \hat m} {2 \hat m \over m_s + \hat m} =
\biggl({M_\pi \over M_K}\biggr)^2
{(M_{K^0}^2 - M_{K^+}^2)_{\rm strong} \over M_K^2 -
M_\pi^2}	   \eqno(3.21)	$$
which is related to the strong kaon mass difference and also appears in the
chiral expansion of the decay amplitude $\eta \to \pi^+ \pi^0 \pi^-$. The
latter has been worked out to one loop order by Gasser and Leutwyler [60]. The
matrix element for $\eta \to 3 \pi$ can be parametrized by one single function
of the pertinent Mandelstam variables, denominated $A(s,t,u)$. To leading
order, the chiral expansion of $A(s,t,u)$ reads
$$ A(s,t,u) = - {(m_d - m_u) B \over 3 \sqrt{3} F_\pi^2} \lbrace 1 + { 3 (s -
s_0) \over M_\eta^2 - M_\pi^2} + {\cal O}(q^2 ) + {\cal O}( e^2 {m_s - \hat m
\over m_u - m_d} ) \rbrace
\eqno(3.22) $$
with $s_0 = M_\pi^2 + M_\eta^2 / 3$ the center of the Dalitz plot. While
current algebra (tree level) predicts $\Gamma ( \eta \to \pi^+ \pi^0 \pi^- ) =
66$ eV, the one--loop calculation leads to $(160 \pm 50$) eV (the uncertainty
is an estimate of the higher order terms in the chiral expansion). Notice that
these values are in disagreement with the empirical result of ($ 281 \pm 29$)
eV.
The large enhancement of the one--loop result over the tree level
prediction is essentially an unitarity effect, i.e. an effect of the strong
pionic final state interactions in the $S$-wave. This was originally predicted
by Roiesnel and Truong [15] who found an even larger discrepancy (see the
discussion in ref.[60]). Notice also that these uncertainties essentially drop
out in the ratio $\Gamma ( \eta \to 3 \pi^0 )
/ \Gamma ( \eta \to \pi^+ \pi^0 \pi^-
) = 1.51$, $1.43$ for current algebra and one loop CHPT [60], respectively. The
emiprical value is $1.35 \pm 0.04$, comparable to the one loop prediction. The
authors of
ref.[58],
however, turn the argument around. They use the empirical
rate for $\eta \to \pi^+ \pi^0 \pi^-  $ to fix the strong kaon mass difference
since $\Gamma ( \eta \to \pi^+ \pi^0 \pi^- )$ fixes $R_1$ (3.21). This gives
$$ ( \Delta M_K^2 )_{\rm strong} = ( M_{K^0}^2 - M_{K^+}^2)_{\rm strong} = 2
M_K \cdot (7.0 \, {\rm MeV}) \, \, \, \, .  \eqno(3.23)  $$
This is considerably different from the result based on Dashen's theorem,
$ ( \Delta M_K^2 )_{\rm strong} = 2 M_K \cdot (5.3 \,$MeV) (a 30 per cent
deviation).  To further support this argument, they also consider a model to
independently estimate the corrections to Dashen's theorem. It is based on the
empirical observation that vector meson dominated form factors in the Born
diagrams give a rather accurate description of the  electromagnetic pion mass
difference [16,32,33] (see fig.3). Beyond leading order and making use of the
Weinberg sum rules (which is necessary to cancel the divergent pieces) this
leads to
\midinsert
\vskip 6.0truecm
{\noindent\narrower \it Fig.~3:\quad
Electromagnetic self--energy of pseudoscalars to order
$e^2$ in the resonance exchange picture.
The wiggly lines denote photons, the double solid lines vector or
axial mesons. Tadpole diagrams are not shown.
\smallskip}
\endinsert
$$ \Delta M_\pi^2 = 2 M_\pi \cdot (6.3 \, {\rm MeV})	    \eqno(3.24)  $$
using the physical mass of the $A_1$ meson. The result (3.24) is considerably
larger than the empirical value $ \Delta M_\pi^2 = 2 M_\pi \cdot (4.6 \, {\rm
MeV})$. The lowest order result, however, is within a few per cent of the
empirical value, $ \Delta M_\pi^2
= 2 M_\pi \cdot (4.7 \, {\rm MeV})$ [32]. This
sheds some doubt on the accuracy of this approach.  To leading order, the
formalism used to arrive at (3.24) respects Dashen's theorem but predicts a
strong violation of it	at next--to--leading order, $( \Delta M_K^2 )_{\rm em}
= 1.78 \, ( \Delta M_\pi^2 )_{\rm em}$. This large deviation from one is
essentially due to
the kaon propagator in the Born terms (cf. fig.3) and thus
the strong kaon mass difference is affected,
$$ ( \Delta M_K^2 )_{\rm strong} =  2 M_K \cdot
(6.3 \pm 0.1 \, {\rm MeV})
\eqno(3.25) $$
which agrees within  10 per cent with the number (3.23) extracted from the
$\eta \to 3 \pi$ decay. Assuming furthermore resonance saturation for the
low--energy constants $L_7$ and $L_{14}^r$ ($\eta '$ exchange), the authors of
ref.[58] can also determine the mass ratios $ 2 \hat m / (\hat m + m_s)$ and
$(m_d -m_u) (m_s + \hat m) / (m_d + m_u) (m_s + \hat m )$ (from the meson
masses and the decays $ \Psi ' \to J/\Psi + \pi , \eta$)
and arrive at
$$ {\hat m \over m_s} = {1 \over 31} \, \, , \, \,
{m_d - m_u \over m_s} = {1 \over 29} \, \, , \, \,
{m_d - m_u \over m_d + m_u} = 0.59 \, \, ,	     \eqno(3.26) $$
or  $$ {m_u \over m_d} = 0.26 \, \, \, .	     \eqno(3.27) $$
This value for the ratio of the up and down quark masses is consistent with
the one given in (3.20). While this set of mass ratios has the virtue of giving
a correct $\eta \to 3 \pi$ amplitude, which is a notorious problem in CHPT,
there are various aspects which have to be looked at in more detail. First,
from (3.26) one deduces $R \simeq 28$ which is four standard deviations off the
value extracted from the baryon spectrum (see also the discussion at the end of
this section). Furthermore, it is not obvious that the one--loop result for the
amplitude $A( \eta \to 3 \pi )$ should be used for fixing the mass ratio $R_1$.
As we will see from the discussion of the scalar form factor in section 4,
strong pionic final state interactions cause the failure of the one loop
approximation close to two--pion threshold. For $\eta \to 3 \pi$, one deals
with $\sqrt s \approx 550$ MeV. Finally, the large next--to--leading order
corrections to the pion mass difference put a question mark on the convergence
of this approach.

{}From the discussion so far it should have become clear that the calculations
of
the quark mass ratios at next--to--leading order involve some model-dependence.
It is therefore necessary to investigate as  many constraints as possible
(meson masses, baryon masses, $\eta \to 3 \pi$, corrections to Dashen's theorem
and alike) to arrive at a
consistent picture of the quark mass ratios at order
$\m^2$. If the modifications to $m_u / m_d$ and $\hat m / m_s$ are indeed as
large as indicated by (3.20) or (3.26), it is not obvious why one should stop
at next--to--leading order. The set of quark mass ratios (3.14) has the virtue
of showing stability against inclusion of the quark masses as perturbations. At
present, I consider it as the most consistent estimation of the light quark
mass ratios.

There is some additional evidence which gives further credit to the constraint
from the baryon sector, i.e. the value of $R$ (3.13). This work is based on
so-called second order "flavor" perturbation theory [61]. It extends the work
of Gell-Mann, Okubo and others [17]. It differs from  CHPT in that the flavor
symmetric part of the quark mass term $(m_u + m_d + m_s)  ( \bar u u + \bar d d
+ \bar s s) / 3$ is already contained in the unperturbed Hamiltonian. The basic
idea is that in this way one includes already most of the (large) $\m^{3/2}$
corrections of CHPT and thus deals with smaller perturbations. If one works out
the flavor perturbations to second order, one finds [61]:
$$ R = { 3 m_\Lambda + m_\Sigma - 2 m_N - 2 m_\Xi \over 2 \sqrt{3} m_T + (m_n -
m_p) + (m_{\Xi^0} - m_{\Xi^-} ) } \, \, \, \, .    \eqno(3.28) $$
Here, $m_T$ is the $\Lambda$--$\Sigma^0$ transition mass. Its physical effect
is that it produces a small mixing $\epsilon$ of the mass--diagonal fields,
$\epsilon = m_T / ( m_\Sigma - m_\Lambda )$. Empirically, it is not well
determined. It could be pinned down from a precision measurement of the
difference in the $p K^- \to \Lambda \eta$ and $n \bar{K}^0 \to \Lambda \eta$
cross sections. At present, one can only deduce a (conservative) lower limit of
$R \ge 38 \pm 11$. A similar result can also be obtained from the splittings in
the decuplet. Finally, if one assumes that the first order result lies as close
as possible to the family of second order solutions, one finds $R = 48 \pm 5$,
quite consistent with the value given in eq.(3.13). It is certainly desirable
to have such extra information to strengthen our understanding of the ratios of
the light quark masses.
Some arguments which seem to support the notion of a much smaller ratio
$m_s / \hat m$ are discussed in section 4.2.
\bigskip
\noindent {\bf IV. THE MESON SECTOR -- SELECTED TOPICS}
\medskip
In this section, I discuss a few selected topics of applying CHPT in the meson
sector. Clearly, space forbids to account for less than a small fraction of the
many existing predictions and their comparison to the data. In section 8,
additional references concerning some of the neglected topics are given.

Here, I will first discuss the classical topic of pion--pion scattering within
the one loop approximation. This is an important topic since, as will be
discussed, it allows to directly test our understanding of the mechanism of
quark condensation in the vacuum. I also discuss briefly the related subject of
$\pi K$ scattering. Then, the question of the range of applicability of the
chiral expansion is discussed. Various schemes to extend the energy range are
critically reviewed. I also sketch the extenstion of the chiral Lagrangian at
next--to--leading order to include the non--leptonic weak interactions with
particular emphasis on the decay modes $K \to 2 \pi$ and $K \to 3 \pi$.
\bigskip
\noindent {\bf 4.1. $\pi \pi$ scattering}
\medskip
The scattering of pions is in a sense the purest reaction to test our
understanding of the low energy sector of QCD. It involves only the
pseudoscalar Goldstone bosons and their dynamics. Also, for the reaction $ \pi
\pi \to \pi \pi$ one can restrict oneself to the sector of $SU(2) \times
SU(2)$, where the quark mass corrections are expected to be very small. To be
more specific, consider the process $\pi^a (p_a) +  \pi^b (p_b) \to \pi^c (p_c)
+ \pi^d (p_d) $, for pions of isospin '$a,b,c,d$' and momenta $p_{a,b,c,d}$.
The conventional Mandelstam variables are defined via $s = (p_a + p_b)^2$, $ t
= (p_a - p_c )^2$ and $u = (p_a - p_d )^2$ subject to the constraint $s+t+u= 4
M_\pi^2$. The scattering amplitudes can be expressed in terms of a single
function, denoted $A(s,t,u)$:
$$ T^{cd;ab} (s,t,u) = A(s,t,u) \, \delta^{ab} \delta^{cd} \, + \,
		       A(t,s,u) \, \delta^{ac} \delta^{bd} \, + \,
		       A(u,t,s) \, \delta^{ad} \delta^{bc} \, \, \,
\eqno(4.1) $$
for the two flavor case. The chiral expansion of $A(s,t,u)$ takes the form
$$ A(s,t,u) = A^{(2)} (s,t,u) +
	      A^{(4)} (s,t,u) + {\cal O} (q^6 )  \, \, \, , \eqno(4.2) $$
where $A^{(m)}$ is of order $q^m$ and the symbol ${\cal O} (q^6 )$ denotes
terms like $s^3, s^2 t, s t^2 , \ldots$. The explicit form of the tree level
amplitude was first given by Weinberg [62]. It can be easily read off from
eq.(2.9) by expanding in powers of the pion field and collecting the terms
proportional to $\pi^4$,
$$  A^{(2)} (s,t,u) = {s - M_\pi^2 \over F_\pi^2} +  {\cal O} (q^4) \, \, \, ,
\eqno(4.3) $$
where it is legitimate to use the physical values of the pion mass and decay
constant since the differences to their lowest order values is of order $q^4$.
The next-to--leading order term has been worked out by Gasser and Leutwyler
[13,63] (cf. fig.2). One splits it into unitarity corrections (denoted
$B(s,t,u)$) and tree and tadpole contributions (denoted $C(s,t,u)$). The
explicit form reads (I use the one given in ref.[64]):
$$\eqalign{
A^{(4)} (s,t,u) =&B (s,t,u) + C (s,t,u) 	   \cr
B(s,t,u) =&{1 \over 6 F_\pi^4} \bigl [ 3 (s - M_\pi^4) \bar{J} (s) +
\bigl\lbrace [ t (t-u) - 2 M_\pi^2 t + 4 M_\pi^2 u - 2 M_\pi^4 ] \bar{J} (t) +
( t \leftrightarrow u ) \bigr\rbrace \bigr]	    \cr
C(s,t,u) =& {1 \over 96 \pi^2 F_\pi^4} \bigl\lbrace 2 ( \bar{\ell}_1 -	4/3 )
(s - 2 M_\pi^2 )^2 + ( \bar{\ell}_2 - 5/6)[ s^2 + (t-u)^2] \cr
& + 12 M_\pi^2 s
(\bar{\ell}_4 -1) - 3 M_\pi^4 ( \bar{\ell}_3 + 4 \bar{\ell}_4 - 5 )
\bigr\rbrace	  \cr
  \bar{J} (z) =&- {1 \over 16 \pi^2} \int_0^1 dx \ln [ 1 - z x (x-1)/ M_\pi^2 ]
\, \, \, .    \cr}   \eqno(4.4) $$
In the chiral limit, one recovers the form of $A(s,t,u)$ first given by Lehmann
[65].  Unitarity and analyticity force the appearance of the loop terms which
contain two unknown scales. These can be expressed in terms of the low energy
constants $\bar{\ell}_{1,2}$ [13]. The polynomial term $C(s,t,u)$ contains two
further constants related to the shift of $F_\pi$ and $M_\pi$ away from their
lowest order values. An explicit expression of $A^{(4)} (s,t,u)$ in terms of
the $\su3$ representation can be found in ref.[66].

For comparison with the data, one decomposes $T^{cd;ab}$ into amplitudes of
definite total isospin ($I=0,1,2$) and projects out partial--wave amplitudes
$T_l^I (s)$,
$$ T_l^I (s) = {\sqrt{1 - 4 M_\pi^2 / s} \over 2i} \bigl[ \exp \bigl\lbrace
2 i [ \delta_l^I (s) + i \eta^I_l (s) ] \bigr\rbrace - 1 \bigr]   \eqno(4.5) $$
with $s = 4 ( M_\pi^2 + q^2 )$ in the c.m system, $l$ denotes the total angular
momentum of the two--pion system. The phase shifts $\delta_l^I (s)$ are real
and the inelasticities $\eta_l^I (s)$ set in at four--pion threshold. They are,
however, negligible below $\bar K K$ threshold (below which they are a
three--loop effect of order $q^8$). Near threshold ($s = 4 M_\pi^2$) the
partial--wave amplitudes take the form
$$ {\rm Re} \, T_l^I =	q^{2l} \lbrace a_l^I + q^2 \, b_l^I + {\cal O}(q^4)
\rbrace \, \, \, . \eqno(4.6) $$
The coefficients $a_l^I$ are called scattering lengths, the $b_l^I$ are the
range parameters. It was already observed by Weinberg [62] that the scattering
lengths resulting from the tree level were much smaller than naively expected.
This is also borne out in the one loop calculation [13]. Let me now concentrate
on the isospin--zero $S$--wave scattering length $a_0^0$. The tree level
prediction is $a_0^0 = 7 M_\pi^2 / 32 \pi F_\pi^2 = 0.16$ [62] and to one loop
order, one can derive the following low energy theorem [13],
$$a_0^0 = { 7 M_\pi^2 \over 32 \pi F_\pi^2} \biggl\lbrace 1 + {1 \over 3}
M_\pi^2 <r^2>_S^\pi - { M_\pi^2 \over 672 \pi^2 F_\pi^2} (15 \bar{\ell}_3 -
353) + {25 \over 4} M_\pi^4 ( a_2^0 + 2 a_2^2 ) + {\cal O} (q^6) \biggr\rbrace
\eqno(4.7) $$
which involves the scalar radius of the pion, $<r^2>_S^\pi \simeq 0.7$ fm$^2$
(see also section 4.4), the low energy constant $\bar{\ell}_3$ and the
$D$--wave scattering lenghts $a_2^0$ and $a_2^2$. Taking the latter from
Petersen [67] and allowing for conservative errors on the scalar radius and
$\bar{\ell}_3$, one arrives at [13,63]
$$ a_0^0 = 0.20 \pm 0.01 \, \, \, \, .	 \eqno(4.8)  $$
One makes two important observations. First, there is a 25 per cent correction
to the tree result (which is due to the fact that the non--analytic term of the
type $M_\pi^2 \ln M_\pi^2$ has a large coefficient). Second, the theoretical
uncertainty is rather small. This is mostly due to the fact that the poorly
known  $\bar{\ell}_3$ is multiplied by a tiny factor. We will come back to this
important result in section 4.2. For a more detailed account of the threshold
behaviour of the other partial waves, see refs.[13,63].

\midinsert
\vskip 20.0truecm
{\noindent\narrower \it Fig.~4:\quad
$\pi \pi$ scattering phase shifts $\delta_0^0$ (a),
$\delta_1^1$ (b) and $\delta_0^2$ (c), in order.
The dashed line gives the tree result and the dashed--dotted  the
one--loop prediction. Also shown is the Roy equation band as discussed
in the text. The data can be traced back from ref.[64].
In (a), the double--dashed line corresponds to the one--loop result
based on eq.(4.10).
\smallskip}
\endinsert
If one now increases the energy, there are two main issues to address. First,
how does the truncated chiral series (4.2) compare with the data and, second,
at what energy do the one--loop corrections become so large that one loop
approximation can not be trusted any more? Before addressing these issues, I
have to discuss an ambiguity which arises in the extraction of the phase shifts
from eq.(3.6). One can either use [64]
$$ \delta_l^I = {\rm Re} \, T_l^{I(2)} + {\rm Re} \, T_l^{I(4)} +
{\cal O} (q^6)			\eqno(4.9)  $$
or
$$ \delta_l^I = {\rm arctan} \, ( {\rm Re} \, T_l^I ) +
{\cal O} (q^6) \, \, \, \, .   \eqno(4.10) $$
The difference between these two forms only shows up at order $q^6$ and
therefore gives an estimate about the relative importance of these higher order
terms (in what follows, I will mostly use the definition (4.9)). In fig.4a,b,c
the phase shifts $\delta_0^0 (s)$, $\delta_1^1 (s)$ and $\delta_0^2 (s)$ are
shown, respectively [64]. For comparison, the tree result and the existing data
are shown together with a band of Roy equation fits [68,69] (the latter impose
unitarity and analyticity and, obviously, some of the data are in conflict with
these fundamental principles). At the time the Roy equation program was carried
out [67,69], the $S$--wave scattering length, which is a fundamental input, was
not known.  The band indicated in the figures corresponds to $0.17 \le a_0^0
\le 0.30$. Before discussing the results, let me point out that for
$\delta_0^0$ one has additional information in the threshold region from
$K_{\ell 4}$ decays. The energy where the one--loop contribution
$\delta_l^{I(4)}$ is half as big as the tree level prediction (in magnitude) is
indicated by the shaded vertical line. Notice that this critical energy is
different for the various channels, $\sqrt s = 450$ MeV, 480 MeV and 470 MeV,
in order. Beyond these energies, the truncated chiral expansion becomes
unreliable. In the isospin--zero $S$--wave (fig.4a), the data below $\sqrt s =
600$ MeV are poor (with the exception of the ones from $K_{\ell 4}$ decays).
Notice that both representations (4.9) and (4.10) stay within the Roy equation
band up to the critical energy. Below 400 MeV, they are virtually identical. In
the $P$--wave, the well--known $\rho$ resonance shrinks the Roy equation band
to a narrow line. The chiral prediction follows it up to 500 MeV. Clearly, it
is not possible to describe the resonance behaviour by this method. This is the
natural barrier to the theory of pseudoscalars only already mentioned before.
In the exotic $S$--wave ($I=2$),
the tree and one--loop contributions are of
opposite sign. This leads to a stronger sensitivity of the chiral predictions
to the actual values of the low--energy constants (see the shaded area in
fig.4c, which corresponds to a change of $\ell_4$ by one unit. The
central value of $\ell_4$ gives the lower rim of this area.)

It was noted in ref.[64] that the phase of the parameter $\epsilon '$, which
measures direct CP--violation in the kaon system, can nevertheless be extracted
with rather good accuracy (see also refs.[70,71]). It is defined via [72]
$$\Phi (\epsilon ') = (\delta_0^2 - \delta_0^0) \biggl|_{s=M_{K^0}^2} \, + \,
{\pi \over 2}	   \eqno(4.11) $$
(see also section 4.5). One therefore needs the $S$--wave phase shifts at
$\sqrt s = M_{K^0} = 494$ MeV, i.e. at an energy where the one--loop
coorections are already large. However, in the difference these corrections
largely cancel and one finds
$$\Phi (\epsilon ') = (45 \pm 6)^\circ \, \, \, \, ,   \eqno(4.12) $$
to be compared with the tree result of $53^\circ$. The uncertainty of $6^\circ$
is a combination of the uncertainties in the low energy constants and the
higher order contributions. Ochs [73] has recently reanalyzed results of phase
shifts	from $\pi^- p \to n \pi^- \pi^+$ for energies above 600 MeV. Using Roy
equation methods, he finds $\Phi (\epsilon ') = (43 \pm 6)^\circ$, in good
agreemnet with the chiral prediction. This is an important result since it
essentially pins down this parameter. Attempts to extract
$\Phi (\epsilon ')$ from the various $K \to 2 \pi $ modes are hampered by the
fact that there are large corrections from isospin breaking which have not yet
been investigated in full detail (some references can be traced back from
ref.[64]).

Donoghue et al. [74] have taken a different point of view on the one--loop
results. They used the available data on the $S,P,D$ partial--wave amplitudes
in the range of energies from threshold to 1 GeV to determine the low energy
constants $\bar{\ell}_{1,2}$ by a best global fit to these data. This leads to
values which are somewhat different form the standard ones [13,14]. However,
from the discussion before it should be clear that such a fit is not quite
legitimate since the one--loop corrections are too large beyond 500 MeV. While
the authors of ref.[74] find a generally satisfactory description of the data
up $700 \ldots 800$ MeV, it is not clear what one learns from such a
comparison. Notice also that the partial--wave amplitudes are bounded in
magnitude by unitarity and therefore a much less sensitive probe of deviations
than the phase shifts.

There is furthermore a whole set of articles in which the one--loop results are
extended to higher energies based on unitarization schemes or explicit
resonance exchanges. These proposals will be discussed in section 4.4. Let me
now return to the importance of the $S$--wave scattering lengths.
\bigskip
\noindent {\bf 4.2. Testing the mode of quark condensation}
\medskip
So far, we have assumed the standard scenario in which	$<0 | \bar u u | 0>$ is
the order parameter of the spontaneous chiral symmetry breaking and the
constant $ B = -<0 | \bar u u | 0> / F^2$  is so large ($\sim 1$ GeV) that the
first term in the quark mass expansion of the pseudoscalar masses ($B \m$)
dominates (this is also the point of view I subscribe, it is backed by lots of
phenomenological and some theoretical arguments, see below). However, this
notion has been challenged already in the seventies [75,76]. It was claimed
that the possibility $M_P^2 \sim \m^2$ fits the data as well (in that case,
$B=0$ in the chiral limit). Some recent work along these lines has been done by
Fuchs et al. [77], who argued that the Goldberger--Treiman discrepancies in the
baryon octet (the deviations from the GT relations, which are exact in the
chiral limit) lead to an upper bound for the quark mass ratio $m_s / \hat m \le
10.6 \pm 4.2$. This is more than 3 standard deviations off the standard value
(3.14) (for earlier work on this problem, see the review by Dominguez [78]).
The same authors have recently proposed a modified CHPT [79] to account for
this. Their starting point is the claim that a scenario in which $B$ is small,
$ B \sim \Lambda_{\rm QCD} \sim F_\pi \sim 100 \ldots 200$ MeV is as natural as
the large $B$ case. Then, the ratio $m_s / B$ is of order one and one has to
modify the chiral power counting. Apart from the small external momenta, one
has to consider the quark mass matrix $\m$ and the parameter $B$ simultaneously
as small quantities, with the ratio $m_s / B$ kept fixed. It is then natural to
assign the external field $\chi$ the dimension one. The effective Lagrangian
takes the form:
$$ \leff = \sum_{l,m,n} B^l \, {\cal L}^l_{m,n} = \sum_N {\cal L}^{(N)}
\, \, \, ,    \eqno(4.13)  $$
with $m$ covariant derivatives, $n$ quark mass insertions and $l$ factors of
$B$ so that $l+m+n = N$ (with $l,m,n$ integers). In the standard case, $m +2n =
N$. This modifies the leading term of order $q^2$ and new terms at order $q^3$
appear (for explicit expressions, see ref.[79]). The leading non--analytic
contributions are still of order $q^4$, i.e. at next--to--next--to leading
order in the modified chiral expansion.

The best place to test this scenario is indeed $\pi \pi $ scattering in the
threshold region. Remember that the $S$--wave scattering lengths vanish in the
chiral limit and therefore directly  measure the symmetry breaking of QCD. This
is why these scattering lengths are of such fundamental importance. In the
modified CHPT, the $\pi \pi$ scattering amplitude reads (up to and including
terms of order $q^3$)
$$ A(s,t,u) = \alpha  {M_\pi^2 \over 3 F_\pi^2} + \beta { s - 4 M_\pi^2 / 3
\over F_\pi^2} \, \, \, \, . \eqno(4.14)    $$
In the standard scenario, $\alpha = \beta = 1$ and one recovers eq.(4.3). In
the extreme case that $B$ vanishes in the chiral limit, one has $\alpha = 4$
and $\beta = 1$. In general, one can interpolate between these two cases via
(to leading order)
$$ \alpha^{(2)} = 1 + 6 {(2 M_K^2 / M_\pi^2 -1) - (m_s / \hat m ) \over (m_s /
\hat{m} )^2 -1 } \, (1 + 2 \xi ) \, \, ; \, \, \beta^{(2)} = 1 \, \, \, ,
\eqno(4.15) $$
where $\xi$ accounts for the Zweig rule violation in the $0^{++}$ channel and
is expected to be small ($\xi = 0$ in what follows). For the value $m_s /
\hat{m} = 10$ advocated in ref.[77], one has $\alpha^{(2)} = 1.96$. This leads
to $a_0^0 = 0.19$ and $a_0^2 = -0.031$ to be compared with the Weinberg result
of $a_0^0 = 0.16$ and $a_0^2 = -0.045$. The effect is maximal in the
combination $a_0^0 + 2 a_0^2$ [79]. However, one also has to know the
corrections of order $q^3$ and $q^4$ to these predictions. They have not yet
been worked out in detail. The terms of order $q^3$ lead to a small increase in
$\beta$ and the non--analytic contributions  are not known (see below for an
estimate). It is, however, clear that this scenario could lead to large value
of $a_0^0$, say $a_0^0 = 0.27$ ($m_s / \hat{m} = 6$
at tree level). Such a value
could not be accomodated by the standard scenario, cf. eq.(4.8).
At present, the
data are not accurate enough to differentiate between these two cases. Pion
production experiments as recently analyzed by Ochs [73] lead to
$a_0^0 = 0.23 \pm 0.08$  and
the $K_{\ell 4}$ data of Rosselet
et al. [80]
tied togther with Roy equation constraints  lead to $a_0^0 = 0.26 \pm 0.05$
[67]. Therefore, a better determination of the $S$--wave scattering lengths is
urgently called for. In fact, recent work at Serpuhkov [81] has established the
existence of short-lived $\pi^+ \pi^-$ atoms. In case they decay from the
ground state, their life time is directly proportional to $(a_0^0 - a_0^2)^2$
[82]. Therefore, a measurement of the life time with an accuracy of 10 per cent
would pin down this combination of the $S$--wave scattering lenghts within 5
per cent. A letter of intent for such an experiment at CERN has been presented
[83]. In case this experiment is approved and reaches the projected accuracy,
it will give strong indications about which value of $B$ is favored. Also, from
the $\Phi$  factory ${\rm DA\Phi NE}$
at Frascati one will get very much improved
statistics on $K_{\ell4}$ decays and thus better information on the phase
$\delta^0_0$ in the threshold region.

Finally, we should mention that Crewther [84] has already investigated this
problem some time ago. He allowed for a general case $M_\pi^2 \sim \hat{m}^d$,
with $d$ an integer. $d=1$ gives the standard scenario and $d=2$ the extreme
small $B$ case. Taking into account the leading non--analytic terms (chiral
logarithms) with a scale of $\mu = 800^{+800}_{-400}$ MeV, he compared the
predictions for the $\pi \pi$ threshold  parameters with the available data.
Although neither $d=1$ nor $d=2$ describes the data very well, a slight
preference for the standard case is found. This result should only be
considered indicative,	a complete one--loop calculation in the modified CHPT
scheme has to be performed so that one has predictions of comparable accuracy
to the ones already existing for the standard case (like eq.(4.8)). It is my
believe that when the dust settles, the small $B$ alternative will be excluded.
Also, the Gell-Mann--Okubo relation for the pseudoscalar masses does not follow
naturally in the modified scheme, it requires some parameter fitting.
This is a rather unappealling
feature. Furthermore, there are indications from lattice
gauge calculations supporting the GMOR scenario [85].
\bigskip
\noindent {\bf 4.3. $\pi K$ scattering}
\medskip
The scattering process $\pi K \to \pi K$ is of particular interest because it
is the simplest reaction of the Goldstone bosons involving strangeness and
unequal meson (quark) masses. Furthermore, since the low energy constants are
already
determined from other processes, $\pi K$ scattering can serve as a test
of the large $N_c$ predictions for some of these couplings. It also gives an
idea about our understanding of the symmetry breaking in the strange sector.
Obviously, corrections to the tree results are expected to be larger than in
the two flavor case since the small expansion parameter $(M_\pi / 4 \pi F_\pi
)^2 = 0.014$ is substituted by $(M_K / 4 \pi F_\pi )^2 = 0.18$.

First, some kinematics has to be discussed. In the $s$ channel, there are two
independent amplitudes with total isospin $I = 1/2$ and $I = 3/2$. The latter
is given by the specific process $\pi^+ K^+ \to \pi^+  K^+$, i.e.
$$ T_{\pi K}^{3/2} (s,t,u) = T (\pi^+ (p_1) K^+(p_2) \to \pi^+ (p_3)  K^+(p_4)
) \, \, \, \, , \eqno(4.16) $$
with $s+t+u = 2 (M_\pi^2 + M_K^2 )$. Using crossing symmetry, the isospin--1/2
amplitude follows via [86]
$$ T_{\pi K}^{1/2} (s,t,u) = {3 \over 2} T_{\pi K}^{3/2} (u,t,s)
- {1 \over 2} T_{\pi K}^{3/2} (s,t,u) \, \, \, \, .  \eqno(4.17) $$
The one--loop result for $T_{\pi K}^{3/2} (s,t,u)$ has been given in
refs.[66,86],
$$\eqalign{
T^{3/2} (s,t,u) &=  T_2 (s,t,u) + T^T_4 (s,t,u) + T^P_4 (s,t,u) +
T^U_4 (s,t,u)	\cr
T_2 (s,t,u) &= {1 \over 2 F_\pi^2} ( M_\pi^2 + M_K^2 - s )	  \cr
T_4^T (s,t,u)&= {1 \over 16 F_\pi^2}   \left\lbrace  \mu_\pi
 [10s - 7M_\pi^2 -13M_K^2] + \mu_K [2M_\pi^2 + 6 M_K^2 -4s] \right. \cr
& \left. + \mu_\eta
[5 M_\pi^2 + 7 M_K^2 -6s] \right\rbrace \cr
T_4^P (s,t,u)
& ={2 \over F_\pi^2 F_K^2 }\left\lbrace 4L_1^r (t-2M_\pi^2)
(t-2M_K^2)
\right. \cr
&\left.
+ 2 L_2^r \bigl[ (s - M_\pi^2 - M_K^2)^2
		 + (u - M_\pi^2 - M_K^2)^2 \bigr]
\right. \cr
&\left.
  + L_3^r  \bigl[ (u - M_\pi^2 - M_K^2)^2
		 + (t - 2 M_\pi^2) (t - 2 M_K^2)   \bigr]
\right. \cr
&\left.
  +  4 L_4^r \bigl[ t(M_\pi^2 + M_K^2) - 4M_\pi^2 M_K^2
\bigr]
\right. \cr
&\left.
  +  2L_5^r M_\pi^2 (M_\pi^2 - M_K^2 - s)
 + 8 (2L_6^r +L_8^r) M_\pi^2 M_K^2
\right\rbrace
\cr} $$
$$\eqalign{
T_4^U (s,t,u)&=\cr
&{1 \over 4 F_\pi^2 F_K^2 }\left\lbrace t(u-s)
\left[ 2M_{\pi \pi}^r(t) + M_{KK}^2(t)
\right]
 + {3 \over2}
\left[(s-t)
\left\lbrace L_{\pi K}(u) + L_{K \eta}(u) \right.
\right. \right. \cr
& \left. \left. \left.
- u(M_{\pi K}^r(u)
+ M_{K \eta}^r(u))\right\rbrace
+  (M_K^2 - M_\pi^2)   (M_{\pi K}^r(u) + M_{K \eta}^r(u))
\right] \right. \cr
& \left. + { 1 \over 2} (M_K^2 - M_\pi^2)
\left[
K_{\pi K} (u)
(5u - 2M_\pi^2
 - 2 M_K^2)
 + K_{K \eta} (u)
(3u - 2 M_\pi^2
 - 2M_K^2) \right]
\right. \cr
& \left.
+ J_{\pi K}^r (s)  ( s - M_\pi^2 - M_K^2)^2
+ {1 \over 8}J_{\pi K}^r (u)
\bigl[ 11u^2
- 12 u ( M_\pi^2 + M_K^2)
+ 4 (M_\pi^2 + M_K^2)^2 \bigr]
\right. \cr
&\left.
+ { 3 \over 8}	J_{K \eta}^r (u)
\left( u - {2 \over 3} (M_\pi^2 + M_K^2)\right)^2
 + { 1 \over 2} J_{\pi \pi}^r (t) \,\,t\,\, ( 2t - M_\pi^2)
\right. \cr
&\left.
+ {3 \over  4} J_{K K}^r (t) t^2
+ { 1 \over 2} J_{\eta \eta}^r (t)
M_\pi^2
(t - { 8 \over 9} M_K^2)
 \right\rbrace	  \cr}
\eqno(4.18) $$
The explicit form of the functions $M_{PQ}^r$ and $J_{PQ}^r$ is given in [14].
The tree level result agrees with the one of Weinberg [62] and Griffith [87].
Notice that the amplitude dependes, like the $\pi \pi$ amplitude (4.4), on the
low energy constants $L^r_{1,2,3,4}$ but also on $L^r_{5,6,8}$.

\midinsert
\vskip 10.0truecm
{\noindent\narrower \it Fig.~5:\quad
Theoretical predictions and empirical data on the $S$--wave scattering
lengths. We show the current algebra point (CA) and the one--loop
chiral perturbation theory result (CHPT). The data can be traced back
from ref.[66].
\smallskip}
\endinsert
In fig.5, the one--loop prediction for the $S$--wave scattering lengths
$a_0^{1/2}$ and $a_0^{3/2}$ is shown in comparison with the data (see
ref.[66]), the result of a Roy equation study [88] and a dispersive
analysis [89].
The uncertainty in the prediction comes from the uncertainties of
the $L^r_i$ added in quadrature (which is a conservative estimate).  It is
clear from that figure that a better determination of these scattering lengths
would  be very much needed. At energies above threshold, the bulk of the data
for the phase shift $\delta_0^{1/2} (s)$ is described fairly well
up to 850 MeV (at
800 MeV, the one-loop contribution is half as large as the one from the tree
level). For the $I=3/2$ phase, only few very inaccurate data exist in the
threshold region. Similar to the $I=2$ $\pi \pi$ $S$--wave, the tree and
one--loop contributions cancel leading to a strong sensitivity to the values of
the low energy constants. In the $P$--wave, the nearby $K^* (892)$ resonance
forces the one--loop prediction to fail rather quickly. It should be pointed
out that the threshold of $\pi K$ scattering is at 633 MeV and therefore the
loop corrections can become large rather quickly. It was, however, stressed in
ref.[66] that one can continue the amplitudes into the unphysical region and
expand around the point $\nu = (s-u) / M_K = t = 0$. This leads to a
satisfactory agreement with existing dispersion--theoretical results [90].
Also, the isospin--even amplitude $T^{(+)} (\nu , t)$ is most sensitive to the
actual values of the $L_i^r$. A better determination of it would allow to test
some of the large $N_c$ predictions for these couplings. In any case, a more
accurate determination of the $\pi K$ scattering amplitudes would serve as a
good test of our understanding of the symmetry breaking in the $\su3$ sector of
QCD.
\bigskip
\noindent {\bf 4.4. Beyond one loop}
\medskip
Up to now, we have considered CHPT calculations in the one--loop approximation.
In the two flavor case, higher order corrections are generally small at low
energies because the expansion parameters are $M_\pi^2 / 16 \pi^2 F_\pi^2 =
0.014$ and $E_{\rm pion}^2 / F_\pi^2$. Furthermore, the amount of work to
perform a multi--loop calculation is substantial and no systematic study of the
accompanying low energy constants is available. As particularly stressed by
Truong and collaborators [91,92] pions in the isospin--zero $S$--state produce
strong final state interactions. As a result of this, the one--loop
approximation to e.g. the scalar form factor of the pion or to the phase shift
$\delta_0^0$ becomes inaccurate at surprisingly low energies.

\midinsert
\vskip 7.0truecm
{\noindent\narrower \it Fig.~6:\quad
Unitarity relation for the scalar form factor of the pion. Only
two--pion intermediate states contribute to order $q^4$, tadpole
diagrams are not shown.
\smallskip}
\vskip -1.0truecm
\endinsert
The simplest object to study these effects is the scalar form factor
(ff) of the pion,
$$ <\pi^a (p') \pi^b (p) \, \, {\rm out} | \hat{m} (\bar u u + \bar d d) | 0 >
= \delta^{ab} \, \tilde{\Gamma}_\pi (s)   \, \, \, ,   \eqno(4.19)  $$
with $s = (p'+p)^2$. It is not directly measurable, but has been determined for
energies from threshold to 1 GeV by a dispersive analysis in ref.[40] via
solution of the Muskhelishvili-Omn\`es equations of the $\pi \pi / K \bar K$
system (see also [49]). In ref.[93], an explicit calculation of the scalar ff
beyond one loop was performed. Consider the normalized scalar ff, $\Gamma_\pi
(s) = \tilde{\Gamma}_\pi (s) /
\tilde{\Gamma}_\pi (0)$. This quantity
is of order one to tree and of order $q^2$ at one--loop level. To perform the
two--loop calculation, one considers the unitarity relation depicted in fig.6.
The key point is that if one has the imaginary part at order $q^4$ (to two
loops), one can use dispersion relations to get the real part also at two--loop
order. To arrive at the imaginary part of the normalized scalar ff to order
$q^4$, one only has to keep two--pion intermediate states, i.e.
$$ {\rm Im} \, \Gamma_\pi  = \sigma \lbrace T^0_{0,2} ( 1 + {\rm Re} \,
\Gamma_{\pi,2} ) + {\rm Re} \, T^0_{0,4} \rbrace \Theta (s-4m_\pi^2 )
+ {\cal O} (q^6)  \, \, \, ,   \eqno(4.20)    $$
with $T^0_{0,2}$ and $T^0_{0,4}$ the
$\pi \pi$ scattering amplitude at tree
and one--loop level and $\Gamma_{\pi , 2}$ the one--loop expression for the
scalar ff (its explicit form can be found in refs.[13,93]). Since these
quantities are known from the work of Gasser and Leutwyler [13], Im
$\Gamma_\pi$ is fixed at order $q^4$. The real part can then be evaluated
by making use of Cauchy's theorem,
$$\Gamma_\pi (s) = 1 + {1 \over 6} <r^2>_S^\pi s + c_S^\pi s^2 + {s^3 \over
\pi} \int_{4 M_\pi^2}^\infty
{ds' \over {s'}^3 } {{\rm Im} \, \Gamma_\pi (s')
\over s' - s - i \epsilon} {\cal \, + \, O} (q^6)  \, \, \, ,   \eqno(4.21) $$
where the scalar radius of the pion and the curvature $c_S^\pi$ are given by
$$\eqalign{
<r^2>_S^\pi &= {3 \over 8 \pi^2 F_\pi^2} \bigl\lbrace (\bar{\ell}_4 -
{13 \over 12} ) +
{ \bar{d}_1 M_\pi^2 \over 16 \pi^2 F_\pi^2 }
\bigr\rbrace + {\cal O}(M_\pi^4) \cr
 c_S^\pi &= {1 \over 16 \pi^2 F_\pi^2} \bigl\lbrace {19 \over 120 M_\pi^2} +
 {\bar{d}_2 \over 16 \pi^2 F_\pi^2}
\bigr\rbrace + {\cal O}(M_\pi^4) \, \, \,
. \cr} \eqno(4.22) $$
There appear two unknown parameters, these are the low energy constants
$\bar{d}_{1,2}$ related to the effective Lagrangian at two--loop order. They
can be fixed by requiring the values of $<r^2>_S^\pi$ and $c_S^\pi$  to agree
with the ones of the dispersive analysis. This leads to $\bar{d}_1 = 22$ and
$\bar{d}_2 = 8.9$. Notice also that the coefficient $\bar{d}_2$ contains
infrared logarithms so that $\Gamma_\pi$ stays finite in the chiral limit.

\midinsert
\vskip 12.0truecm
{\noindent\narrower \it Fig.~7:\quad
Scalar form factor of the pion. The curves labelled
'1', '2', 'O' and 'B' correspond to the chiral prediction to one--loop,
to two--loops, the modified Omn\`es representation and the result of the
dispersive analysis, respectively. The real part is shown in (a) and
the imaginary part in (b).
\smallskip}
\vskip -1.0truecm
\endinsert
In fig.7a,b, the real and the imaginary part of the scalar ff to two loops are
shown, respectively, in comparison to the one--loop prediction and the
empirical curve based on the phase shift analysis of Au, Morgan and Pennington
[94]. For the real part, the one--loop approximation reproduces Re $\Gamma_\pi
(s)$ within 10 percent for energies below 400 MeV but it fails above 450 MeV.
In contrast to the bend down of the empirical solution, it rises steadily. The
two--loop curve shows the proper behaviour, the broad enhancement around
$\sqrt	\simeq 550$ MeV will be discussed below. At two--pion threshold, the
two--loop correction to the one--loop enhancement is already substantial. For
the imaginary part, the one--loop result fails already very close to threshold.
This is not surprising since, quite generally, the imaginary part calculated
order by order in the energy expansion, only reflects the real part at one
order less. The two--loop prediction is close to the
empirical result up to 560
MeV. However, since two--loop corrections are large, one might wonder what
happens with contributions of yet higher order.
I will come back to this point.
First, however, a few more comments on the real part of the scalar ff. The
reason for the turnover at $\sqrt s \simeq 550$ MeV can be understood from the
fact that one can write Re $\Gamma_\pi$ in an Omn\`es (exponential) form,
$$ {\rm Re} \, \Gamma_\pi (s)
= P(s) \, {\rm e}^{{\rm Re} \, \Delta_0 (s)} \,
\cos \delta^0_0 (s) + {\cal O} (q^6) \, \, \, ,    \eqno(4.23)	   $$
where the Watson final--state theorem [95] is fulfilled at next--to--leading
order, Im $\Delta_0 (s) = \delta_0^0 (s) + {\cal O} (q^6)$. The explicit form
of $P(s)$ and $\Delta_0 (s)$ are given in ref.[93]. Although
this representation is
not unique, the appearance of the factor $\cos \delta_0^0$ explains the
turnover. The phase $\delta_0^0$, which enters at one--loop level, goes through
$\pi / 2$ at $\sqrt s \simeq 680$ MeV forcing the real part to vanish at this
energy. This is clearly a two--loop effect since $\cos \delta_0^0 = 1 -
(\delta_0^0 )^2 + \ldots = 1 + {\cal O}(s^2 / F_\pi^4 )$ and explains why the
one--loop prediction increases monotonically.
Furthermore, in ref.[96] the
physical interpretation of the broad enhancement in Re $\Gamma_\pi$, which is
reminiscent of a resonance structure, was given. Consider the imaginary part
(which does not show any resonance behaviour)
and subtract from it the uncorrelated
two--pion background. This subtracted imaginary part has a bell shape peaked
around 600 MeV with a width of 320 MeV.
The strong final state interactions mock
up a broad, low--lying scalar--isoscalar "particle" which is indeed needed to
furnish the intermediate--range attraction in the boson--exchange picture of
the nucleon--nucleon interaction. The low--lying scalar used in this type of
approach  is therefore nothing but a convenient representation.

The exponential (Omn\`es) form (4.23) suggests that higher loops in CHPT can be
summed in an easy manner using the final--state theorem. However, since
$\Delta_0 (s)$ behaves badly at high energies and the form (4.23) is not
unique, in ref.[93] a modified Omn\`es form was proposed which cures these
shortcomings. It reads
$$ \eqalign{
\bar{\Gamma} (s) &= \Gamma^\Lambda {\rm e}^{\Delta_\Lambda (s) } = ( 1+ b \cdot
s)  {\rm e}^{\Delta_\Lambda (s) }	    \cr
\Delta_\Lambda (s) &= {s \over \pi}
\int_{4 M_\pi^2}^\infty {ds' \over s'} {\Phi
(s') \over s' - s - i \epsilon}      \cr
\tan \Phi &= {\sin 2 \delta_0^0 {\rm e}^{-2 \eta} \over 1 + \cos 2 \delta_0^0
{\rm e}^{-2 \eta} }    \cr}    \eqno(4.24)  $$
where the cut off $\Lambda$ is chosen around 1 GeV (below $\bar{K} K$
threshold). The reduced Omn\`es function $\Delta_\Lambda (s)$ takes into
account
the two--pion cut and therefore the reduced ff $\Gamma^\Lambda$ can be well
represented by a polynomial linear in $s$ (the slope parameter $b$ is adjusted
to the empirical
value of the scalar radius). The representation (4.24) is very
accurate at low energies. In the chiral limit, $\bar{\Gamma} (s)$ contains
the leading and and next--to--leading singularities in $s$. In addition, at
each order $N$ in the chiral expansion, the final--state theorem is obeyed,
$$ \bar{\Gamma}_N = {\rm e}^{2 i \Phi_N} \bar{\Gamma}_N^* \, \, \, \,
\eqno(4.25) $$
The curves labelled "O" in fig.7a,b show the result of the modified Omn\`es
representation with $\Lambda = 1$ GeV. The real part follows the exact solution
closely up to 550 MeV. Beyond this energy, it falls off slightly too steeply.
However, would one have a more accurate representation of the phase $\Phi$, a
better description of the scalar ff would result. The imaginary part
reproduces the exact solution within 15 per cent up to 700 MeV. The
representation (5.7) allows furthermore to investigate the importance of higher
loops in the following sense. Expanding $\bar{\Gamma}$ in powers of $q^2$ and
collecting pieces from the polynom and the exponential, one has $\bar{\Gamma} =
c_0 + c_2 + c_4 + c_6 + \ldots$ with $c_0 =1$ and $c_{2n} = {\cal O} (q^{2n})$.
For the real part, one finds that three loop contributions  cannot be neglected
beyond energies of 500 MeV. For the imaginary part the three loop contributions
are less than half of the two--loop ones (in magnitude)  below 600 MeV and
higher loops are negligible. This explains why the already good two--loop
result is only mildly affected in this energy range. One can perform the same
study for the vector ff of the pion [93]. Below 500 MeV, the data have large
error bars which makes an exact comparison difficult. However, one finds that
the two--loop result describes the data below 500 MeV. The polynomial pieces
are completely dominant,
{\it i.e.} the unitarity corrections are small (in contrast
to the scalar ff). Clearly, the nearby $\rho$ resonance starts to play a
dominant
role beyond this energy, indicating that explicit resonance degrees of
freedom should be included.

An attempt to incorporate the resonances in EFT was made in ref.[71] in the
context of $\pi \pi$ and $\pi K$ scattering. The resonance Lagrangian of Ecker
et al. [32] was used and the momentum dependence of the resonance propagators
was fully taken into account. This leads to a good description of the data up
to energies of 1 GeV and, by construction, the $P$--waves are in perfect
agreement with the data. The phase of $\epsilon '$ comes out to be $\Phi
(\epsilon ') = 44.5^\circ$, consistent with the value (4.12). However, this
calculation is not complete since not all effects at order $q^6$ and higher
have been taken into account (only the ones related to the various resonance
propagators). The whole program of carrying out the loop expansion in an
effective theory of pseudoscalar Goldstone bosons coupled to resonances has yet
to be performed.

There are
some other proposals to enlarge the range of applicability of the
truncated chiral expansion. Truong [91] has investigated the scalar and vector
ff of the pion. To enforce the final--state theorem to all orders, he considers
an Omn\`es--Muskhelisvhili integral equation for the inverse ff. This is
equivalent to summing an infinite series of bubble graphs and formally nothing
but the Pad\'e approximant [0,1] (notice that this notation differs from the
one
used by Truong) of the chiral series. This means that $F = 1 + F^{(2)} +
F^{(4)}$ is substituted by $F_{[0,1]} = F^{(2)} / ( 1 - F^{(4)} / F^{(2)})$
with
$F$ a generic symbol for a form factor and the subscript gives the chiral
dimension.  While good agreement with the data is claimed [91], the method was
criticized
in ref.[93]. First, it was shown that the Pad\'e approximant sums
next--to--leading order chiral logarithms with the wrong weight, i.e. it does
not respect the constraints from the chiral symmetry. Furthermore, the modified
Omn\`es representation discussed before gives a better description of the
scalar
ff than the [0,1] approximant (and respects the next--to--leading order logs).
Also, the Pad\'e [0,1] apparently describes the data in the resonance region of
the $P$--wave well. However, the position, width and
heigth of the resonance are
very sensitive to the vector radius (which is used to fix the coupling $L^r_9$
as discussed in section 2.5). Dobado, Herrero and Truong [92] have also applied
this method to unitarize the $\pi \pi$ scattering amplitudes.  Dobado and
Pelaez [97] have recently investigated the inverse amplitude method which is
formally equivalent to the Pad\'e [0,1] series. Their philosophy is to fix the
couplings $L^r_{1,2}$ and   $L_3$ in some partial waves. Having done that, they
find a good description of all phase shifts up to energies of 1 GeV in $\pi
\pi$ and $\pi K$ scattering. Notice, however, that the data they claim to fit
well for the phase $\delta_0^2 (s)$ are outside the Roy equation band discussed
before. Clearly, all these unitarization methods are build on the notion of
imposing strict (elastic) unitarity on the expense of a controlled chiral
expansion. Choosing a particular unitarization procedure induces an unwanted
model--dependence and thus violates the strict requirements of CHPT. It is
true, however, that higher loop calculations will definitively have to be done
by unifying dispersion theoretical methods with chiral low--energy constraints
as exemplified in the calculation of the scalar ff discussed before. It is
always preferrable to have more loops evaluated than using some unitarization
prescription [98].
\bigskip
\noindent {\bf 4.5. The decays $K \to 2 \pi$ and $K \to 3 \pi$}
\medskip
The non--leptonic weak interactions are a wide field were CHPT methods can be
applied, in particular for the many decay modes of the kaons. I will focus here
on some recent work in connection with next--to--leading order calculations of
the reactions  $K \to 2 \pi$ and $K \to 3 \pi$ (one reason being that it
relates
back to the $\pi \pi$  phase shifts discussed before). There are many other
articles on kaon decays. Here, let me just mention the nice series of papers by
Ecker, Pich and deRafael [99] on decays like $K \to \pi e^+ e^- , K \to \pi
\mu^+ \mu^- , K^0 \to \pi^0 \gamma  \gamma$ and so on. I refer to these papers
and the upcoming review by these authors [100] for a comprehensive treatment of
these and other $K$--decays (and the list of references therein).

Let us now concentrate on the decays $K \to 2 \pi , 3 \pi$ at next--to--leading
order. To lowest order, the Lagrangian of the strangeness-changing
non--leptonic
weak interactions reads
$$ \leff^{(2)} (\Delta S = 1) = - c_2 Tr ( \lambda_6 U^\dagger \partial_\mu U
U^\dagger \partial^\mu U ) - c_3 t_{ik}^{jl} {\rm Tr} ( Q_j^{i}
U^\dagger \partial_\mu U ) {\rm Tr} ( Q_l^k U^\dagger \partial_\mu U )
\eqno(4.26)   $$
with $(Q^{i}_j )_{kl} = \delta_{il} \delta_{jk}$ $3 \times 3$ matrices in
flavor
space, $\lambda_6 = Q_2^3 +Q_3^2$ projects out the octet and the $t_{ik}^{jl}$
the 27-plet of the interactions. This follows from the fact that the chiral
transformation properties of the weak interactions are characterized by a
decomposition  into left--chiral representation as $(8_L , 1_R)
\oplus (27_L , 1_R
)$. The so-called weak mass term can be neglected. To lowest order, the
constants $c_2$ and $c_3$ of the $\Delta I = 1/2$ and $\Delta I = 3/2$
operators
can be fixed from $K \to 2 \pi$ decays [101],
$$ c_2 / F_\pi^2 = 0.95 \cdot 10^{-7} \, \, ,
 c_3 / F_\pi^2 = -0.008 \cdot 10^{-7} \, \, .	\eqno(4.27)  $$
The fact that $c_2 >> c_3$ constitutes the so-called $\Delta I = 1/2$ rule. At
next--to--leading order, the chiral non--leptonic Lagrangian has been worked
out
by Kambor, Missimer and Wyler [102]. Using chirality and $CPS$ symmetry [103]
as
the basic principles, one arrives at:
$$\leff^{(4)} (\Delta S = 1) = \sum_{i=1}^{47} E_i^r \, Q_i^8 \, + \,
\sum_{j=1}^{33} D_j^r \, Q_j^{27} \,  \,  \, \, . \eqno(4.28)	 $$
The explicit form of the octet ($Q_i^8$) and the 27--plet ($Q_j^{27}$)
operators
can be found in ref.[102]. Not all of these counterterms are independent, but
they serve as a useful basis. If one switches off the external fields, one is
left with 15 (13) independent contact terms for the octet (27--plet). Clearly,
a
complete determination of the renormalized weak low--energy constants $E_i^r$
and $D_i^r$
is not available (this is one place were the model estimates
discussed in section 2.5 are very useful).

Kambor, Missimer and Wyler [104] have considered the decays $K \to 2 \pi$ and
$K
\to 3 \pi$ at next--to--leading order. The appearing counterterms can be fitted
from
decay rates and slope parameters.  For doing that, one has to consider the
isospin decomposition of the pertinent amplitudes. Consider first the
CP--conserving $K \to 2 \pi$ decays,
$$\eqalign{
A(K_s \to \pi^0 \pi^0 )&= i \sqrt{2 / 3} \, a_{1/2} {\rm e}^{i
\delta_0^0} \, \, -  i \sqrt{4 / 3} \, a_{3/2} {\rm e}^{i \delta_0^2}	 \cr
A(K_s \to \pi^+ \pi^- )&= -i \sqrt{2/ 3} a_{1/2} {\rm e}^{i
\delta_0^0} \, \, -  i \sqrt{4 / 3} \, a_{3/2} {\rm e}^{i \delta_0^2}	 \cr
A(K^+ \to \pi^+ \pi^0 )&= -i \sqrt{3 / 4} \,a_{3/2} {\rm e}^{i
\delta_0^2}	   \cr}      \eqno(4.29)    $$
where the coefficients $a_{1/2}$ and $a_{3/2}$ are real. The two pions in the
final state are in an $S$--wave with total isopsin zero or two and therefore
the corresponding $\pi \pi$ phases appear (see the discussion after eq.(4.12)).
The five $K \to 3 \pi$ amplitudes are decomposed in terms of two intercepts
$(\alpha_1 , \alpha_3 )$,
three linear slope ($ \beta_1 , \beta_3 , \gamma_3 $)
and five quadratic slope ($\chi_1 , \chi_3 , \xi_1 , \xi_3 , \xi_3 '$)
parameters. These twelve quantities can be expressed in terms of seven
combinations of the weak counterterms plus the strong $L^r_{1, \ldots , 5}$
together with loop and tree level contributions, the latter involving the
parameters $c_2 , c_3 , F_\pi$ and $F_K$ (neglecting terms of order $M_\pi^2 /
M_K^2$). In ref.[104] the available data were refitted and then a least--square
fit was used to determine the values of $c_2 , c_3$ and the $K_i \,   (i=1
\ldots, 7)$ from the next--to--leading order expressions. The most prominent
result of this analysis is that the value for $c_2$ is diminished by 30 per
cent,
$$ c_2 / F_\pi^2 = 0.66 \cdot 10^{-7}	   \eqno(4.30)	$$
to one loop and the 27--plet coupling $c_3$ remains unchanged. This reduction
of $c_2$ implies that a factor 1.5 in the $\Delta I = 1/2$ enhancement is due
to long distance effects (see also refs.[105]).{\footnote{*}{In ref.[106] it
was argued that no suppression of the $\Delta I = 3/2$ piece due to
long--distance effects exists.}} Notice that the corresponding phase shift
difference $\delta_0^0 - \delta_0^2$ comes out to be $29^\circ$, which slightly
less than the tree level result of $37^\circ$. This is an effect of the fitting
procedure since to the order the amplitudes are determined, the imaginary parts
should come out at their tree level values. To recover the result (4.12) one
would have to go further in the loop expansion. In ref.[104]
it was also pointed
out that there are two dominating $\Delta I = 1/2$ counterterms (which have
previously been discussed in ref.[107]) which are considerably larger than
the two well--determined  $\Delta I = 3/2$ counterterms. This means that the
$\Delta I = 1/2$  enhancement is preserved at next--to--leading order, a
welcome feature. For a more detailed account of these topics, the reader should
consult refs.[102,104].

Kambor et al. [108] have further investigated these decay modes. They point out
that to order $M_\pi^2 / M_K^2$ one can derive relations which are independent
of the weak counterterms $K_1 , \ldots , K_7$ at the four derivative level
(besides five already known relations at the two derivative level). To be more
precise, there are  2 (3) relations in the  $\Delta I = 1/2$ (3/2) sector
 between various
of the slope parameters. These conditions follow directly from the CHPT
 calculation
and thus are a highly non--trivial test. The two $\Delta I = 1/2$ relations are
in good agreement with the data where as the comparison in the $\Delta I = 3/2$
sector is less favorable. However, it is stressed in ref.[108] that large
electromagnetic corrections in the $\Delta I = 3/2$  case are possible and that
the analysis in this sector is more sensitive to small errors in the analysis
due to cancellation of large numbers. Clearly, a better empirical determination
of the decays $K_L \to 3 \pi , K_L \to \pi^+ \pi^0 \pi^-$ and $K^+ \to 3 \pi$
would lead to a precision test of CHPT in the $\Delta I = 3/2$	sector. As a
final comment, let me point out that in $\Delta I = 1/2$  amplitudes one has
large loop corrections which are due to the strong pionic final state
interactions in the isospin--zero $S$--wave. It is conceivable that a
resummation technique as discussed in section 4.4 will allow to sum these
rescatterig diagrams in harmony with chiral symmetry.
\bigskip
\noindent {\bf 5. FINITE TEMPERATURES AND SIZES}
\bigskip
CHPT allows to make precise statements about the low temperature behaviour of
the strong interactions. Furthermore, Goldstone boson induced finite size
effects can be calculated in a controlled fashion which is of importance for
lattice gauge calculations. In this section, I will first discuss some
aspects of finite temperatures, mostly the melting of the quark and gluon
condensates with increasing temperature. Then, I will cover some developments
concerning finite size effects, in particular the comparison of recent Monte
Carlo studies of $\sigma$ models and the chiral predictions. Technical
details will generally be omitted and the reader should consult the pertinent
references.
\bigskip
\noindent {\bf 5.1. Effective theory at finite temperature}
\medskip
In the chiral limit, the pions are massless and dominate the spectrum.
At sufficiently low temperatures,
they do not interact forming a Bose gas with the pressure directly
proportional to the fourth power of the temperature. Interactions among the
pions generate power-like corrections which are controlled by the parameter
$T^2 / 8 F_\pi^2$ and the contributions of the heavy particles are
exponentially suppressed (see below for details). If the temperature is
sufficiently small, a perturbative analysis of these effects can be performed
making use of CHPT techniques as developed by Gerber, Gasser and Leutwyler
[109,110,111] (see these papers for references on earlier work on this
subject). Instead of the S--matrix, one deals with the generating functional
$Z$ defined via
$$ Z ={\rm Tr} \, [{\rm e}^{-H/T} ]
\eqno(5.1)    $$
with $T$ the temperature and $H$ the Hamiltonian. The quantity in the square
brackets is nothing but the analytic continuation of the time evolution
operator to the point $t = -i / T$. This allows to write
$$ ( U_2 | {\rm e}^{-t H} | U_1 ) = \int [dU] \exp \bigl\lbrace - \int d^4 x
\tilde{{\cal L}}_{{\rm eff}} \bigr\rbrace
\eqno(5.2)   $$
with $U_1 = U(\vec{x},0)$, $U_2 = U(\vec{x},t)$ and
$\tilde{{\cal L}}_{{\rm eff}}$ can be
obtained from $\leff$ by replacing the Minkowski metric $g_{\mu \nu}$ by $-
\delta_{\mu \nu}$ and changing the overall sign. To perform the trace, one
sets $U_1 = U_2$ and integrates over $U_1$ over an interval of length $1/T$.
The functional integral then extends over all field configurations which are
periodic in the time direction,
$$\eqalign{
Z &= \int [dU] \exp \lbrace - \int_M d^4 x
\tilde{{\cal L}}_{{\rm eff}} \rbrace \cr
			       U( \vec{x} , x_4 )  &=
U ( \vec{x} , x_4 + \beta )			  \cr}
\eqno(5.3)  $$
with $\beta = 1 / T$. The pions live on the manifold $M$ which is nothing but
the torus $R^3 \times S^1$. The circumference of $S^1$ is given by the
inverse temperature $\beta$. It is important to realize that the coupling
constants like $F , B, L_1^r , L_2^r , \ldots$ remain unaffected (are
temperature independent). The boundary conditions in the direction of $x^4$
are dictated by the trace that defines the partition function (strictly
speaking, the trace operation only makes sense at finite volume. The finite
size effects generated by the box are discussed in ref.[112]). The only
change induced by the temperature is a modification of the pion propagator.
In the chiral limit and in euclidean space, the $T$--dependent propagator is
given by
$$ G(x) = \sum_{n=-\infty}^{\infty} \Delta (\vec{x} , t + {n \over T})
\eqno(5.4) $$
where $\Delta (x) = [4 \pi^2 ({\vec{x}}^2 + t^2 )]^{-1}$ is the $T=0$
propagator. Any Green function
can be evaluated at finite $T$ along the
same lines. Denote by $A$ an arbitrary operator (like e.g. the product of two
currents), its thermal expectation value is simply given by
$$ <A>_T = {{\rm Tr} [{\rm e}^{-\beta H} \, A ] \over
{\rm Tr}  [{\rm e}^{-\beta H}
]\, }
\eqno(5.5) $$
In particular, the free energy density $z$ reads
$$ z = \epsilon_0 \, - \, P \, = \, -T \,      \lim_{V
\to \infty}
   { \ln \, Z \over V}
\eqno(5.6) $$
\midinsert
\vskip 5.0truecm
{\noindent\narrower \it Fig.~8:\quad
Typical two and three loop diagrams for calculating the
free energy density. The black dots ($\bullet$) denote insertions
from $\leff^{(2)}$.
\medskip
\smallskip}
\vskip -1.0truecm
\endinsert \noindent
with $\epsilon_0$ the energy density of the vacuum and $P$ the pressure.
Clearly, to evaluate the partition function to some given order in the
temperature, one has to evaluate CHPT to the corresponding order in the loop
expansion. To make this statement more transparent, consider some typical
Feynman diagrams contributing to $z$ (fig.8). They have no external legs,
however, instead of the small external momenta one now has to deal with
temperature insertions. The following remarkable feature emerges [111,113].
Tree diagrams from $\leff^{(2,4,6,\ldots )}$ are temperature independent and
therefore only contribute to the vacuum energy. If one now evaluates the free
energy density to order $q^n$, counterterms from $\leff^{(n-2)}$ enter
through one--loop graphs and thus give rise to a $T$--dependent contribution.
This can, however, be absorbed in the renormalization of the pion mass [111].
Therefore, after subtracting the vacuum energy density and expressing the
remaining contributions in terms of the physical pion mass, the free energy
density evaluated to order $q^n$ only involves low--energy constants from the
effective Lagrangian up to and including order $n-4$. This makes a three loop
calculation feasible since  only the known one--loop couplings enter [111].
 The calculations are performed in dimensional regularization. In that case,
the temperature independent part of the pion propagagtor generates all the
singularities as $d \to 4$ (see ref.[114] for details). Furthermore, most
thermodynamic quantities can be expressed in terms of powers of the thermal
propagator at $x=0$ and the functions $g_r$ associated to the
noninteracting
$d$--dimensional Bose gas,
$$ g_r (M,T) = 2 \int_0^\infty {d \lambda \over (4 \pi \lambda )^{d/2}
}\lambda^{r-1} \exp ( - \lambda M^2 ) \sum_{n=1}^\infty \exp ( -n^2 / 4
\lambda T^2 )
\, \, \, .
\eqno(5.7)$$
This essentially specifies all the tools one needs for finite temperature
calculations (for a more detailed account, see ref.[111]).
\bigskip
\noindent {\bf 5.2. Melting condensates}
\medskip
The quark condensate is the thermodynamic variable conjugate to the mass of
the light quarks. It is given by
$$ < \bar{q} q >_T = {{\rm Tr} \, [\bar{q} q \exp(-H/T)] \over
{\rm Tr} \, [\exp(-H/T) ] }
= { \partial z \over \partial m_q}
\eqno(5.8)  $$
where $m_q$
denotes the quark mass. The quark condensate plays the same role as
the magnetization in a ferromagnet. Following this analogy, it is supposed to
decrease as temperature increases and eventually to disappear. Disorder takes
over and the symmetry is restored. In the case of QCD at low temperatures,
the pions dominate the properties of the system since the contributions of
the massive states to the partition function are exponentially suppressed.
The free energy density admits an expansion in powers of $T^2$ and ${\rm log}
\, T$ much like the $\pi \pi$ scattering amplitude in powers of the external
momenta and logs thereof:
$$ z = \sum_{m,n} c_{m,n} (T^2)^m \, (T^2 \, {\rm log} T )^n \, \, +
\, \, {\cal O} ({\rm e}^{- \bar{ M} / T} )
\eqno(5.9)$$
with $m,n = 0,1,2, \ldots$ and $\bar M$ denotes the mass of the lightest
massive state (like the kaon if one works in the two--flavor case). Clearly,
eq.(5.9) is an asymptotic series, the exponentially suppressed contributions
from the massive states can never be accounted for by powers of $T^2$ or
logarithms. This is completely analogous to perturbative QCD series, where
instantons generate non--perturbative corrections of the type $\exp (-8 \pi /
g^2 )$. From eq.(5.8) it follows immediately that the thermal expectation
value of the quark condensate has a similar representation. For $N_f$
massless quarks, the expansion of $< \bar{q} q >_T$ reads
$$ < \bar{q} q >_T  = < 0 | \bar{q} q | 0 > \biggl\lbrace 1 - c_1 \bigl( {T^2
\over 8 F_\pi^2} \bigr) - c_2 \bigl( {T^2 \over 8 F_\pi^2} \bigr)^2 - c_3
\bigl( {T^2 \over 8 F_\pi^2} \bigr)^3 \ln ({\Lambda_q \over T} ) \, \, + \,
{\cal O} (T^8) \biggr\rbrace
\eqno(5.10) $$
with
$$ c_1
= {2 \over 3} { N_f^2 - 1 \over N_f} \, \, \, , \, \,
   c_2
= {2 \over 9} { N_f^2 - 1 \over N_f^2} \, \, \, , \, \,
   c_3
= {8 \over 27} ( N_f^2 + 1 ) \, N_f  \, \, . \eqno(5.11)   $$
The coefficients $c_1  , c_2$ and $c_3$ were first worked out by Bin\'etruy and
Gaillard [115], Gasser and Leutwyler [109]  and Gerber and Leutwyler [111],
respectively. For massless quarks, this result is exact. The first two terms
in the expansion are entirely given in terms of the parameter $F_\pi$ which
characterizes the lowest order effective Lagrangian. Notice also the
appearance of the characteristic temperature $T_c = \sqrt 8 F_\pi \simeq 250$
MeV. Only at next--to--next--to--leading order a logarithm appears. The
unknown scale can be related to the $SU(2)$ low energy constants
$\bar{\ell}_{1,2}$ of the next--to--leading order effective Lagrangian. These
couplings are related to the isospin--zero $D$--wave scattering length
$a_2^0$ via
$$ a_2^0 = {1 \over 144 \pi^3 F_\pi^4 } \bigl\lbrace \ln ({\Lambda_q \over
M_\pi}) - 0.067 + {\cal O} (\m ) \bigr\rbrace
\eqno(5.12) $$
\midinsert
\vskip 9.0truecm
{\noindent\narrower \it Fig.~9:\quad
The quark condensate at finite temperature (after ref.[111]).
The solid, dashed and dashed--dotted line represent the one, two and
three loop calculations. The error bar at the three loop curve at
150 MeV represents the uncertainty in the scale of the chiral
logarithm.
\smallskip}
\vskip -1.0truecm
\endinsert \noindent
which leads to $\Lambda_q = 470 \pm 110$ MeV [113]. In figure 9, the one, two
and three loop results
are shown for temperatures below 150 MeV (using the
central value of $\Lambda_q$). At higher temperatures, the corrections to the
tree result become so large that the expansion (5.10) does not make sense any
more. In particular, CHPT can not predict the temperature of the chiral phase
transition, at that point the corrections eat up the leading term completely.
The case of non--zero quark masses has also been discussed by Gerber and
Leutwyler [111]. As expected, a finite pion mass slows down the melting of
the quark condensate. Even though the quark masses are tiny, the temperature
dependence is strongly affected because the corresponding Boltzmann factor
involves $M_\pi$ rather than $M_\pi^2 \sim \hat{m}$. Furthermore, these
authors also discuss the influence of the massive states ($K, \eta , \rho , N
, \ldots$) making use of a dilute gas approximation. Below $T = 150$ MeV, the
massive states influence the melting of the condensate very little because of
the Boltzmann  factors. Beyond this temperature, one can  not neglect the
interactions between the massive states and the ones to the pions any more.
To summarize, chiral symmetry  predicts the temperature dependence of the
quark condensate. As the temperature increases, it gradually melts. At a
temperature of approximately 150 MeV, the loop corrections (in the three loop
approximation) have decreased the condensate about a factor of two, rendering
the perturbative analysis useless beyond this point. However, if one ignores
that for a moment and follows the result to higher temperatures, the
condensate vanishes at a temperature of $170 \ldots 190$ MeV.

In a similar fashion, one can study the temperature dependence of the gluon
condensate [116,117] (which is expected to be much weaker than the one of the
quark condensate as indicated by some models [118]). For that, one considers
the trace of the energy--momentum tensor,
$$ \Theta_\mu^\mu = - {\beta (g) \over 2 g^3} G_{\mu \nu}^a G^{\mu \nu , a}
\, + \, { \lbrace 1+ \gamma (g) \rbrace } \, \bar{q} \m q
\, + \, c \, {\bf 1}
\eqno(5.13) $$
Here, $\beta (g)$ is the QCD $\beta$--function, $\mu dg / d \mu = \beta (g)$.
The first term in (5.13) accounts for the fact that the strong coupling
constant is scale dependent (conformal anomaly) and the second one is due to
the explicit scale breaking from the quark masses. The last term is fixed by
the normalization condition that the trace of the energy--momentum tensor is
zero in the vacuum, i.e. $c = <0 | G^2 |0>$, with $G^2 = - \beta
G_{\mu \nu}^a G^{\mu \nu , a} / 2 g^3$. At finite temperature, one has
$$ < G^2 >_T =
<0 | G^2 |0> - < \Theta_\mu^\mu >_T
\eqno(5.14) $$
Two comments are in order. First, the gluon condensate is not an order
parameter, conformal symmetry is also broken in the high temperature phase of
QCD. Second, the vev $<0 | G^2 |0>$ can not be determined very accurately,
but its value drops out if one considers the temperature  dependence of the
gluon condensate. The quest is now to calculate $< \Theta_\mu^\mu >_T$. One
can make use of the thermodynamic relations $ s= dP/dT$ and $\epsilon = T s
-P$, with $s$ the entropy density. Therefore, the knowledge of the equation
of state $P = P(T)$ is all one needs [111,112,116],
$$< \Theta_\mu^\mu >_T
= \epsilon - 3 P = T^5 {d \over dT} \bigl( {P \over
T^4} \bigr)
\eqno(5.15)   $$
Since the correction to the free Bose gas for the massless Goldstone bosons
starts out at order $T^8$, one finds a weak temperature dependence of the
gluon condenstae. For two massless quark flavors, one has   [116]
$$< \Theta_\mu^\mu >_T
= {\pi^2 \over 270} {T^8 \over F^4_\pi} \bigl\lbrace
\ln \bigl({\Lambda_p \over T}\bigr) - {1 \over 4} \bigr\rbrace \, + \, \ldots
\eqno(5.16)  $$
where the ellipsis stands for higher terms in the temperature expansion and
the contribution from the exponentially suppressed massive states. One
notices that the gluon condensate indeed admits a very weak temperature
dependence. The physical reason is that the operators $\Theta_\mu^\mu$ and
$G^2$ are chiral singlets and thus single pion emision/absorption is not
allowed. Furthermore, $< \Theta_\mu^\mu >_T$ vanishes up to and including two
loops in CHPT. This leads to the leading $T^8$ term in (5.16). Clearly, since
the pion contribution is so much suppressed, the massive states play a more
important role already at lower temperatures. For a detailed discussion of
these topics, see Gerber and Leutwyler [117].

Finally, one can also study the kinetics of the hot pion gas and its relation
to the assumed quark/gluon plasma transition. This topic will not be
addressed here, I refer to the article by Goity [119] which also includes
references to pertinent work in that field.
\bigskip
\noindent{\bf 5.3. Effective theory in a box}
\medskip
\goodbreak
One can also address the following question: What happens if the pion fields
are
confined in a box of finite volume? At first sight, this might sound
academical since one knows that there is no spontaneous symmetry breaking in
a finite volume. However, CHPT can be used to understand how the infinite
volume limit is approached. This is of particular relevance for lattice
simulations. On the lattice, the world is a four-dimensional box. In general,
one has to deal with finite size effects induced by the particles of mass $M$
which are the lowest excitations in the spectrum. A general analysis of these
finite size effects and the approach to the continuum limit has been
performed by L{\"u}scher [120] for the case $M L >> 1\, (L$ denotes the size
of the box). Matters are, however, different in the presence of Goldstone
bosons (massless excitations). No matter how large one choses the lattice
size, the parameter $M L$ will always be small and finite size effects	will
be large. This can be most easily understood in the case of the quark
condensate. For fixed quark masses and zero temperature, the vev $<\bar{q}
q>$ is always zero  as long as $L$ is finite. Only when $L \to \infty$,
symmetry breaking occurs and $<\bar{q} q> \ne 0$. Obviously, the finite size
effect $<\bar{q} q>_L -  <\bar{q} q>_\infty$ is as big as the quantity itself
(see also the discussion by Leutwyler in ref.[116]). As will be shown later
on, CHPT can be used to systematically calculate the Goldstone boson induced
finite size effects in an expansion in powers of $1 / L^2$ (in four
dimensions). Ultimately, this will become a tool to control finite size
effects in the study of lattice QCD with reasonably light quarks. At present,
these methods are tested versus Monte Carlo data on $O(N)$ $\sigma$
models as will be
discussed below.

The effective Lagrangian at finite volume $V = L_1 \times L_2 \times L_3
\times L_4 =  L_1 \times L_2 \times L_3 \times \beta$ has been studied in
detail by Gasser and Leutwyler [109,112] and also by Neuberger [121].
The lowest order term in the effective theory which generates the leading
low energy contributions to the Green functions has the standard form
involving the two coupling constants $F$ and $B$. The pion field is subject
to periodic boundary conditions,
$$ U(x) = U(x+n)  \, \, \, , \, \, n = (n_1 L_1 , n_2 L_2 , n_3 L_3 , n_4
\beta )    \eqno(5.17) $$
with the $n_i$ integer numbers. The constants $F$ and $B$ are not
affected by the finite volume [112]. As before, the only modification appears
in the pion propagator, it takes the form
$$ G(x) = \sum_n G_0 (x+n)    \eqno(5.18) $$
with $G_0 (x)$ the free ($T = 0, L = \infty$) propagator. The quantities
$M_\pi$, $T$ and $1 / L_i$ are all booked as ${\cal O}(p)$ at fixed ratios
$M_\pi / T$ and $L_i T$ (the so--called "$p$--expansion"). Effectively, one
seeks an expansion of the generating functional in powers of these small
quantities (a complication arises due to zero modes as discussed below). At
next--to--leading order, matters are more complicated. Since the box breaks
Lorentz invariance, one expects additional terms related to this phenomenon
and also boundary terms which are sensitive to the boundary conditions on the
fields. There exists, however, a tremendous simplification if one choses a
rectangular box and imposes the same  boundary conditions on the fields on
the walls of the box as in the time direction proportional to $L_4$. Then,
the partition function is invariant under permutations of the walls of the
four--dimensional box. Since we had already seen that the low--energy
constants do not depend on the temperature, one can use this symmetry to show
that they also do not depend on the finite volume. Furthermore, no Lorentz
non--invariant vertices and surface terms appear in this case. Therefore, the
effective Lagrangian is just the one used at  zero temperature in the
infinite volume limit.

Let us now consider $N_f$ quarks of the same mass and denote the associated
Goldstone boson mass by $M$. The partition function takes the form (for $L_1
= L_2 = L_3 = L) [109]$
$$\eqalign{ \quad \quad
& {\rm	Tr \, e}^{- \beta H}  = {\rm e}^{- \beta L^3 z}  \cr
 z &= \epsilon_0 - {N_f^2 - 1 \over 2} g_0 (M^2 , T , L) +
{N_f^2 - 1 \over 4 N_f} {M^2 \over F^2} [ g_1 (M^2 , T , L)]^2 +
{\cal{O}}(p^8) \cr}
\eqno(5.19) $$
with $\epsilon_0$ the energy density of the groundstate. The functions $g_0$
and $g_1$ are defined via
$$ g_r (M^2 , T , L) = {1 \over 16 \pi^2} \int_0^\infty d\lambda
\lambda^{r-3} \sum_{n \ne 0} \exp { (-M^2 \lambda - n^2 / 4 \lambda )}
\eqno(5.20)    $$
These take the familiar form known from the relativistic Bose gas in the
infinite volume limit. It is important to omit the term with $n_1 = n_2 =
n_3 = n_4 = 0$ from the sum in (5.20). In the chiral limit, the system
develops zero modes and consequently the function $g_1$ exhibits a pole,
$$  g_1 (M^2 , T , L) = {1 \over M^2 L^3 \beta} + \bar{g}_1 (M^2 , T , L)
\eqno(5.21)   $$
where $\bar{g}_1$ is finite when $M$ vanishes. This pole is due to the
propagation of the zero modes. Their contribution to the functional integral
is not of the Gaussian type as can be seen from
$$ \int d^4 x {\cal{L}}^{(2)} = -{1 \over 2} N_f F^2 M^2 V + {1 \over 4} F^2
\int d^4 x \Tr (\partial_\mu \phi \partial_\mu \phi + M^2 \phi^2 ) +
{\cal{O}}(p^4)
\eqno(5.22)  $$
The zero modes can therefore not be treated perturbatively.  To deal with
them, one introduces collective  coordinates and reorders the chiral expansion
by enhancing the zero--mode propagators ("$\epsilon$--expansion")
$$ T = {\cal O}(\epsilon ) \, \, , \, \,
1/L = {\cal O}(\epsilon ) \, \, , \, \,
 M = {\cal O}(\epsilon^2 )
\eqno(5.23) $$
Expanding now the partition function in powers of $\epsilon$, the graphs
involving exclusively zero modes are of order one whereas graphs with
non--zero modes in the propagators are suppressed by at least $\epsilon^2$
with the exception of the one--loop graphs which enter in the normalization
$A$ of the final result [109]
$$ {\rm  Tr \, e}^{- \beta H} =
A \, X_{N_f} (s) \, \, , \, s = {1 \over 2}
F^2 M^2 V     \eqno(5.24)   $$
with $X_{N_f} (s) $ the integral over the flavour group
$$X_{N_f} (s) = \int_{SU(N_f)} d\mu (U) \exp {(s{\rm{Re}} \Tr U) }
\eqno(5.25) $$
and
$d\mu (U)$ is the Haar measure. One arrives at this result by the
standard procedure introducing collective coordinates, $U = u {\rm{e}}^{i \xi
(x)} u$. Here, $\xi (x)$ collects the non--zero modes and the zero modes are
described by the constant matrix $u \in SU(N_f)$. For details of this
procedure, see Gasser and Leutwyler [109], Hasenfratz and Leutwyler [112] or
the original work by Polyakov [123]. In the framework of the
$\epsilon$--expansion, one can discuss the symmetry restoration when the
quark masses are much smaller than 1/volume [109,112,116]. With this, we have
all tools at hand to calculate finite size effects. In the following section,
we will compare some theoretical predictions with the results of Monte Carlo
studies.
\bigskip
\noindent{\bf 5.4. Finite size effects: CHPT versus Monte Carlo}
\medskip
\goodbreak
Here I will briefly discuss the comparison of theoretical predictions of
Goldstone boson induced finite size effects with recent
numerical simulations. For all technical details, I refer the reader to the
pertinent articles. First, consider models with broken $O(N)$ symmetries in
more than two dimensions. These are of relevance in the study of the Higgs
mass bounds, two--flavour QCD  or finite size effects in ferromagnets close
to the critical point.
Hasenfratz and Leutwyler [112] have given
theoretical predictions for the free energy, magnetization, susceptibilities
and two--point functions up to and including order $(1/L^{d-2})^2$ corrections
for $d=3$ and $d=4$. In the first case, only two low--energy constants enter
whereas in four dimensions, one has to deal with three additional ones. They
use the linear $\sigma$--model in $d$ dimensions making use of "magnetic"
language,
$${\cal L} = {1 \over 2} \partial_\mu \phi^a \partial_\mu \phi^a
+ {1 \over 2} m^2
\phi^a \phi^a + {1 \over 4} \lambda ( \phi^a \phi^a )^2
- H \phi^0
\eqno(5.26) $$
The real field $\phi$ has $N$ ($a = 0, \ldots , N-1 )$ components and the
external magentic field $H = ( H , 0 , \ldots , 0)$ breaks the $O(N)$
symmetry explicitely. As $H$ tends to zero, we are interested in the coupling
constant regime where spontaneous symmetry breaking $O(N) \to O(N-1)$ occurs
signaled by a non--vanishing value of the magnetisation $\Sigma$,
$$\lim_{H \to 0} < \phi^0 > = \Sigma \, \, , \, \, < \phi^i > = 0
\eqno(5.27)    $$
Translated to QCD language, $\Sigma$ is the vev of the scalar quark density
and $H$ is equivalent to the quark mass matrix. The corresponding effective
theory is formulated in terms of a vector field $S^a (x)$ which embodies the
$N-1$ Goldstone fields and is subject to the constraint $S^a (x) S^a (x) =
1$. The low energy properties of this model and the finite size effects at
next--to--leading order are discussed in detail in ref.[122]. In three
dimension, there are two low energy constants entering and in $d=4$ there are
five, the extra three being related to scales of logarithmic corrections.

For the $d=3$ classical $O(3)$ Heisenberg model in the broken phase near the
critical point, some of these predictions were confronted with Monte Carlo
data by Dimitrovic et al. [124]. They found that the finite size behaviour of
the magnetisation and the correlation functions are in good agreemnet with
the CHPT predictions. Furthermore, they could determine the critical indices
for the correlation length and magnetisation (in agreement to previous
studies using other methods). In the case of the $O(4)$ model in four
dimensions, numerical studies have been performed  by Hasenfratz et al.
[125]. Good agreement  with the theoretical predictions is found and it is
shown that in situations where the Goldstone bosons control the dynamics of
the system, one can indeed determine the infinite volume, zero external
source quantities from the finite volume simulations in a controlled way.
Another quantity of interest is the so--called constrained effective
potential which in the context of the non--linear $\sigma$--model has been
studied by G{\"o}ckeler and Leutwyler [126]. A numerical simulation has been
presented by Dimitrovic et al. [127]. They found good agreement for the shape
of the constrained effective potential in the vicinity of its minimum, but
also noted that this method is not as
accurate in determining the low energy
constants
$\Sigma$ and $F$ than the one presented in [125].

Hansen [128] has extended this analysis to theories with $SU(N) \times SU(N)$
broken symmetries and Hansen and Leutwyler [129] have studied the charge
correlations and topological susceptibility of QCD at finite volume and
temperature. It is conceivable that these predictions will become useful when
one will be able to perform QCD simulations with dynamical fermions of
reasonably small masses. At present, many Monte Carlo studies are done
in the quenched approximation (i.e. suppressing fermion loops). In that case,
the theoretical predictions are not applicable. Effective theories in the
quenched approximation are being developed by Sharpe [130] and Bernard and
Golterman [131]. Finally, Leutwyler and Smilga [132] have discussed the
spectrum of the Dirac operator and the role of the winding number in QCD.
The distribution of winding number and the spectrum of the Dirac operator
at small eigenvalues are related to the quark condensate at infinite volume,
which implies that the formation of the quark condensate is connected to the
occurrence of eigenvalues of the order $\lambda_n \sim 1/V$. What their
analysis can not provide is a reason why a condensate arises. For a more
detailed discussion of these and other topics, the reader should consult
ref.[132].
\bigskip
\noindent{\bf 5.5. An application to high--$T_c$ superconductivity}
\medskip
\goodbreak
To demonstrate the universality of the CHPT methods, let me briefly discuss
some work by Hasenfratz and Niedermayer [133] on the correlation length of
the anti--ferromagnetic (AF) $d=2+1$ Heisenberg model at low temperatures.
One of their motivations is the recent discovery of quasi--two--dimensional
anti--ferromagnets like the undoped $La_2 Cu O_4$ compound. In this crystal,
a single unpaired electron at each $Cu^{++}$ site makes the material
antiferromagnetic with a quasi--two--dimensional structure. This means that
the forces in the plane are much stronger than the forces between
neighbouring planes. As it is well known, in doped $La_2 Cu O_4$ one observe
high $T_c$ superconductivity. Doping {\it e.g.}
with $Sr$ atoms renders the planes
metallic and leads to the superconductivity. To understand this phenomenon
one has to understand first the physics of the undoped $La_2 Cu O_4$ system.
Due to the particular structure of this crystal, one can use CHPT methods to
understand its low temperature behaviour. It should be well described by the
$d = d_s + 1 = 2 + 1$ dimensional quantum Heisenberg model,
$$ {\cal{H}} = J \sum_{<ij>} \vec{S}_i \vec{S}_j \, \, \, \, (J > 0)
\eqno(5.28)   $$
where $\vec{S}_i$ is the spin operator at site $i$ and $\vec{S}_i^2 =
S(S+1)$. In the case of $La_2 Cu O_4$, one has $S = 1/2$. The groundstate is
assumed to be antiferromagnetically ordered (which is strictly proven only
for $|S| \ge 1$). In the case of a non--vanishing value of the coupling
strength $J$, the rotational symmetry is broken and one has two (almost)
massless Goldstone bosons (magnons). These modes pick up a small mass since
the Mermin-Wagner-Coleman theorem [134] forbids massless Goldstone bosons in
two dimensions at finite temperature. These magnons interact weakly at low
temperatures, $T << J$. This makes an analytic study feasible. Note, however,
that at very low energies (much smaller than $T$) the magnon interaction starts
to become stronger again which leads to a finite mass gap in the theory. The
exact result for the mass gap of the $d=2$ O(3) non-linear $\sigma$--model
has been given in ref.[135] and can also be used in the present case. The
temperature dependence of the correlation length $\xi$ can now be calculated
from an effective Lagrangian which shares the properties of the quantum AF
model (5.28). The details are given in ref.[133] and the final result reads:
$$ \xi (T) = {e \over 8} {\hbar c \over 2 \pi F^2} \exp(2 \pi F^2 / T) \biggl[
1 - {1 \over 2}{T \over 2 \pi F^2} + {\cal{O}}\bigl( ({ T \over 2 \pi
F^2})^2 \bigr)	 \biggr]
\eqno(5.29)  $$
\midinsert
\vskip 12.0truecm
{\noindent\narrower \it Fig.~10:\quad
Inverse correlation length of undoped $La_2CuO_4$. The
solid line is the CHPT prediction, eq.(5.29), and the dashed one accounts
only for the first term
in the square brackets. The data are taken from
refs.[138,139].
\smallskip}
\vskip -1.0truecm
\endinsert \noindent
with $F^2$ the spin--stiffness and $c$ the spin--velocity (notice that in
$d=3$, $F$ has dimension $[{\rm energy}]^{1/2}$). The result (5.29) applies for
the temperature range $T_N < T << 2 \pi F^2$, with $T_N \simeq 300 \, K$ the
temperature until which the  $La_2 Cu O_4$ compound remains in the broken
phase. A recent measurement of the spin--velocity gives $\hbar c = (850 \pm
30) {\rm meV} \,
\AA$ [136]. Notice that the temperature dependence of $\xi (T)$ is
essentially given by the exponential prefactor (which was already predicted
in ref.[137]), at 600 K  the correction $T / 4 \pi F^2$ is $\sim 17$ percent.
The important prefactor $e / 8$ comes from the exact form of the mass gap. In
fig.10, a one--parameter fit to the recent data is shown, one finds for the
spin--stiffness $2 \pi F^2 = 150 \, \, {\rm meV} =
1740 \AA$ [138]. Clearly, the fit is
not perfect, but this can not be expected. Anisotropies in the spin--spin
interactions might lead to such deviations. Phenomenologically, it was	found
that for doped samples the following rule holds: $\xi^{-1} (x,T) =
\xi^{-1} (x,0) + \xi^{-1} (0,T)$ with $x$ the doping concentration and the
formula (5.29) applies to $x = 0$. For a further discussions, see refs.[137]
and the talk by Hasenfratz [139]. Wiese and Ying [140] have recently
performed a Monte Carlo simulation of the 2-$d$ antiferromagnetic
quantum Heisenberg model. They compare the finite size and temperature
effects with the ones given by the chiral perturbation theory
calculation
of Hasenfratz and Niedermayer [140]. They find $\hbar c = (845 \pm
32)$ meV$\AA$ and $2 \pi F^2 = (157 \pm 7)$ meV, in good agreement
with the data and the one--parameter fit described above. They also
show results for the ground state energy density and the staggered
magnetisation. This again demonstrates the usefulness of the EFT
approach.
\bigskip
\vfill
\eject
\noindent{\bf VI. BARYONS}
\medskip
\noindent{\bf 6.1. Relativistic formalism}
\medskip
\goodbreak
In this section, I will be concerned with the inclusion of baryons in
the effective field theory. The relativistic formalism dates back in the
early days, see {\it e.g.} Weinberg [10], Callan  et al.     [11],
Langacker and Pagels [14] and
others (for reviews, see Pagels [8] and
the book by Adler and Dashen [4]).
The connection to QCD Green functions
was performed in a systematic fashion by Gasser, Sainio and
${\rm \check S}$varc [142] (from here on referred as GSS)
and Krause [143]. As done in the GSS paper, I will
outline the formalism
in the two--flavor case, {\it i.e.} for the pion--nucleon ($\pi N$)
system. The extension to full flavor $SU(3)$ is spelled out in
appendix B.

There is a variety of ways to describe the transformation properties
of the spin--1/2 baryons under chiral $SU(2) \times SU(2)$. All
of them lead to the same physics. However, there is one most convenient
choice (this is discussed in detail in Georgi's book [21]). Combine the
proton $(p)$ and the neutron $(n)$ fields in an isospinor $\Psi$
$$ \Psi = \pmatrix { p \cr n \cr}  \eqno(6.1)$$
The Goldstone bosons are collected in the matrix--valued field $U(x)$.
In the previous sections, we had already seen that the self--interactions
of the pions are of derivative nature, {\it i.e.} they vanish at
zero momentum transfer. This is a feature we also want to keep for the
pion--baryon interaction. It calls for a non--linear realization of the
chiral symmetry. Following Weinberg [10] and Callan  et al. [11], we
introduce a matrix--valued function $K$.
It not only depends on the group
elements $L,R \in SU(2)_{L,R}$, but also on the pion field (parametrized
in terms of $U(x)$) in a highly non--linear fashion, $K = K(L,R,U)$.
Since $U(x)$ depends on the space--time coordinate $x$, $K$ implicitely
depends on $x$ and therefore the transformations related to $K$
are local. To be more specific, $K$ is defined via
$$ L u = u' K \eqno(6.2) $$
with $u^2 (x) = U(x)$ and $U'(x) = R U(x) L^\dagger = {u'}^2(x)$.
The transformation properties of the pion field induce a well--defined
transformation of $u(x)$ under $SU(2) \times SU(2)$. This defines $K$
as a non--linear function of $L$, $R$ and $\pi (x)$. $K$ is a realization
of $SU(2) \times SU(2)$,
$$ K = \sqrt{L U^\dagger R^\dagger} R \sqrt U	\eqno(6.3) $$
and the baryon field transforms as
$$ \Psi \to K(L,R,U) \Psi   \eqno(6.4) $$
The somewhat messy object $K$ can be understood most easily in terms of
infinitesimal transformations. For $K = \exp (i \gamma_a \pi_a)$,
$L = \exp (i \alpha_a \pi_a)
     \exp (i \beta_a \pi_a)$ and
$R = \exp (-i \alpha_a \pi_a)
     \exp (i \beta_a \pi_a)$ (with $\gamma_a$, $\alpha_a$, $\beta_a$
real) one finds,
$$ \vec \gamma = \vec \beta + i [\vec \alpha , \vec \pi] / F_\pi
+ {\cal O}({\vec \alpha}^2 ,  {\vec \beta}^2, {\vec \pi}^2)
\eqno(6.5) $$
which means that the nucleon field is multiplied with a function
of the pion field. This gives some credit to the notion that chiral
transformations are related to the absorption or emission of pions.
To construct now the lowest order effective Lagrangian of the pion--nucleon
system, one has to assemble all building blocks. These are the mesons
and baryon fields, $U(x)$ and $\Psi (x)$, respectively as well as the
appropriate covariant derivatives (I restrict myself here to the case
with only scalar and vector external fields. The more general
expressions are given in Appendix B),
$$ \eqalign{
\nabla_\mu U &= \partial_\mu U - i e {\cal A}_\mu
[ Q , U ]      \cr
D_\mu \Psi&= \partial_\mu  \Psi  + \Gamma_\mu \Psi \cr
\Gamma_\mu&= {1 \over 2} \biggl\lbrace
  u^\dagger (\partial_\mu
- ie{\cal A}_\mu Q) u	+
u   (\partial_\mu
- ie{\cal A}_\mu Q)
  u^\dagger
\biggr\rbrace \cr}
\eqno(6.6) $$
with $Q = {\rm diag}(1,0)$ the nucleon charge matrix and $v_\mu =
e (1 + \tau_3) {\cal A}_\mu /2$ the external vector field (${\cal A}_\mu$
denotes the photon field). $D_\mu$ transforms homogeneously under
chiral transformations, $D_\mu' = K D_\mu K^\dagger$.
The object $\Gamma_\mu$ is the so--called chiral connection. It is
a gauge field for the local transformations
$$ \Gamma_\mu' = K \Gamma_\mu K^\dagger + K \partial_\mu K^\dagger
\eqno(6.7) $$
The connection $\Gamma_\mu$ contains one derivative. One can also
form an  object of axial--vector type with one derivative,
$$ {1 \over 2} u_\mu = {i \over 2} (u^\dagger \nabla_\mu u -
u \nabla_\mu u^\dagger) = {i \over 2} \lbrace u^\dagger , \nabla_\mu
u \rbrace = i u^\dagger \nabla_\mu U u^\dagger
\eqno(6.8) $$
which transforms homogeneously, $u_\mu' = K u_\mu K^\dagger $.
The covariant derivative $D_\mu$ and the axial--vector object $u_\mu$
are the basic building blocks for the lowest order effective theory.
Before writing it down, let us take a look at its most general form.
It can be written as a string of terms with an even number of external
nucleons, $n_{ext} = 0, 2, 4, \ldots$. The term with $n_{ext} = 0$
obviously corresponds to the meson Lagrangian (2.9) so that
$${\cal L}_{eff} [\pi, \Psi, \bar \Psi]
= {\cal L}_{\pi \pi}
 + {\cal L}_{\bar \Psi \Psi}
 + {\cal L}_{\bar \Psi \Psi \bar \Psi \Psi}
 + \ldots
\eqno(6.9) $$
Typical processes related to these terms are pion--pion, pion--nucleon
and nucleon--nucleon scattering, in order. In what follows, we will
mostly be concerned with processes with two external nucleons,
$$ {\cal L}_{\bar \Psi \Psi} =
	  {\cal L}_{\pi N}  =	\bar \Psi (x) {\cal D}(x) \Psi (x)
\eqno(6.10) $$
The differential operator ${\cal D}(x)$ is now subject to a chiral
expansion. Its explicit form to lowest order follows simply by
combining the connection $\Gamma_\mu$ and the axial--vector $u_\mu$
(which are the objects with the least number of derivatives)
with the appropriate baryon bilinears
$$ \eqalign{
{\cal L}_{\pi N}^{(1)} &= \bar \Psi {\cal D}^{(1)} \Psi \cr
&= \bar \Psi (i \gamma_\mu D^\mu - \krig m + {i \over 2} \krig g_A
\gamma^\mu \gamma_5 u_\mu) \Psi \cr}
\eqno(6.11) $$
The effective Lagrangian (6.11) contains two new parameters. These are
the baryon mass $\krig m$ and the axial--vector coupling $\krig g_A$
in the chiral limit,
$$\eqalign{
m&= \krig m [1 + {\cal O} (\hat m) ] \cr
g_A&= \krig g_A [1 + {\cal O} (\hat m) ] \cr}
\eqno(6.12) $$
Here, $m = 939$ MeV denotes the physical nucleon mass and $g_A$
the axial--vector strength measured in neutron $\beta$--decay,
$n \to p e^+ \bar \nu_e$, $g_A \simeq 1.26$. The fact that $\krig m$
does not vanish in the chiral limit (or is not small on the typical
scale $\Lambda \simeq M_\rho$) will be discussed below. Furthermore,
the actual value of $\krig m$, which has been subject to much recent
debate, will be discussed in the context of pion--nucleon scattering.
The occurence of the constant $\krig g_A$ is all but surprising.
Whereas the vectorial (flavor) $SU(2)$ is protected at zero momentum
transfer, the axial current is, of course, renormalized. Together
with the Lagrangian ${\cal L}_{\pi \pi}^{(2)}$ (2.9), our lowest
order pion--nucleon Lagrangian reads:
$$ {\cal L}_1 = {\cal L}_{\pi N}^{(1)}
	      + {\cal L}_{\pi \pi}^{(2)}
\eqno(6.13) $$
To understand the low--energy dimension of ${\cal L}_{\pi N}^{(1)}$,
we have to extend the chiral counting rules of section 2.3 to the
various operators and bilinears involving the baryon fields. These
are:
$$\eqalign{
& \krig m = {\cal O}(1) \,\, , \, \, \Psi, \, \bar \Psi =
{\cal O}(1) \, \, , \, \,
D_\mu \Psi = {\cal O}(1) \cr
& \bar \Psi \gamma^\mu \gamma_5 \Psi
= {\cal O}(1) \, , \, \, \,
 \bar \Psi \sigma^{\mu \nu} \Psi = {\cal O}(1) \, , \, \, \,
 \bar \Psi \sigma^{\mu \nu} \gamma_5 \Psi = {\cal O}(1) \cr
& (i \barre D - \krig m) \Psi = {\cal O}(p) \, , \, \, \,
 \bar \Psi \gamma_5 \Psi = {\cal O}(p) \cr}
\eqno(6.14) $$
Here, $p$ denotes a generic nucleon $\underline {{\rm three}}$--momentum.
Since $\krig m$ is of order one, baryon four--momenta can never be
small on the typical chiral scale. Stated differently, any time derivative
$D_0$ acting on the spin--1/2 fields brings down a factor $\krig m$.
However, the operator
$(i \barre D - \krig m) \Psi$ counts as order ${\cal O}(p)$. The proof
of this can be found in ref.[143] or in the lectures [144].

To exhibit the physics content of the Lagrangian ${\cal L}_{\pi N}^{(1)}$
(6.11), let me parametrize the pions in the
``$\sigma$--model'' gauge
(which is a convenient choice),
$$ U =(\sigma + i \pi )/F \, , \, \, \, \sigma^2 + \pi^2 = F^2
\eqno(6.15) $$
and expand the connection $\Gamma_\mu$ as well as the axial--vector
$u_\mu$ in powers of the pion fields:
$$\eqalign{
i \Gamma_\mu & = v_\mu +
{i \over 8 F^2 } [ \pi \, , \, \partial_\mu \pi ]
 - { 1 \over 8 F^2} [\pi \, , \, [ \pi \, , \, v_\mu ]] + \ldots \cr
{1 \over 2} u_\mu & = - {1 \over 2 F} \partial_\mu \pi +
{i \over 2 F } [ v_\mu \, , \, \pi]
 - { 1 \over 16 F^3}
\lbrace \pi \, , \, \lbrace \pi \, , \, \partial_\mu \pi \rbrace \rbrace
+ \ldots \cr}
\eqno(6.16) $$
\midinsert
\vskip 21.0truecm
{\noindent\narrower \it Fig.~11:\quad
Feynman rules in the $\sigma$--model gauge. Nucleons,
pions and photons are denoted by solid, dashed and wiggly lines,
respectively.
\smallskip}
\vskip -1.0truecm
\endinsert
The corresponding vertices are shown in fig.11. Clearly, one recovers
some well--known vertices like the pseudovector (derivative) $\pi N$
coupling or the Kroll--Ruderman term which plays an important role
in pion photoproduction. Notice that all prefactors are given in
terms of the four lowest--order constants, $F$, $B$, $\krig g_A$
and $\krig m$. To lowest order, the Goldberger--Treiman relation
[145] is exact, $\krig g_A \krig m = \krig g_{\pi N} F$,
which allows us to write the
$\pi N$ coupling in the more familiar form $\sim \krig g_{\pi N}
\partial_\mu \pi^a$.

It goes without saying that we have to include pion loops, associated
with ${\cal L}_1$ given in (6.13). The corresponding generating
functional reads [142]
$$\eqalign{
\exp \lbrace i {\cal Z} \rbrace & = N \int [dU] \exp \biggl\lbrace
i \int dx {\cal L}_{\pi \pi}^{(2)} + i \int dx \bar \eta S^{(1)}
\eta \biggr\rbrace  \cr
{\cal D}^{(1)ac} S^{(1)cb} & = \delta^{ab} \delta^{(4)}(x - y)
\cr} \eqno(6.17) $$
with $\eta$, $\bar \eta$ external Grassmannian sources and $S^{(1)}$
the inverse nucleon propagator related to ${\cal D}^{(1)}$ (6.11).
$a$, $b$ and $c$ are isospin indices. This generating functional
can now be treated by standard methods. The details are spelled out
by GSS [142]. I will now concentrate on the low--energy structure of
the effective theory which emerges. Pion loops generate divergences,
so one has to add counterterms. This amounts to
$$\eqalign{
{\cal L}_1 & \to {\cal L}_1 + {\cal L}_2 \cr
{\cal L}_2 & =	\Delta {\cal L}_{\pi N}^{(0)}
	     +	\Delta {\cal L}_{\pi N}^{(1)}
	     +	       {\cal L}_{\pi N}^{(2)}
	     +	       {\cal L}_{\pi N}^{(3)}
	     +	       {\cal L}_{\pi \pi}^{(4)} \cr} \eqno(6.18) $$
Let me first discuss the last three terms on the r.h.s. of
eq.(6.18). These are the expected ones. The structure of the $\pi N$
interaction allows for odd powers in $p$, so starting from
${\cal L}_{\pi N}^{(1)}$ to one--loop order one expects couterterms
of dimension $p^2$ and $p^3$. A systematic analysis of all these terms
has been given by Krause [143]. The first two terms,
$ \Delta {\cal L}_{\pi N}^{(0)}$
and $ \Delta {\cal L}_{\pi N}^{(1)}$, are due to the fact that the
lowest order coefficients $\krig m$ and $\krig g_A$ are renormalized
(by an infinite amount) when loops are considered. This is completely
different from the meson sector, the constants $B$ and $F$ are
$\underline {{\rm not}}$ renormalized by the loops. The origin of this
complication lies in the fact that the nucleon mass does not vanish in
the chiral limit. To avoid any shift in the values of $\krig m$
and $\krig g_A$ one thus has to add
$$ \Delta {\cal L}_{\pi N}^{(0)} = \Delta \krig m
\biggl( { \krig m \over F} \biggr)^2 \bar \Psi \Psi \, \, \, , \, \, \,
   \Delta {\cal L}_{\pi N}^{(1)} = \Delta \krig g_A
\biggl( { \krig m \over F} \biggr)^2 {1 \over 2}
\bar \Psi i \gamma^\mu \gamma_5
u_\mu \Psi  \eqno(6.19) $$
\midinsert
\vskip	8.0truecm
{\noindent\narrower \it Fig.~12:\quad
Chiral expansion for the $\pi N$ scattering amplitude,
$T_{\pi N} \sim p^N$. Tree graphs contribute at $N$ = 1, 2, 3 $\ldots$,
$n$--loop graphs at $N$= 2,3, $\ldots$ (after mass and coupling constant
renormalization). The contributions from 2,3,$\ldots$ loops are analytic
in the external momenta at order $p^3$ (here, $p$ is a pion four-,
nucleon three--momnentum or the pion mass).
\smallskip}
\vskip -1.0truecm
\endinsert
The first term in (6.19) can be easily worked out when one considers
the nucleon self--energy $\Sigma_{N}(p)$ related to the nucleon
propagator via $S(p) = [\krig m -\barre p - \Sigma_N (p)]^{-1}$
in the one--loop approximation [142]. The low--energy structure
of the theory in the presence of baryons is much more complicated
than in the meson sector. This becomes most transparent when one
compares the $\pi \pi$ and $\pi N$ scattering amplitudes, $T_{\pi \pi}$
and $T_{\pi N}$, respectively. While $T_{\pi \pi}^{{\rm tree}}\sim
p^2$ and
$T_{\pi \pi}^{{\rm n-loop}} \sim (p^2)^{n+1}$, the corresponding
behaviour for $T_{\pi N}$ is shown in fig.12  [142]. Here, $p$ denotes
either a small meson four--momentum or mass or a nucleon three--momentum.
Tree graphs for $T_{\pi N}$ start out at order $p$ followed by a
string of higher order corrections $p^2$, $p^3$, $\ldots$.
One--loop graphs start out at order $p^2$ (after appropriate
mass and coupling constant renormalization) and are non--analytic
in the external momenta at order $p^3$ (in the chiral limit $\hat m =0$).
Higher loops start out at $p^2$ and are analytic to orders
${\cal O}(p^2 \, , \, p^3)$. This again means that the low--energy
constants associated to ${\cal L}_{\pi N}^{(2,3)}$ will
get renormalized.
Evaluation of one--loop graphs associated with ${\cal L}_1$
therefore produces all non--analytic terms in the external momenta
of order $p^3$ like {\it e.g.} leading threshold or branch point
singularities. Let us now consider the case $\hat m \ne 0$.
Obviously, the $\pi N$ amplitude also contains terms which are
non--analytic in the quark masses. A good example is the Adler--Weisberger
relation in its differential form -- it contains a factor $
F_\pi^{-2}$ and therefore a term which goes like
$\hat m \ln \hat m$. Due to the complicated low--energy structure of the
meson--baryon
system, it has never been strictly proven that one--loop graphs
generate all leading infrared singularities, in particular the ones
in the quark masses. However, in all calculations performed so far the
opposite has never been observed. In any case, the exact one--to--one
correspondence between the loop and small momentum expansion is not
valid in the meson--baryon system if one treats the baryons fully
relativistically. This can be overcome, as will be discussed in the
next section, in an extreme non--relativistic limit. Here, however, I
wish to point out that the relativistic formalism has its own advantages.
Two of them are the direct relation to dispersion theory and the
inclusion of the proper relativistic kinematics in certain processes.
These topics will be discussed later on. Let me make a few remarks
concerning the structure of the nucleons (baryons) at low energies.
Starting from a structureless Dirac field, the nucleon is surrounded
by a cloud of pions which generate {\it e.g.} its anomalous magnetic
moment (notice that the lowest order effective Lagrangian (6.11) does
only contain the coupling of the photon to the charge). Besides the
pion loops, there are also counterterms which encode the traces of
meson and baryon excitations contributing to certain properties of the
nucleon. Finally, one point which should be very clear by now: One can
only make a firm statement in any calculation if one takes into account
$\underline {{\rm all}}$ terms at a given order. For a one--loop
calculation in the meson--baryon system, this amounts to the tree terms
of order $p$, the loop contributions of order $p^2$, $p^3$ and the
counterterms of order $p^2$ and $p^3$. This should be kept in mind
in what follows.
\bigskip
\noindent{\bf 6.2. Non--relativistic formalism}
\medskip
\goodbreak
As we saw, the fully relativistic treatment of the baryons leads to
severe complications in the low--energy structure of the EFT.
Intuitively,
it is obvious how one can restore the one--to--one correspondence
between the loop and the small momentum expansion. If one considers
the baryons as extremely heavy, only the relative pion momenta will
count and these can be small. The emerging picture is that of a very
heavy source surrounded by a cloud of light (almost massless) particles.
This is exactly the same idea which is used in the so--called heavy
quark effective field theory methods used in heavy quark physics. To
be precise, consider a meson of the type $Q \bar q$ (or $\bar Q q)$,
with $Q = b$, $t$ the heavy and $q = u$, $d$ the light quark.
The light quarks and the gluons are a cloud around the heavy source
$Q$, they form the so--called ``brown mug''. This is the picture
underlying the so--called heavy quark symmetries in QCD [146] which
allows {\it e.g.} to relate a set of form factors to one invariant
function (sometimes called the ``Isgur--Wise function'').
Therefore, it appears natural to apply the insight gained from
heavy quark EFT's to the pion--nucleon sector. Jenkins and Manohar
[147,148,149] have given a new formulation of baryon CHPT based on
these ideas. It amounts to taking the extreme non--relativistic
limit of the fully relativistic theory and an expansion in powers of the
inverse baryon mass. Obviously, the relativistic theory and its extreme
non--relativistic limit  are connected by a series of
Foldy--Wouthuysen transformations [150].

Let me first spell out the underlying ideas before I come back to the
$\pi N$ system. Our starting point is a free Dirac field with mass $m$
$${\cal L} = \bar \Psi (i \barre \partial - m) \Psi    \eqno(6.20)$$
Consider the spin--1/2 particle very heavy. This allows to write its
four--momentum as
$$p_\mu = m v_\mu + l_\mu	  \eqno(6.21) $$
with $v_\mu$ the four--velocity satisfying $v^2 = 1$ and $l_\mu$
a small off--shell momentum, $v \cdot l \ll m$ (for the present
discussion we can set $l_\mu = 0$).
One can now construct eigenstates of the velocity projection
operator $P_v = ( 1 + \barre v )/2$ via
$$\eqalign{
& \Psi = {\rm e}^{-imv\cdot x} \, (H + h) \cr
& \barre v H = H \, , \, \, \, \ \barre v h = - h
\cr} \eqno(6.22) $$
which in the nucleon rest--frame $v_\mu = (1,0,0,0)$ leads to the
standard non--relativistic reduction of a spinor into upper
and lower components. Substituting (6.22) into (6.20) one finds
$${\cal L} = \bar H(i v \cdot \partial) H
	   - \bar h(i v \cdot \partial + 2 m ) h
	   + \bar H i \barre \partial^\perp h
	   + \bar h i \barre \partial^\perp H
\eqno(6.23) $$
with $\barre \partial^\perp$ the transverse part of the Dirac operator,
$\barre \partial = \barre v (v \cdot \partial) + \barre \partial^\perp$.
Form eq.(6.23) it follows immediately that the large component
field $H$ obeys a free Dirac equation
$$v \cdot \partial H = 0    \eqno(6.24) $$
modulo corrections which are suppressed by powers of  $1/m$.
A more elegant path integral formulation is given by Mannel et al. [151].
There is one more point worth noticing. In principle, the field $H$
should carry a label '$v$' since it has a definite velocity. The Lorentz
invariant Lagrangian ${\cal L}$ follows from ${\cal L}_v = {\cal L}
[\bar H_v, H_v]$ via a suitable integration,
$${\cal L} = \int {d^3 v \over 2v_0} {\cal L}_v
\eqno(6.25)$$
This is of importance when the incoming and outgoing baryons do not
have the same velocity. For most of the purposes to be discussed we
do not need to worry about the label '$v$' and will therefore drop it.

Let me now return to the $\pi N$ system. The reasoning is completely
analogous to the one just discussed. I follow here the systematic
analysis of quark currents in flavor $SU(2)$ of Bernard et al. [152].
Starting from eq.(6.11) we write $\Psi$ in the form of eq.(6.22)
and eliminate the small component field $h$ via its equation of motion,
  $$h = {1\over2}(1 - \barre v) {1\over 2 \krig m} \bigl(i \barre D + {\krig
  g_A\over 2} \barre u \gamma_5 \bigr) H + {\cal O}({1/ \krig m^2})
\eqno(6.26)$$
This leads to
  $$\eqalign{
  {\cal L}^{(1)}_{\pi N} =& \bar H \bigl( i v \cdot D + \krig g_A S\cdot u
  \bigr) H   \cr & + {1\over 2 \krig m} \bar H \bigl(i\barre D + {\krig g_A
 \over
  2} \barre u \gamma_5 \bigr) {1 - \barre v\over 2} \bigl( i \barre D + {\krig
  g_A\over 2} \barre u \gamma_5 \bigr) H + {\cal O}({1 / \krig m^2}) \cr}
\eqno(6.27)$$
As advertised, the nucleon mass term has disappeared. Furthermore,
any Dirac bilinear $\bar \Psi \Gamma_\mu \Psi$ ($\Gamma_\mu = 1$,
$\gamma_\mu$, $\gamma_5$, $\ldots$) can be expressed in terms of the
velocity $v_\mu$ and the spin--operator $2 S_\mu = i \gamma_5
\sigma_{\mu \nu} v^\nu$. It obeys the relations
  $$S\cdot v = 0, \, \,  S^2 = -{3\over 4}, \, \,  \bigl\{ S_\mu , S_\nu
\bigr\}
   = {1\over 2} \bigl( v_\mu v_\nu - g_{\mu \nu}\bigr), \,
  \, \,  [S_\mu , S_\nu]
   = i \epsilon^{\mu \nu \alpha \beta} v_\alpha S_\beta
\eqno(6.28)$$
Using the convention $\epsilon^{0123} = -1$, we can rewrite the
standard Dirac bilinears as [147, 152]:
$$\eqalign{
\quad  &\bar H \gamma_\mu H = v_\mu \bar
  H H, \,\, \bar H \gamma_5 H = 0 ,\,\,
  \bar H \gamma_\mu \gamma_5 H = 2 \bar H
  S_\mu H \cr
\quad & \bar H\sigma^{\mu\nu} H = 2 \epsilon^{\mu\nu\alpha\beta}
 v_\alpha
  \bar H S_\beta H , \, \, \bar H \gamma_5 \sigma^{\mu \nu} H =
  2i(v^\mu \bar H S^\nu H - v^\nu \bar H S^\mu H) \cr}
\eqno(6.29) $$
Therefore, the Dirac algebra is extremely simple in the extreme
non--relativistic limit. The disappearance of the nucleon mass term to
leading order in $1/ \krig m$ now allows for a consistent chiral power
counting. The tree level Lagrangian is of order ${\cal O}(q)$ and
one--loop diagrams contribute to order ${\cal O}(q^3)$, with $q$
now a genuine small momentum (for the baryons, only the small
$l_\mu$ counts). The power counting scheme is discussed in detail
by Jenkins and Manohar [149] based on the chiral quark model approach
of Manohar and Georgi [20]. In essence, since all momenta can be
considered small compared to the chiral scale $\Lambda \simeq M_\rho$
one has a scheme as in the meson sector, with the main difference
that odd powers in $q$ are allowed here. An important point is that
all vertices now consist of a string of operators with increasing
powers in $1/m$. For the calculation of one--loop diagrams, one needs,
however, only the mass independent propagator and vertices,
we have for example
$$\eqalign{
\noindent &{\rm Nucleon \, \, propagator}:
\quad	S(\omega) = {i \over v \cdot l + i
\eta} \, , \, \, \eta >0 \cr
&{\rm Photon-nucleon \, \, vertex}:
\quad	i e { 1 + \tau_3 \over 2}
\epsilon \cdot v + {\cal O}(1/\krig m) \cr
&{\rm Pion-nucleon \, \, vertex}:
\quad	(\krig g_A / F) \tau^a S \cdot q
 + {\cal O}(1/\krig m) \cr}
 \eqno(6.30) $$
with $\omega = v \cdot l$. For the contact terms, one therefore has to
go further in the $1/m$ expansion to achieve the same accuracy.
This means that in case of a one--loop calculation one has to take
into account all local contact terms which contribute to the order
the loops do. A consequence is
that the effective Lagrangian written down
by Jenkins and Manohar [147] is not complete since it only contains
terms with one derivative or one quark mass insertion, i.e.
${\cal L}_{\pi N}^{(1)}$ and parts of
${\cal L}_{\pi N}^{(2)}$. However one should also account for
${\cal L}_{\pi N}^{(3)}$. To make this statement more transparent,
consider the isovector anomalous magnetic moment of the nucleon. The
isovector--vector quark current can be expressed in terms of two
form factors. In the heavy mass limit, this is most conveniently
done in the Breit frame where the photon transfers no energy since it
allows for a unique translation of Lorentz--invariant matrix elements
into non--relativistic ones. For the case at hand, this gives
  $$\eqalign{
  <N(p')| \bar q \gamma_\mu \tau_3 q| N(p)> =&  \bigl\{ F_1^V(t) + {t\over 4
  m_N^2} F_2^V(t) \bigr\} \, v_\mu \bar H \tau_3 H \cr & +
  {1\over m_N} \bigl\{ F_2^V(t) + F_1^V(t) \bigr\} \,\bar H [S_\mu,
 S\cdot(p'-p)]
  \tau_3 H \,.\cr } \eqno(6.31)$$
Consider now the Pauli form factor $F_2^V (t)$. At zero momentum
transfer, it gives the isovector anomalous magnetic moment,
$F_2^V (0) = \kappa_V = \kappa_p - \kappa_n = 3.71$.
The calculation of $F_2^V(t)$ is straightforward [152], at $t=0$ one
finds
  $$\kappa_V = c_6' - {\krig g_A^2 M \krig m \over 4 \pi F^2 }
  \eqno(6.32) $$
where the second term on the r.h.s. of (6.32) is the one--loop
contribution (which is non--analytic in the quark mass since $M \sim
\sqrt { \hat m}$ [155]) and the constant $c_6'$ has to be identified
with the anomalous isovector magnetic moment in the chiral limit.
It originates from a counterterm of ${\cal L}_{\pi N}^{(2)}$,
  $${\cal L}^{(2)}_{\pi N} = {c'_6+1\over 4 \krig m}
 \epsilon^{\mu\nu\alpha\beta}
  v_\mu \bar H S_\nu f^+_{\alpha \beta} H \eqno(6.33)$$
  where the field strength tensor $f_{\alpha \beta}^+$ is defined via
 $f_{\alpha \beta}^+ = u^\dagger F_{\alpha \beta} u^\dagger +
  u F_{\alpha \beta} u$ with $F_{\alpha \beta}$ the conventional
  photon fields strength tensor. Choosing $c_6' = 5.62$, one
reproduce the empirical value of $\kappa_V$. This value of
$\krig \kappa_V = c_6'$ is close to what one would estimate if one
generates the contact term (6.33) from $\rho$--exchange since the
tensor coupling of the $\rho$ is about $\kappa_\rho \simeq 6$.
Other observables and the interplay between the loop and counterterm
contributions are discussed in ref.[152].

Another topic is the matching to the relativistic formalism. This
has been addressed in ref.[152]. The underlying assumption is that
the extreme non--relativistic formulation should match to the relativistic
one (this point of view is not shared by everybody). In that case,
one can derive matching conditions between the various low--energy
constants. This allows one to use previously determined values of
these coefficients also in the non--relativistic approach. In fact,
these low--energy constants depend on the scale of dimensional
regularization, called $\lambda$ here. To be specific, consider the
wave--function renormalization ($Z$--factor) of the nucleon. It is
defined by the residue of the propagator at the physical mass pole,
  $${i\over p\cdot v - \krig m - \Sigma(\omega) } \to {iZ_N \over p\cdot v -
 m_N}
\eqno(6.34) $$
with $\Sigma (\omega)$ the nucleon self--energy in the one--loop
approximation. One finds
  $$Z_N = 1 + \Sigma'(0) = 1 - {3\krig g_A^2 M^2 \over 32 \pi^2 F^2} \biggl[
 3\ln
  {M\over \lambda} +1 \biggr] - {4M^2 \over F^2} {b'}^r_{12}(\lambda)
\eqno(6.35)$$
where the coupling constant $b_{12}'$ has eaten up the infinity from
the loop [142,152].
If one compares (6.35) to the expression of GSS
[142] expanded in powers of $\mu = M/\krig m$, one finds
  $$Z_N^{rel} = 1 - {3\krig g_A^2 M^2 \over 32 \pi^2 F^2} \biggl[ \ln {M\over
  \lambda} +2 \ln \mu + 2 \biggr]
  - {4M^2 \over F^2} b^r_{12}(\lambda) + {\cal
  O}(\mu^3)\eqno(6.36)$$
 Therefore, the heavy mass and the relativistic theory only match if one
sets
$$\eqalign{
\lambda & = \krig m \cr
     b{'}_{12}&= b{'}^r_{12}(\lambda)
  + \gamma{'}^b_{12} L, \qquad \gamma{'}^b_{12}  = -
  {9\over 8} \krig g_A^2 \cr}  \eqno(6.37)$$
The result $\lambda = \krig m$ is not quite unexpected since one has
integrated out the field $h$ of mass $2 \krig m$. Therefore, if one
chooses to retain only the leading non--analytic contributions
from the loops (as it is often done), one has to set $\lambda =
\krig m$. However, as stressed many times before, an accurate CHPT
calculation at a given order has to account for all loop and
counterterm contributions to the order one is working. A more detailed
discussion of the interplay between the relativistic and the heavy mass
formulation is given in ref.[152].

One particular advantage of the heavy mass formulation is the fact that
it is very easy to include the	baryon decuplet, i.e. the spin--3/2
states. This has been done in full detail by Jenkins and Manohar [148].
The inclusion of the $\Delta(1232)$ is motivated mostly by the
following argument: The $N \Delta$ mass--splitting $m_\Delta - m_N$
is only about twice as much as the pion mass, so that one expects
significant contributions from this close--by resonance (the same
holds true for the full decuplet in relation to the octet). This
expectation is borne out in many phenomenological models. However,
it should be stressed that if one chooses to include this baryon
resonance (or the full decuplet), one again has to account for all
terms of the given accuracy one aims at, say ${\cal O}(q^3)$ in a
one--loop calculation. This tends to be overlooked in the presently
available literature. Furthermore, the mass difference $m_\Delta -
m_N$ does not vanish in the chiral limit thus destroying the consistent
power counting (as it is the case with the baryon mass in the
relativistic formalism discussed in section 6.1).
I will not consider the full baryon octet
coupled to the decuplet  but will
only focus on the physics of the $\pi N \Delta$ system. The decuplet
fields are described by Rarita--Schwinger spinor $(\tilde T^\mu)_{abc}$
with $a$, $b$, $c \in \lbrace 1$, $2$, $3 \rbrace$. This spinor
contains both spin--1/2 and spin--3/2 components. The spin--1/2 pieces
are projected out by use of the constraint $\gamma_\mu \tilde T^\mu = 0$.
Under $SU(3) \times SU(3)$ $\tilde T^\mu$ transforms as $(\tilde T^\mu)_{abc}
\to (K)_a^d (K)_b^e (K)_c^f (\tilde T^\mu)_{def}$ and one defines a
velocity dependent field via
$$ \tilde T^\mu = {\rm e}^{-i m_\Delta v \cdot x} \, (T + t)^\mu
\eqno(6.38) $$
with $m_\Delta$ the $\Delta(1232)$mass (in general one should use the
average decuplet mass). In terms of the physical states we have
$T_{111} = \Delta^{++}$,
$T_{112} = \Delta^{+}/\sqrt 3$,
$T_{122} = \Delta^{0}/\sqrt 3$,
$T_{222} = \Delta^{-}$. The effective $\Delta N \pi$ Lagrangian to
leading order reads
$${\cal L}_{\Delta N \pi}^{(1)} = - i \bar T^\mu v \cdot D T_\mu
+ \Delta \bar T^\mu T_\mu + {3 \krig g_A \over 2 \sqrt{2} }
(\bar T^\mu u_\mu
H + \bar H u_\mu T^\mu)
\eqno(6.39)$$
with $\Delta = m_\Delta - m_N$. Clearly one is left with some residual
mass dependence from the $N \Delta$--splitting (notice that the average
octet--decuplet splitting is smaller than $\Delta$). In the language
of ref.[148] we have set ${\cal C} = 3 \krig g_A /2$  which is nothing
but the coupling constant relation $g_{\Delta N \pi} = 3 g_{\pi N} /
\sqrt 2 = 1.89$. Empirically this relation is fulfilled within a
few per cent. From the width of the decay $\Delta \to N \pi$ one has
${\cal C} = 1.8$ [149], consistent with the value given before (if one
uses the full decuplet the value of ${\cal C}$ reduces to 1.5). The
propagator of the spin--3/2 fields reads
$$ S_\Delta (\omega) = i { v^\mu v^\nu - g^{\mu \nu} - {4 \over 3}
S^{\mu} S^\nu \over \omega - \Delta}
\eqno(6.40) $$
For all practical purposes, it is most convenient to work in the
rest--frame $v_\mu = (1,0,0,0)$. In that case, one deals with the
well--known non--relativistic isobar model which is discussed in
detail in the monograph by Ericson and Weise [154]. We will come back to
some calculations involving the $\Delta (1232)$ (or the full decuplet)
in the following section.

A last comment on the heavy fermion EFT is necessary. While it is an
appealing framework, one should not forget that the nucleon (baryon)
mass is not extremely large. Therefore, one expects significant
corrections from $1/m$ suppressed contributions to many observables.
This will become more clear in the discussion of threshold pion
photo-- and electroproduction. It is conceivable  that going to
one--loop order ${\cal O}(q^3)$ is not sufficient to achieve a very
accurate calculation. This means that even more than in the meson sector
higher--loop calculations should be performed to learn about the
convergence of the chiral expansion. Ultimately, one might want to
include more information in the unperturbed Hamiltonian. At present,
it is not known how to do that but it should be kept in mind.
\bigskip
\noindent{\bf 6.3. Photo--nucleon processes}
\medskip
\goodbreak
In this section, I will be concerned with reactions involving photons,
nucleons and pions. There has been much impetus from the experimental
side in the recent years through precise measurements of the nucleons
electromagnetic polarizabilities or the recent data on threshold pion
photo-- and electroproduction.

Let us first consider nucleon Compton scattering at low energies.
The energy expansion of the spin--averaged
Compton amplitude in the rest
frame takes the form
$$T(\gamma N \to \gamma N) = - { e^2 Z^2 \over 4 \pi m}
+ \bar \alpha \, \omega'\omega \, \svec
\epsilon'\cdot \svec \epsilon + \bar \beta\,
( \svec \epsilon'\times
\svec k')\cdot ( \svec \epsilon \times \svec k) + {\cal O}(\omega^4)
\eqno(6.41)$$
with ($\omega, \svec k, \svec \epsilon$) and
($\omega', \svec k', \svec \epsilon'$) the frequencies, momenta and
polarization vectors of the incoming and outgoing photon, respectively.
The first term, which is energy--independent, is nothing but the
Thomson scattering amplitude as mandated by gauge invariance. It
constitutes a low--energy theorem as the photon energy tends to zero.
At next--to--leading order in the energy expansion, the photon probes
the non--trivial structure of the spin--1/2 particle (here, the
nucleon) it scatters off. This information is encoded in two structure
constants, the so--called electric ($\bar \alpha$) and magnetic
($\bar \beta$) polarizabilities. These have been determined rather
accurately over the last years [155]
$$\eqalign{
\bar \alpha_p & = 10.8 \pm 1.0 \pm 1.0 \, , \, \, \,
\bar \beta_p   = 3.4 \mp 1.0 \mp 1.0  \cr
\bar \alpha_n & = 12.3 \pm 1.5 \pm 2.0 \, , \, \, \,
\bar \beta_n   = 3.1 \mp 1.6 \mp 2.0  \cr}
\eqno(6.42) $$
all in units of $10^{-4}$ fm$^{-3}$ which I will use throughout. The
errors in (6.42) are anticorrelated since one imposes the well--established
dispersion sum rule results $(\bar \alpha+ \bar \beta)_p = 14.2 \pm 0.3$
and $(\bar \alpha+ \bar \beta)_n = 15.8 \pm 0.5$[156]. So the proton
and the neutron behave essentially as electric dipoles and their
respective sum of the polarizabilities is approximately equal.
For calculating the em polarizabilities, it is most convenient to
consider the spin--averaged Compton tensor in forward direction
[157,158]
$$\eqalign{ \quad
\theta_{\mu\nu}&= {1\over 4} \Tr \lbrace ( \barre p + m ) \,\,
		    \int d^4x \, e^{ik\cdot x} <N(p) | T\,J^{em}_\mu (x )
J^{em}_\nu(0) | N(p) > \, \,
\rbrace \cr
& =e^2\,\lbrace g_{\mu\nu} \, A(s) + k_\mu k_\nu\, B(s) + (p_\mu k_\nu +
p_\nu k_\mu) \, C(s) + p_\mu p_\nu \, D(s) \rbrace   \cr }
\eqno(6.43) $$
The pertinent Mandelstam variables are $s = (p+k)^2$ and $u=2m^2 -s$
since $t = 0$ in forward direction. Gauge invariance
($k_\mu \Theta^{\mu \nu} =0$) and $u \leftrightarrow s$ crossing
symmetry reduce the number of independent scalar functions in (6.43)
to two. It is straightforward to read off the polarizabilities
$$
\bar \alpha + \bar \beta =
-{e^2 m \over 2 \pi } \,\, A''(m^2) \,  , \, \,
\bar \beta = - {e^2 \over  4 \pi m } \, B(m^2)
\eqno(6.44) $$
where the prime denotes differentiation with respect to $s$. The
non--renormalization theorem of the electric charge leads to the
additional constraint $A(m^2) = Z^2$.

To one--loop in CHPT, the em polarizabilities are pure loop effects.
This means that there are no contributions from the contact terms
${\cal L}_{\pi N}^{(2,3)}$. Therefore, no undetermined low--energy
constants enter the final results and one tests the loop structure
of chiral QCD. In the relativistic formalism, the calculation of
$\bar \alpha_{p,n}$ and
$\bar \beta_{p,n}$ has been performed by Bernard et al. in refs.[157,158].
Although one has to evaluate 52/22 Feynman diagrams for the
proton/neutron, the final results are simple one dimensional integrals
due to the particular kinematics. In chiral $SU(2)$, the expansion
of the polarizabilities in powers of $\mu = M_\pi / m$ (ratio of the
pion and the nucleon mass) takes the form
$$\eqalign{
\bar \alpha_p&= C
\biggl\lbrace {5\pi \over 2\mu
} + 18\, {\rm ln} \mu + {33 \over 2}
+ {\cal O}(\mu) \biggr\rbrace \cr
\bar \alpha_n&= C
\biggl\lbrace {5\pi \over 2\mu
} + 6\,{\rm ln}\mu - {3\over 2}
+ {\cal O}(\mu) \biggr\rbrace \cr
\bar \beta_p&= C
\biggl\lbrace {\pi \over 4\mu
} + 18\, {\rm ln} \mu + {63 \over 2} + {\cal O}(\mu) \biggr\rbrace \cr
\bar \beta_n&= C
\biggl\lbrace {\pi \over 4\mu
} + 6\,{\rm ln}\mu + {5\over 2}
+ {\cal O}(\mu) \biggr\rbrace \cr
} \eqno(6.45)$$
with $C= e^2
g_{\pi N}^2 / 192 \pi^3 m^3  \simeq 0.26 \cdot 10^{-4}$ fm$^3$.
To leading order, the polarizabilities diverge as $1/M_\pi$ as
$M_\pi$ tends to zero. This is an expected result since the pion
cloud becomes long--ranged in the chiral limit, i.e. the Yukawa
suppression factors turn into a simple power law fall--off. In fact,
in the heavy mass formalism, it is only this most singular term
which results from the one--loop calculation [152]. Working in the
Coulomb gauge $\epsilon \cdot v = 0$, it is easy to filter out the
diagrams which lead to this singularity. These diagrams were also
found in the relativistic formalism, however, after tedious calculations.
The most singular term reproduces the data fairly well
$$\eqalign{
\bar \alpha_p & = \bar \alpha_n = {5 e^2 g_{\pi N}^2 \over 384 \pi^2
m^2 } { 1 \over M_\pi} = 13.6 \cr
\bar \beta_p & = \bar \beta_n = \bar \alpha_p /10 = 1.4 \cr}
\eqno(6.46) $$
i.e.
$(\bar \alpha + \bar \beta)_p =
 (\bar \alpha + \bar \beta)_n = 14.9$ close to the empirical values
and, furthermore,
$\bar \alpha_{p,n} \gg \bar \beta_{p,n}$. Some of the $1/m$
corrections modifying these leading order results are resummed
in the relativistic calculation, they tend to decrease the polarizabilities
and lead to too small values for the sums [157,158].However, there
is also the contribution from the $\Delta(1232)$ which is important
for $\bar \beta_{p,n}$ in many phenomenological models. Clearly, the
$1/m$ suppressed terms lead to isospin breaking ($\bar \alpha_p \ne
\bar \alpha_n$,
$\bar \beta_p \ne
\bar \beta_n$).
In a recent paper    [159] the $\ln \mu$--term in (6.45) was criticised
as incorrect since the nucleon mass appears in the logarithm. This is,
however, a false statement -- from the discussion of the matching
condition $\lambda = m$ it should be clear that such terms  are
indeed possible. Clearly, they are not the whole story at order $q^4$.
When one extends the calculation to flavor $SU(3)$, one has to consider
pion and kaon loops. This was for the first time done in ref.[160], where
also predictions for the electromagnetic polarizabilities of the
hyperons are given. For typical values of the $SU(3)$ $F$ and $D$
couplings (see appendix B), the kaon loops tend to increase $\bar \alpha_p$
over $\bar \alpha_n$ by about 15 per cent. A similar observation
was also made in ref.[159]. In that paper, in addition the contribution
from the decuplet ($\Delta$, $\Sigma$, $\Sigma^\star$) in the intermediate
states were considered (the $\Delta$--loop contribution had previously
been given by Kaiser [161]). The authors of ref.[159] finds
$\bar \alpha_p = 14.1$, $ \bar \alpha_n = 13.4$,
$\bar \beta_p = 0.2$ and $ \bar \beta_n = 0.1$ not too different
from the leading order results (6.46) (clearly, the magnetic
polarizabilities are most affected by the decuplet contribution).
However, unrealistically small values for the $F$ and $D$ coupling
constants are used in this calculation.
What has to be done here is to perform a full--fledged calculation
to include all terms up--to--and--including ${\cal O}(q^4)$. This
is possible in the heavy mass formalism and is the only way to get a
clean handle at the terms modifying the leading order ($q^3$) result
(6.46). In ref.[152] a prediction was also made for the spin--dependent
Compton amplitude $f_2(\omega)$ at low $\omega$ and in ref.[162],
the small momentum expansion of the Drell--Hearn--Gerasimov sum rule
was investigated. These predictions become relevant when the
scattering of polarized photons on polarized nucleons will be performed.
For further details, the reader is referred to refs.[152,162].

Next, let me consider threshold pion photo-- and electroproduction. In
the threshold region, i.e. when the real
photon has just enough energy to
produce the pion at rest, the differential
cross section for the process $\gamma (k)
+ p(p_1) \to \pi^0 (q)+ p(p_2)$ can be expressed in terms of the
electric dipole amplitude $E_{0+}$, $d \sigma / d\Omega = (E_{0+})^2$
as $\vec q$ tends to zero. This multipole is of particular interest
since in the early seventies a low--energy theorem (LET) for
neutral pion production was derived [163]. The recent measurements
at Saclay and Mainz [164] seemed to indicate a gross violation of
this LET, which predicts $E_{0+} = -2.3 \cdot 10^{-3} / M_{\pi^+}$
at threshold. However, the LET was reconsidered (and rederived)
and the data were reanalyzed, leading to
$E_{0+} = (-2.0 \pm 0.2) \cdot 10^{-3} / M_{\pi^+}$ in agreement
with the LET prediction. These developments have been subject
of a recent review by Drechsel and Tiator [165].
Therefore, I will focus here  on the additional insight gained from CHPT
calculations. In fact, the LET for neutral pion photoproduction at threshold
is an expansion in powers of $\mu = M_\pi / m \sim 1/7$ and predicts the
coefficients of the first two terms in this series, which are of order $\mu$
and $\mu^2$, respectively, in terms of measurable quantities like the
pion--nucleon coupling constant $g_{\pi N}$, the nucleon mass $m$ and the
anomalous magnetic moment of the proton,
$\kappa_p$.
In ref.[166] it was, however, shown that a certain class of loop
diagrams modifies the LET at next--to--leading order ${\cal O}(\mu^2 )$. This
analysis was extended to threshold pion electroproduction $\gamma^\star p \to
\pi^0 p$ in ref.[167]. In the electroproduction case, there is another
$S$--wave multipole called $L_{0+}$ due to the longitudinal coupling of the
virtual photon to the nucleon. The resulting expressions for $E_{0+}$  and
$L_{0+}$ to order ${\cal O}(\mu^2 )$ and ${\cal O}(\nu )$, with $\nu = -k^2 /
m^2 > 0$ ($k^2 < 0$ in the electroproduction reaction) read (notice that
$\mu$ is of order $q$ whereas $\nu$ is of order $q^2$):
$$\eqalign{ \quad
E_{0+} =			 C \biggl\{ &- \mu + \mu^2 \biggl[ {1\over 2}
(3 +
\kappa_p) + {m^2 \over 4\pi F_\pi^2 }\, \xi_2(-\nu / \mu^2) \biggr ] \cr
  &-\nu \biggl[ {1\over 2} (1 + \kappa_p) + {m^2\over 8\pi F_\pi^2}\, \bigl[
  2\xi_2(-\nu / \mu^2) - \xi_1(-\nu / \mu^2) \bigr ] \biggr]
  + {\cal O}(q^3) \biggr\} \cr
L_{0+} =			C  \biggl\{ &- \mu + \mu^2  \biggl[ {3\over 2}
+ {m^2 \over 8\pi F_\pi^2 }\, \xi_1(-\nu / \mu^2) \biggr]
  -{\nu \over 2} +
  + {\cal O}(q^3) \biggr\}  \cr  }
\eqno(6.47)$$
with $C= e g_{\pi N} / 8 \pi m = 0.17$ GeV$^{-1}$. The transcendental
functions $\xi_{1,2} (\alpha )$, $\alpha =   \nu / \mu^2$ can not be further
expanded in $\mu$ and $\nu$ separately. They are given by:
$$\eqalign{
\xi_1(\alpha) &= {1\over 1 + \alpha} + {\alpha \over 2(1+ \alpha)^{3/2} }
\biggl[ {\pi \over 2} + {\rm arcsin} {\alpha \over 2 + \alpha} \biggr] \cr
\xi_2(\alpha) &= {3\alpha\over 4( 1 + \alpha)^2} + {4 + 4\alpha +3 \alpha^2
\over 8(1+ \alpha)^{5/2} }
\biggl[ {\pi \over 2} + {\rm arcsin} {\alpha \over 2 + \alpha} \biggr] \cr }
\eqno(6.48)$$
\midinsert
\vskip 13.0truecm
{\noindent\narrower \it Fig.~13:\quad
a) Total cross section for $\gamma p \to \pi^0 p$ from the
one--loop calculation of Ref.[169].
The tree and
one-loop results are indicated by the dashed and solid line, respectively.
The data are from the Saclay
(closed circles) and
Mainz
(open circles) groups [164].
b) Same  with the inclusion
of isospin--breaking in comparison
with the Mainz data. Notice the weak cusp at $\pi^+ n$ threshold
($E_\gamma = 151.4$ MeV). After the $\pi^+ n$ threshold, the predicted
total
cross section agrees with the one given in Fig.13a.
\smallskip}
\vskip -1.0truecm
\endinsert \noindent
At the photon point $k^2 = 0$, one has $\nu = 0$ and $\xi_1 (0) = 1$,
$\xi_2 (0) = \pi / 4$. Therefore, to order $\mu^2$ one has a term $\sim (m/4
F_\pi )^2$ which modifies the LET of ref.[163]. To order $\mu^2$ and $\nu$,
the form (6.48) is the correct one as given by QCD. The origin of these novel
contributions proportional to $\xi_{1,2} (\alpha )$ can    most easily be
traced back in the heavy mass formalism. In ref.[152] it was shown that
working in the Coulomb gauge and the proton rest frame, only four diagrams
are non--vanishing for $\pi^0$ threshold production. These are the so--called
triangle and rescattering diagrams and their crossed partners. The
contribution of the triangle diagram is non--analytic in the quark mass,
$\Delta A_1 \sim M_\pi \sim \sqrt{\hat m}$ [166]. This non--analyticity is
generated by the loop integral associated to this diagram, it develops a
logarithmic singularity in the chiral limit. Therefore, the naive arguments
based on $s\leftrightarrow u$ crossing which lead to $\Delta A_1 \sim
M_\pi^2$ are invalid. To see this effect, one has to calculate loop diagrams
(this was known to Vainsthein and Zakharov [163], but they
unfortunately
calculated the harmless rescattering diagram). Notice also that the anomalous
magnetic moment $\kappa_p$ is generated by loops and counterterms in CHPT. It
would therefore be surprising if that would be the only such contribution --
and it is not. As a final comment on eq.(6.47) let me point out that the LETs
contain no photon--nucleon form factors [168]. These are build up order by
order in the chiral expansion and to order ${\cal O} (q^2)$, one only sees
their leading terms. In ref.[169], a complete analysis of threshold pion
photoproduction was given based on the relativistic formalism. This is
necessary since in the static limit one only gets the terms including order
$\mu^2$ in $E_{0+}$. A quick look at (6.47) reveals that the coefficient of
the $\mu^2$ term is large and so the leading order term $\sim \mu$ is
cancelled. This means that the expansion in $\mu$ is slowly converging and
one has to go beyond the extreme non--relativistic limit. The total cross
section for $\gamma p \to \pi^0 p$ in the threshold region is shown in
fig.13a. It agrees reasonably well  with the data. In that paper, one can
also find a prediction for the cross section of $\gamma n \to \pi^0 n$. This
process is even more sensitive to the chiral symmetry breaking since for the
electric dipole amplitude at threshold, $E_{0+} \sim  {\cal O}(\mu^2 )$. A
measurement of this reaction is intended using a deuteron target [170]. In
the one--loop approximation used in these calculations, one has no
isospin--breaking and therefore  $M_{\pi^0} = M_{\pi^+}$ and $m_p = m_n$. To
study some of the effects of isospin--breaking, in ref.[171] a calculation
was presented in which the physical masses for the charged and neutral pions
as well as for the proton and neutron were used (this is not a complete
calculation since at this order there are many other effects). In that case,
one has two thresholds, $\pi^0 p$ and $\pi^+ n$, separated by $\simeq 5$ MeV.
The cross section in this energy regime is shown in fig.13b. It is clearly
improved compared to the one--loop prediction. This indicates that a
calculation beyond the isosymmetric one--loop approximation
should be performed. Very recently, new accurate
data very close to threshold have become available for the electroproduction
process $\gamma^\star p \to \pi^0 p$ [172]. In particular, the $S$--wave
cross section $a_0$, which is a particular combination of $E_{0+}$ and
$L_{0+}$, was analyzed (for the first time in this kinematical regime). A one
loop CHPT calculation was performed by Bernard et al. [173] and it was shown
that loop effects are indeed necessary to explain the data as shown in
fig.14.
\midinsert
\vskip 8.0truecm
{\noindent\narrower \it Fig.~14:\quad
The $S$--wave component of the neutral pion electroproduction
cross section, calculated from CHPT
(solid line) and PV (dotted line) are compared. The kinematics is
$W = 1074$ MeV and $\epsilon = 0.58$ [173].
       The data extracted in ref.[172] are also shown.
\smallskip}
\vskip -1.0truecm
\endinsert

Consider now the photoproduction of charged pions. In that case, there exists
the LET due to Kroll and Ruderman [174] for the electric dipole amplitude
(for $\gamma p \to \pi^+ n$ and $\gamma n \to \pi^- p$).
The predictions are in
agreement with the data. The loop corrections are much more tame than in the
case of neutral pion production, they start at order ${\cal O}(\mu^2 , \mu^2
\ln \mu^2 )$ whereas  the leading term is of order one. In the case of
charged pion electroproduction, an interesting observation was made in
ref.[175]. The starting point is the venerable LET due to Nambu, Luri{\'e}
and Shrauner [176] for the isospin--odd electric dipole amplitude
$E_{0+}^{(-)}$ in the chiral limit,
$$
E_{0+}^{(-)}(M_\pi=0, k^2) ={e g_A \over 8 \pi
F_\pi} \biggl\lbrace 1 +{k^2 \over 6} r_A^2 + { k^2 \over 4m^2} (\kappa_V
+ {1 \over 2}) + {\cal O} (k^3) \biggr\rbrace
\eqno(6.49)$$
Therefore, measuring the reactions
$\gamma^\star p \to \pi^+ n$ and
$\gamma^\star n \to \pi^- p$ allows to extract
$E_{0+}^{(-)}$ and one can determine the axial radius of the nucleon, $r_A$.
This quantity measures the distribution of spin and isospin in the nucleon,
i.e. probes the Gamov--Teller operator $\vec \sigma \cdot \vec \tau$. A
priori, the axial radius is expected to be different from the typical
electromagnetic size,
$r_{{\rm em}} \simeq 0.8$ fm. It is customary to parametrize
the axial form factor $G_A (k^2)$ by a dipole form, $G_A (k^2) = (1 - k^2 /
M_A^2 )^{-2}$ which leads to the relation $r_A = \sqrt{12} / M_A$. The axial
radius determined from electroproduction data is typically $r_A = 0.59 \pm
0.04$ fm [177] whereas (anti)neutrino-nucleon reactions lead to somewhat
larger values, $r_A = 0.65 \pm 0.03$ fm. This discrepancy is usually not
taken seriously since the values overlap within the error bars. However, it
was shown in ref.[175] that pion loops modify the LET (6.49) at order $k^2$
for finite pion mass. In the heavy mass formalism, the coefficient of the
$k^2$ term reads
$$ {1 \over 6} r_A^2 + {1 \over 4 m^2}(\kappa_V +{1 \over2}) +
{1 \over 128 F_\pi^2} (1 - {12 \over \pi^2})
\eqno(6.50)$$
where the last term in (6.50) is the new one. This means that previously one
had extracted a modified radius, the correction being $3 (1 - 12/\pi^2 ) / 64
F_\pi^2 \simeq -0.046$ fm$^2$. This closes the gap between the values of
$r_A$ extracted from electroproduction and neutrino data. It remains to be
seen how the $1 / m$ suppressed terms will modify the result (6.50). Such
investigations are underway.
\bigskip
\noindent {\bf 6.4. Baryon masses and the $\sigma$--term}
\medskip
\goodbreak
In this section, I will mostly be concerned with some recent work on the
so--called pion--nucleon $\sigma$--term related to low--energy $\pi  N$
scattering. As an entr{\'e}e, however, it is
mandatory to briefly discuss the
chiral expansion of the baryon masses. As we will see later, the
$\sigma$--term is nothing  but a nucleon mass shift related to the finiteness
of the quark masses.

Gasser [55] and Gasser and Leutwyler [3] were the first to systematically
investigate the baryon masses at next--to--leading order. The quark mass
expansion of the baryon masses takes the form
$$m_B = m_0 + \alpha {\cal M} + \beta {\cal M}^{3/2} + \gamma {\cal M}^2 +
\ldots
\eqno(6.51)$$
The constant $m_0$  reminds us that the baryon masses do not vanish in	the
chiral limit. The non--analytic piece proportional to ${\cal M}^{3/2}$ was
first observed by Langacker and Pagels [141]. If one retains only the terms
linear in the quark masses, one obtains the Gell-Mann--Okubo relation
$m_\Sigma + 3 m_\Lambda = 2 (m_N + m_\Xi )$ (which is fulfilled within 0.6
per cent in nature) for the octet and the equal spacing rule for the
decuplet,
$m_\Omega- m_{\Xi^*} = m_{\Xi^*} - m_{\Sigma^*}
= m_{\Sigma^*} - m_\Delta$
(experimentally, one has 142:145:153 MeV). However, to extract quark mass
ratios
from the expansion (6.51), one has to work harder. This was done in
refs.[3,55]. The non--analytic terms were modelled by considering the baryons
as static sources surrounded by a cloud of mesons and photons -- truely the
first calculation in the spirit of the heavy mass formalism.  The most
important result of this analysis was already mentioned in section 3.2,
namely that the ration $R = (\hat m - m_s ) / (m_u - m_d )$ comes out
consistent with the value obtained from the meson spectrum.  Jenkins [179]
has recently repeated this calculation using the heavy fermion EFT of
refs.[147,148,149], including also the spin--3/2 decuplet fields in the EFT.
She concludes that the success of the octet and decuplet mass relations is
consistent with baryon CHPT as long as one includes the decuplet. Its
contributions tend to cancel the large corrections from the kaon loops like
$M_K^2 \ln M_K^2$. The calculation was done in the isospin limit $m_u = m_d =
0$ so that nothing could be said about the quark mass ratio $R$. One would
like to see such a refined analysis.

Let me now turn to the $\pi N$ $\sigma$--term. It is defined via
$$ \sigma = {\hat m \over 2 m } < p| \bar u u + \bar d d| p>
= \hat m {\partial m \over \partial {\hat m}}
\eqno(6.52)$$
making use of the Feynman--Hellmann theorem and the proton state $|p>$ is
normalized via $< p' | p > = 2 (2 \pi )^3 p_0 \delta ( \vec p - \vec p' )$.
The quantity $\sigma$ can be calculated from the baryon spectrum. To leading
order in the quark masses, one finds
$$ \eqalign{
\sigma&= { \hat m \over m_s - \hat m} {m_\Xi + m_\Sigma - 2 m_N
\over 1 - y} + {\cal O}({\cal M}^{3/2}) \cr
y&= { 2<p| \bar s s | p> \over <p| \bar u u + \bar d d| p> }
\cr}
\eqno(6.53)$$
where $y$ is a measure of the strange quark content of the proton. Setting
$y=0$ as suggested by the OZI rule, one finds $\sigma = 26$ MeV. However,
from the baryon mass analysis it is obvious that one has to include the
${\cal O}({\cal M}^{3/2})$ contributions and estimate the
${\cal O}({\cal M}^2 )$ ones. This was done by Gasser [55] leading to
$$\sigma = {35 \pm 5 \, {\rm MeV} \over 1 - y }
= {\sigma_0 \over 1 - y }
\eqno(6.54)$$
However, in $\pi N$ scattering one does not measure $\sigma$, but a
quantity called $\Sigma$ defined via
$$\Sigma = F_\pi^2 \bar D^+ (\nu = 0, t = 2 M_\pi^2)
\eqno(6.55)$$
with  the bar on $D$ denoting that the pseudovector Born terms have been
subtracted, $\bar D = D - D_{\rm pv}$. The amplitudes $D^\pm$ are related to
the more conventional $\pi N$ scattering amplitudes $A^\pm$ and $B^\pm$ via
$D^\pm = A^\pm + \nu B^\pm$, with $\nu = (s-u) / 4m$. The superscript '$\pm$'
denotes the isospin  (even or odd). $D$ is useful since it is related to the
total cross section via the optical theorem. The kinematical choice $\nu = 0
, t = 2 M_\pi^2$ (which lies in the unphysical region) is called the
Cheng--Dashen point [180]. The relation between $\Sigma$ and the $\pi N$
scattering data at low energies is rather complex, see e.g. H{\"o}hler [181]
for a discussion or Gasser [182] for an instructive pictorial (given also in
ref.[144]). Based on dispersion theory, Koch [183] found $\Sigma = 64 \pm 8$
MeV (notice that the error only reflects the uncertainty of the method, not
the one of the underlying data). Gasser et al. [184] have recently repeated
this analysis and found $\Sigma = 60$ MeV (for a discussion of the errors,
see that paper). There is still some debate about this value, but in what
follows I will use the central result of ref.[184].  Finally, we have to
relate $\sigma$ and $\Sigma$. The relation of these two quantities is based
on the LET of Brown et al. [185] and takes the following form at the
Cheng--Dashen point:
$$\eqalign{
\Sigma&= \sigma + \Delta \sigma + \Delta R \cr
\Delta \sigma&= \sigma(2 M_\pi^2) - \sigma(0) \cr}
\eqno(6.56)$$
$\Delta \sigma$ is the shift due to the scalar form factor of the nucleon,
$<p' | \hat m (\bar u u + \bar d d ) | p> = \bar u (p') u (p) \sigma (t)$,
with $t = (p'-p)^2$ and $\Delta R$ is related to a remainder not fixed by
chiral symmetry. The latter was found to be very small by GSS [142], $\Delta
R = 0.4$ MeV. The one--loop calculation of ref.[142] for $\Delta \sigma$ gave
4.6 MeV, and in the heavy mass formulation one finds 7.4 MeV [152]. Adding
up the pieces, this would amount to $y \approx 0.3 \ldots 0.4$, i.e. a large
strange quark content in the nucleon.  However, since one is dealing with
strong $S$--wave $\pi \pi$ and $\pi N$ interactions, the suspicion arises
that the one--loop approximation is not sufficient, as already discussed in
section 4.4. Therefore, Gasser et al. [184,186] have performed a
dispersion--theoretical analysis tied together with CHPT constraints for the
scalar form factor $\sigma (t)$. The resulting numerical value is
$$\Delta \sigma = ( 15 \pm 0.5 ) \, {\rm MeV}
\eqno(6.57)$$
which is a stunningly large correction to the one--loop result. If one
parametrizes the scalar form factor as $\sigma (t) = 1 + r_S^2 \, t + {\cal
O} (t^2)$, this leads to $r_S^2 = 1.6$ fm$^2$, substantially larger than the
typical electromagnetic size. This means that the scalar operator
$\hat m (\bar u u + \bar d d )$ sees a
more extended pion cloud. Notice that for the pion, a similar enhancement of
the scalar radius was already observed, $(r_S^2 / r^2_{\rm em})_\pi \simeq
1.4$ [13]. Putting pieces together, one ends up with $\sigma = 45 \pm 8$ MeV
[184] to be compared with $\sigma_0 / (1-y) = ( 35 \pm 5 ) \, {\rm MeV} /
(1-y)$. This means that the strange quarks contribute approximately 10 MeV to
the $\sigma$--term and thus the mass shift induced by the strange quarks is
$m_s <p| \bar s s |p> \simeq (m_s / 2 \hat m ) \cdot 10 \, {\rm MeV} \simeq
130$ MeV. This is consistent  with the estimate made in ref.[187] based on
the heavy mass formalism including the decuplet fields. The effect of the
strange quarks is not dramatic. All speculations starting from the first
order formula (6.53) should thus be laid at rest. The lesson to be learned
here is that many small effects can add up constructively to explain a
seemingly large discrepancy like $\Sigma - \sigma_0 \approx \sigma_0$. What
is clearly needed are more accurate and reliable low--energy pion--nucleon
scattering data to further pin down the uncertainties. For an update on these
issues, see Sainio [188].
\bigskip
\noindent{\bf 6.5. Non-leptonic hyperon decays}
\medskip
\goodbreak
Since the early days  of soft--pion techniques, the non--leptonic hyperon
decays have been studied using EFT methods (see e.g. the monograph [189]).
There are seven such decays, $\Lambda \to \pi^0 n\, , \, \Lambda \to \pi^- p
\, , \, \Sigma^+ \to \pi^+ n\, , \, \Sigma^+ \to \pi^0 p\, ,
\Sigma^- \to \pi^- n \, , \, \Xi^0 \to
\pi^0 \Lambda$ and $\Xi^- \to \pi^- \Lambda$. These are compactly written as
$H_a^b$, where $H$ denotes the decaying hyperon, $a = {-,0,+}$ gives the
charge of the outgoing pion and $b = {-,0,+}$ denotes the charge of the
decaying hyperon. For example, $\Lambda_0^0$ stands for $\Lambda \to \pi^0
n$. These decays are produced by the strangeness--changing $\Delta S = 1$
weak Hamiltonian proportional to $\bar{u} \gamma^\mu (1+\gamma_5 ) s \bar{d}
\gamma_\mu~(1+\gamma_5 ) u$
(modulo QCD corrections). Under $SU(3) \times SU(3)$
this operator transforms as $(8_L , 1_R ) \oplus
(27_L , 1_R )$. Experimentally,
the octet piece dominates the decay amplitudes. Therefore, we will neglect
the 27--plet contributions in what follows. The invariant matrix elements
take the form
$$ M(H_a^b ) = G_F M_\pi^2 \bar{u}_{B'} \bigl\lbrace A^{(S)}(H_a^b ) + \gamma_5
A^{(P)}(H_a^b ) \bigr\rbrace u_B
\eqno(6.58)    $$
where $u_B$ denotes the spinor of baryon $B$ and $\bar{u}_B = u_B^+
\gamma^0$. The $S$--wave decay is parity--violating whereas the $P$--wave
decays conserve parity. Isospin symmetry of the strong interactions implies
three relations among the amplitudes, separately for the $S$-- and the
$P$--waves. These are $M(\Lambda_-^0 ) = - \sqrt{2} M(\Lambda_0^0 ) \, , \,
M(\Xi_-^- ) = -\sqrt{2} M(\Xi_0^0 )$ and $\sqrt{2} M(\Sigma_0^+ ) +
M(\Sigma_-^- ) = M(\Sigma_+^+ )$. Thus, only four amplitudes are independent.

To leading order in the effective theory, $\Delta S = 1$ non--leptonic
Hamiltonian reads [20,21]
$$ H_{\rm eff} (\Delta S = 1)
= a \Tr \bigl( \bar{B} {\lbrace u^+ h u , B \rbrace} \bigr)
+ b \Tr \bigl( \bar{B} [u^+ h u , B ] \bigr) + {\rm h.c.}
\eqno(6.59) $$
where the explicit expression for the baryon field $B$ is given in appendix
B and the trace runs over the flavor indices. The $SU(3)$ matrix $h$ is a
spurion field to project out the octet component, $h_i^j = \delta_2^j
\delta_i^3$. From (6.59) it follows for the $S$--waves that
$A^{(S)}(\Sigma_+^+ ) = 0 $, $A^{(S)}(\Sigma_-^- ) = (a-b) / F $,
$A^{(S)}(\Lambda_-^0 ) =-(a+3b) / \sqrt{6} F $ and
$A^{(S)}(\Xi_-^- ) = (3b-a)/ \sqrt{6} F $. These can be combined to give the
Lee--Sugawara relation [190],
$$  A^{(S)}(\Lambda_-^0 ) + 2  A^{(S)}(\Xi_-^- ) + \sqrt{3 \over 2}
A^{(S)}(\Sigma_-^- ) = 0
\eqno(6.60)    $$
which is experimentally quite well fulfilled (see below). No such simple
relation can be derived for the $P$--wave amplitudes since these always
involve pole diagrams in which the baryon changes strangeness before or after
pion emission. It is worth noting that to leading order, both the $S$-- and
the $P$--wave decay amplitudes are independent of the strange quark mass
(setting $m_u = m_d = 0$). Furthermore, it is not possible to achieve even a
descent fit for the $P$--wave amplitudes once the parameters $a$ and $b$ in
(6.59) are fixed to reproduce the $S$--waves. This together with a possible
solution involving higher dimensional operators is discussed in some detail
in refs.[20,21].

Naturally, the question arises what the chiral corrections to these decays
are. Bijnens et al. [191] were the first  to address this issue. Apart from
the standard lowest order chiral meson-baryon Lagrangian (see appendix B)
they used for the symmetry breaking the following term:
$${\cal L}^{(2)} = a_1 \Tr [ {\cal M} (U + U^\dagger ) ]
+ b_1 \Tr [ \bar{B} ( u^\dagger {\cal M} u^\dagger + u {\cal M} u) B ]
+ b_2 \Tr [ \bar{B} B ( u^\dagger {\cal M} u^\dagger + u {\cal M} u) ]
\eqno(6.61)    $$
and set ${\cal M} = {\rm diag} (0,0,m_s )$, i.e. only kaon loops
contribute. The parameters $a_1 \, , \, b_1 $ and $b_2$ were determined from
the mass spectrum. They worked in the one--loop approximation keeping only
the non--analytic terms of the type $m_s \ln m_s \sim M_K^2 \ln M_K^2$. No
local contact terms were considered. This, of course, introduces a scale
dependence, the subtraction point
was chosen to be $\mu = 1$ GeV. Tuning the parameters $a$
and $b$ (6.59) to get a best fit to the $S$--wave amplitudes, the following
results emerged. The one--loop correction was zero for $A^{(S)} (\Sigma_+^+
)$, of the order of 30 per cent for $A^{(S)} (\Sigma_-^- )$ and
$A^{(S)} (\Lambda_-^0 )$ and 75 per cent for $A^{(S)} (\Xi_-^- )$. The
resulting $S$--wave predictions are  no longer satisfactorily agreeing with
the data and the Lee--Sugawara relation breaks down. In the $P$--waves, the
corrections are even larger, in two cases much bigger than the tree result.
None of the $A^{(P)} (H_a^b )$ can be reproduced, the discrepancies to the
empirical numbers are large. This calculation sheds some doubt on the
validity of the chiral expansion in the three flavor sector. Jenkins [192]
has recently reinvestigated this topic. Her calculation differs in various
ways from the one of ref.[191]. First, she uses the heavy mass formalism and
also includes the spin--3/2 decuplet in the effective theory. Second, she
accounts for wave--function renormalization and third, includes     another
higher derivative term. This is nothing but the first term in (4.26) with
$\lambda_6 $ substituted by the spurion $h$. The reason to include this term
is the strong $\Delta I = 1/2$ enhancement. The other assumptions are the
same (no counter terms and only the leading non--analytic pieces from the
loops are retained. Also, $m_u = m_d = 0$ is assumed).
The main result of
this calculation is the large cancellation between the octet and decuplet
pieces in the loops (already observed in the baryon mass calculation [179]),
therefore the total loop contribution is considerably reduced. In units of
$G_F M_\pi^2$, the results of ref.[192] read: $A^{(S)} (\Sigma_+^+ ) = -0.09
\, ( 0.06 \pm 0.01 ) \, , \, A^{(S)} (\Sigma_0^+ ) =
-1.41 \, ( -1.43 \pm 0.05 ) \,
, \, A^{(S)} (\Sigma_-^- ) = 1.90 \, ( 1.88 \pm 0.01 ) \, , \,
A^{(S)} (\Lambda_-^0 ) = 1.44 \, ( 1.42 \pm 0.01 ) \, , \,
A^{(S)} (\Lambda_0^0 ) = -1.02 \, ( -1.04 \pm 0.01 ) \, , \,
A^{(S)} (\Xi_-^- ) = -2.04 \, ( -1.98 \pm 0.01 )$ and
$A^{(S)} (\Xi_0^0 ) = 1.44 \, ( 1.52 \pm 0.02 )$. Good agreement between theory
and experiment is found. Furthermore, the correction to the Lee--Sugawara
relation is small. In case of the $P$--wave amplitudes, the central
parameters do not lead to satisfactory description of the data. Indeed, the
$A^{(P)}$ are very sensitive to some of the input parameters. What is more
important is the observation that in the case of the $P$--waves, the tree
level prediction consists of two terms which tend to cancel to a large
extent. This suppression of the $SU(3)$ symmetric amplitudes effectively
enhances the loop corrections, the terms of the type $m_s \ln m_s$ are as
important as the tree contributions (also they are nominally suppressed).
This means that in the $P$--wave amplitudes one has large $SU(3)$ violation,
but not a breakdown of the chiral expansion. To get the same accuracy on
these $SU(3)$--violating pieces as on the $SU(3)$--symmetric ones which
dominate the $S$--wave amplitudes, one would have to work much harder.
Clearly, the last word is not spoken here since none of the calculations
performed so far accounts for $\underline {{\rm all}}$
terms at one loop order, the
leading non--analytic corrections are just one part of the whole story.

Neufeld [193] has recently considered the
strangeness--changing radiative decays $\Sigma^+ \to p \gamma$, $\Xi^- \to
\Sigma^- \gamma$, $\Sigma^0 \to n \gamma$, $\Lambda \to n \gamma$, $\Xi^0 \to
\Lambda \gamma$ and $\Xi^0 \to \Sigma^0 \gamma$ in the relativistic
formalism. The general form of the amplitude $B_b \to B_a + \gamma$ ($a,b =
1,\ldots ,8)$ takes the form
$$ M_{ab} = \bar{u}_a i \sigma^{\mu \nu} \epsilon_\mu (p_a - p_b)_\nu \,
( A_{ab} + B_{ab} \gamma_5 ) u_b
\eqno(6.62)  $$
The parity--conserving amplitude $A_{ab}$ can be expressed in terms of the
magnetic
moments and therefore the corresponding counterterms can be fixed. The
situation is different for the counterterms related to the parity--violating
amplitude $B_{ab}$. They are not known but restricted by CPS--symmetry [103].
The general structure of these counterterms is discussed in ref.[193]. It is
argued that the one--loop contributions together with the lowest order
contact terms (the pieces which are left for vanishing momenta) are not
sufficient to explain the existing data. However, the imaginary parts are
uniquely determined by the loop graphs. In particular, for the decays
$\Lambda \to n \gamma$ and $\Xi^- \to \Sigma^-
\gamma$ one can give a
prediction for the asymmetry parameter $\alpha_{ab} = 2 {\rm Re} (A_{ab}
B^*_{ab} ) / (|A_{ab}^2| + |B_{ab}^2|)$
based on the recent measurements for
the pertinent branching ratios. These constraints lead to
$$  -0.8 \le \alpha_{n \Lambda} \le 0.7 \, \, , \, \,
    -0.8 \le \alpha_{\Sigma^- \Xi^-} \le 0.8
\eqno(6.63)  $$
One can reduce these ranges to some extent by making some plausible
assumptions as discussed in ref.[193]. The main conclusion of that paper is
that to test the chiral predictions of the asymmetry parameters (6.63) a
measurement is called for.
In the heavy mass formalism, Jenkins et al. [193] have performed a
calculation of the imaginary parts of $A_{ab}$ and $B_{ab}$. They argue
that with the exception of the asymmetry parameter for $\Sigma^+ \to
p \gamma$, good agreement between theory and the available data is
obtained.
\bigskip
\noindent{\bf 6.6. Nuclear forces and exchange currents}
\medskip
\goodbreak
In this section, I will discuss the recent work of Weinberg and collaborators
[194,195,196] on the nature of the nuclear forces and the extension of this
to meson--exchange currents by Rho et al.[197,198].

On the semi--phenomenological level, nuclear forces are well understood in
terms of meson exchange. There is, of course, the long--range pion component
first introduced by Yukawa [199]. The intermediate range--attraction between
two nucleons can be understood in terms of correlated two--pion exchange,
sometimes parametrized in terms of a fictitious scalar--isoscalar
$\sigma$--meson with mass $M_\sigma \approx 550$ MeV. $\omega$--meson
exchange gives rise to part of the short--range repulsion and the $\rho$
features prominently in the isovector--tensor channel, where it cuts down
most of the pion tensor potential. Weinberg [194,195] has recently discussed
the constraints from the spontaneously broken chiral symmetry on the nature
of the nuclear forces. For that, consider the Lagrangian (6.9) including also
the four--nucleon terms ${\cal L}_{\bar{\Psi} \Psi \bar{\Psi} \Psi}$. Heavy
particles are integrated out, their contribution is hidden in the values of
the pertinent low--energy constants.  Also, consider only nucleons with
momenta smaller than some scale $Q$. This induces a $Q$--dependence of the
various coupling constants. Alternatively, one could work in the framework of
dimensional regularization discussed so far, but the "$Q$--language" is more
familiar to nuclear physicists.  Clearly, one must set $Q \ll M_\rho \simeq
m$. Because of this restriction in the nucleon momenta, it is advantageous to
use old--fashioned time--ordered perturbation theory in which nucleon
propagators are substituted by energy denominators and the integrations run
over the three--momenta. The lowest order Lagrangian mandated by chiral
symmetry is (for the power counting, see below)
$$\eqalign{ \quad
{\cal L}  =&{\cal L}_{\pi \pi} +  {\cal L}_{\pi N} + {\cal L}_{\bar N N} \cr
	  =&-{1 \over 2 D^2} (\partial_\mu \vec \pi)^2
- {M_\pi^2 \over 2D} \vec \pi ^2 \cr
&
- \bar \Psi_N \biggl[ i \partial_0 - m
-{ g_A \over D F_\pi} \vec \tau \cdot (\vec \sigma \cdot \vec \nabla )
\vec \pi
- {1 \over 2 D F_\pi^2} \vec \tau \cdot (\vec \pi \times \partial_0)
\vec \pi) \biggr] \Psi_N \cr
	      & - {1 \over 2} C_S (\bar \Psi_N \Psi_N)
(\bar \Psi_N \Psi_N)
    +{1 \over 2}  C_T (\bar \Psi_N \vec \sigma \Psi_N) \cdot
(\bar \Psi_N
\vec \sigma \Psi_N) + \ldots   \cr}
\eqno(6.64)$$
with $\vec t$ the isospin generators and $\Psi_N$ denotes the nucleon field.
We have used stereographic coordinates on $S^3 \sim SO(4)/SO(3) \sim SU(2)
\times SU(2) / SU(2)$ with $D = 1 + \vec{\pi}\, ^2 / 4 F_\pi^2$. $C_S$ and
$C_T$ are unknown constants and the ellipsis stands for terms with more
derivatives, pion mass insertions or nucleon fields. Consider the scattering
of $N$, $N \ge 2$, nucleons. There arises a complication in the power
counting due to the existence of shallow nuclear bound states. One should not
consider the $S$--matrix but rather the effective potential which is defined
via the sum of the connected diagrams without $N$ nucleon intermediate
states. This stems from the fact that the energy denominators of states
involving solely $N$ nucleons are small ($\sim Q^2 / 2m$) and these cause the
perturbation series to diverge at low energies. For a typical pion exchange
ladder diagram, this infrared singularity is discussed in detail in
ref.[195]. If, however, at least one pion exists in the intermediate state,
no such problem arises (for a more general discussion of the $N$ nucleon
case, see Weinberg [195]). So we are seeking an expansion in the small
parameters $Q/m$ and $M_\pi / m$. To avoid large factors from time
derivatives giving the nucleon mass term, one performs a field redefinition
which allows to express the time derivative in terms of other chirally
invariant couplings with redefined coefficients,
$$\bigl[ i \partial_0 - {1 \over 2 D F_\pi^2} \vec{t} \cdot ( \vec{\pi}
\times \partial_0 \vec{\pi} ) \bigr] \, \Psi_N = \bigl[ m + {g_A \over D
F_\pi} \vec{t}
\times ( \vec{\sigma} \cdot \vec{\nabla} ) \vec{\pi} + \ldots \bigr] \, \Psi_N
\eqno(6.65)$$
This is similar to the heavy mass formalism of ref.[147].  The power counting
goes as follows. Consider a vertex of type $i$ with $d_i$ derivatives or pion
mass insertions involving $n_i$  nucleon fields. A graph with $V_i$ vertices
of type $i$ and $L$ loops scales with $\nu$ powers of $Q$ or $M_\pi$
$$ \nu = 2 - N + 2L + \sum_i V_i (d_i + { n_i \over 2} - 2)
\eqno(6.66)$$
which follows after making use of some topological identities. For any
interaction which is allowed by chiral symmetry (remember that the pion mass
counts as order $Q$), the coefficient of $V_i$ is non--negative. The lowest
order diagrams are tree graphs ($L = 0$) with vertices which have
$d_i = 2$ and $n_i = 0$ or $d_i = 1$ and $n_i = 2$ or $d_i = 0$ and $n_i =
4$. This just leads to the effective Lagrangian (6.64). One can derive an
interaction Hamiltonian (for details, see refs.[194,195])
$$\eqalign{
H_{int} & =
H_{int}^{\pi \pi}
+ H_{int}^{\pi N}
+ H_{int}^{N N} \cr
H_{int}^{N N} & = {1 \over 2} C_S (\bar \Psi_N \Psi_N)
(\bar \Psi_N \Psi_N)
    +{1 \over 2}  C_T (\bar \Psi_N \vec \sigma \Psi_N) \cdot
(\bar \Psi_N
\vec \sigma \Psi_N)  \cr}
\eqno(6.67)$$
{}From this Hamiltonian, one can easily derive the lowest--order static
potential between two nucleons,
$$V_{12} (\vec{r}_1 - \vec{r}_2 ) = [ C_S + C_T \vec{\sigma}_1 \cdot
\vec{\sigma}_2 ] \delta (\vec{r}_1 - \vec{r}_2 ) - \bigl( {g_A \over F_\pi}
\bigr)^2 (\vec{t}_1 \cdot \vec{t}_2 ) (\vec{\sigma}_1 \cdot \vec{\nabla}_1 )
(\vec{\sigma}_2 \cdot \vec{\nabla}_2 ) \, Y ( | \vec{r}_1 - \vec{r}_2 | )
- (1' \leftrightarrow 2' ) \eqno(6.68)$$
with $Y(r) = \exp (-M_\pi r ) / 4 \pi r$ the standard Yukawa function. This
potential contains the long range one--pion exchange, the pion pair exchange
leading to the intermediate range attraction is hidden in the constant $C_S$.
In fact, to accomodate the small nuclear binding energies, one has to choose
$C_S$ large and $C_T$ small. This is discussed in detail in ref.[195]. The
lowest order approximation of the chirally constrained two--nucleon potential
can not explain the mysterious intermediate--range attraction and also shows
no sign of the repulsive hard core.  Ordonez and van Kolck [196] have
considered all terms which are suppressed up to and including ($Q / m$)$^2$
with respect to the leading potential (6.64). These include one--loop
diagrams and local contact terms with coefficients of relative order
($Q / m$) and ($Q / m$)$^2$. Again, intermediate states with nucleons only
are excluded. The low--energy constants accompanying the new contact terms
have not yet been determined, so that the quantitative value of the approach
can not be assessed at present. However, some interesting qualitative
features emerge (for explicit expressions of the higher order corrections to
the two--nucleon potential, see ref.[196]). In this framework, the
uncorrelated two--pion exchange together with some of the higher order
contact terms furnishes the intermediate--range attraction, the conventional
correlated two--pion exchange (pion rescattering, intermediate $\Delta$
states) only appears at next--to--next--to--leading order, ($Q / m$)$^4$
(compare
the discussion in ref.[96]). Furthermore, the potential is energy--dependent
and contains all terms known from phenomenological approaches. The connection
to the meson--exchange potentials can be made if one expresses the
low--energy constants in terms of meson couplings and form factors. Weinberg
[194] has also made a very interesting observation concerning the
three--nucleon forces. To lowest order, these fall into two classes, the one
which has only two--body interactions and the other one with genuine
three--body forces. The latter class involves diagrams with exclusively
non--linear pion interactions, i.e. the emission of two pions at one vertex
($ \sim N \pi \pi)$. Summing up all diagrams of this type, the net result is
zero. Thus, to leading order, the three--nucleon potential is entirely build
from two--body forces without additional genuine three--body ones. Such a
behaviour has indeed been observed in exact calculations of the three nucleon
system in the continuum (deuteron break--up just above threshold [200]). In
this reaction, the momenta involved are small and all observables like the
analyzing power etc. can nicely be explained by using only realistic
two--body potentials and including some charge symmetry breaking. No genuine
three--body forces seem to be needed. Beyond leading order, there is a
three--nucleon potential $\sim (Q / m$)$^3$ [196]. It involves some higher
order pion--nucleon interactions and also some contact operators consisting
of six nucleon fields. The quantitative effects of this potential have not
yet been worked out.

Meson exchange currents arise naturally in the meson--exchange picture of the
nuclear forces. An external electromagnetic or axial probe does not only
couple to the nucleons (impulse approximation) but also to the mesons in
flight. The existence of these two--body operators has been verified
experimentally [201]. More than 10 years ago,
the so--called "chiral filter hypothesis" was introduced [202]. It states
that the response of a nucleus to a long--wavelength electroweak probe is
given solely by the soft--pion exchange terms dictated by chiral symmetry.
Stated differently, all the heavier mesons and nucleon excitations are not
seen, even up to energies of the order of 1 GeV. Why this holds true at such
energies has not yet been explained. Rho [197] has given a simple argument
how the "chiral filter" can occur in nuclei. His lowest order analysis
follows closely the one of Weinberg [194]. Any matrix--element $ME$ of the
effective potential $V$ or of a current $J_\mu$ has the form $ME \sim Q^\nu
\, F(Q/m)$, as discussed before. In the presence of a slowly varying external
electromagnetic field $A_\mu$ (or a weak one), the Hamiltonian takes the form
$$\eqalign{
{\rm H_{\rm eff}} &= {\rm H}_{\pi \pi} + {\rm H}_{\pi N} + {\rm H}_{N N} +
{\rm H_{\rm ext}}    \cr
{\rm H_{\rm ext}} &= {e \over D^2} \bigl[ (\vec{\pi} \times
\partial_\mu \vec{\pi} )_3
+ {i g_A \over 2 F_\pi} \bar{\Psi}_N \gamma_5 \gamma_\mu ( \vec{\tau} \times
\vec{\pi} )_3 \Psi_N \bigr] A_\mu + \ldots \cr} \eqno(6.69)$$
and this additional term H$_{\rm ext}$ modifies the power counting.  Since
one derivative is replaced by the external current, the tree graphs ($L=0$)
with the lowest power $\nu$ must fulfill
$$ d_i + {1 \over 2} n_i + 2 = -1 \eqno(6.70)$$
which leads to $d_i = 0$ and $n_i = 2$ or
$d_i = 1$ and $n_i = 0$. In contrast to the case of the nuclear forces, to
leading order no four--nucleon contact term contributes. This means that
there is no short--ranged two--body current (to lowest order), the exchange
current is entirely given in terms of the soft--pion component derived from
(6.70). This justifies the chiral filter hypothesis at tree level. Park et
al.[198] have also investigated one--loop corrections to the axial--charge
operator, the first correction to the pertinent soft--pion matrix--element is
suppressed by $(Q/m)^2$ [195]. The authors of ref.[198] use the heavy mass
formalism which simplifies the calculation considerably. It is shown that the
loop corrections are small for distances $r \ge 0.6 $ fm, which means that
the lowest order argument of Rho [197] is robust to one--loop order. A vital
ingredient in the calculation is that zero--range interactions between the
nucleons are omitted, such interactions are suppressed in nuclei by the
short--range correlations. This is a very nice result. However, what remains
mysterious is why the chiral filter hypothesis works up to so high energies
-- the answer to this lies certainly outside the realm of baryon CHPT.

Finally, Weinberg [203] has recently used these methods to calculate
pion--nucleus scattering lengths.
In particular, the chiral symmetry result
for the $\pi d$ scattering length
does agree with the more phenomenologically
derived ones previously [154] and is thus in agreement with the data. Here,
the strength of the method lies in the fact that the consistent power
counting employed puts the phenomenological result on firm grounds and
justifies the assumptions going into its derivation.
\bigskip
\noindent{\bf VII. STRONGLY COUPLED HIGGS SECTOR}
\medskip
\goodbreak
\noindent{\bf 7.1. Basic ideas}
\medskip
\goodbreak
The standard model of the strong and electroweak interactions is a
spectacularly successful theory. However, we do not understand the dynamics
underlying the electroweak symmetry breaking of  $SU(2)_L \times
U(1)_Y \to U(1)_{\rm em}$. What we know is that this mechanism provides the
$W$ and $Z$ bosons with their mass and therefore with their longitudinal
degrees of freedom. This indicates that the interactions of longitudinally
polarized gauge  bosons will provide us with information about the mechanism
triggering the symmetry breaking. In the minimal standard model (MSM), which
has been studied extensively (see e.g. the monograph by Gunion et al.[204]),
the symmetry breaking proceeds as follows. The Lagrangian
$${\cal L}_{\rm MSM}  = {\cal L}_g - V \eqno(7.1) $$
contains a part which includes Yang--Mills and gauge couplings,
gauge--fixing, ghost and mass terms as well as Yukawa couplings, ${\cal L}_g$,
plus the Higgs potential $V$. The latter can be written in an $O(4)$ notation
as
$$V({\vec w} , H) = {\lambda \over 4} \bigl[ ({\vec w}^2 + H^2 ) - v^2
\bigr]^2    \eqno(7.2)$$
where $\vec w = (w^+ , w^- , z )$ is a triplet of scalars and $H$ is a fourth
scalar field.  In case of spontaneous symmetry breaking, the vacuum is
asymmetric under the $O(4)$, $< \vec w > = 0$ and $<H> = v$. Shifting the
Higgs field so that it has no vacuum expectation value (vev), $H \to H - v$,
one can rewrite (7.2) as
$$V({\vec w} , H) = {\lambda \over 4}  ({\vec w}^2 + H^2 )^2  + \lambda v H
({\vec w}^2 + H^2 )   + \lambda   v^2 H^2
\eqno(7.3)$$
which shows that there are three massless scalars $M_{\vec w} = 0$ and the
Higgs field has the mass $M_H^2 = 2 \lambda v^2$. This is the expected
pattern since the $O(4)$ is broken down to $O(3)$, i.e. there should be
(12-6)/2=3 Goldstone bosons. These are then eaten up by the originally
massless $W^\pm$ and $Z$ vector bosons, providing them with their mass and
longitudinal degrees  of freedom since a massless vector field has only two
transverse components. Therefore, the dynamics of the longitudinal components
$V_L$ (from here on, I collectively denote the $W's$ and $Z's$ by $V$) is
intimately linked to the symmetry breaking sector.  The pattern described
above is the MSM Higgs mechanism. The conventionally used complex
scalar doublet
$\Phi$ is given by $\Phi^T = (-i w^+ , (H + v - iz )/\sqrt{2} )$. From muon
decay we know the scale,
$$<H> = v = ( \sqrt{2} G_F )^{-1/2} = 246 \, \, {\rm GeV}  \eqno(7.4)$$
with $G_F = 1.166 \cdot 10^{-5}$ GeV$^{-2}$ the Fermi constant.
However, the scalar coupling strength $\lambda$ is undetermined and thus the
Higgs mass is a free parameter of the model. If the Higgs boson is light, of
the order of 100 GeV, the system including also the gauge and matter sector
is amenable to perturbative calculations.  This is one possibility. If the
Higgs is heavy, say $M_H \sim 1$ TeV, the couplings become large and the
symmetry breaking sector is strongly coupled [205,206]. Naive perturbation
theory fails and one has to invoke non--perturbative methods.  Alternatively,
there might be no Higgs boson whatsoever but instead the symmetry breaking is
induced by some strongly coupled theory. An example for this scenario would
be the celebrated technicolor [207] which is a scaled up version of QCD with
the replacements $F_\pi \to v$ and $N_C \to N_{TC}$ (the numbers of
techicolors
might not be three). Such theories generally have a rich spectrum  of
low--lying resonances, in the case of technicolor these are the techni--rho
and the pseudo--Goldstone bosons (the ones which do not give mass to the
$V's$). For orientation, the techni--rho is expected to have a mass of
$m_{\rho, TC} = \sqrt{3/N_{TC}} (v/F_\pi) m_\rho = 2.04$ TeV for an
$SU(3)_{TC}$ theory with two technifermion families (for a general
discussion, see ref.[208]).

In the absence of an understanding of the symmetry breaking sector, one has
to resort to symmetry arguments if one indeed deals with a strongly coupled
system. This is what I will assume from here on. In that case, the EFT
methods become useful.
On general grounds, if the underlying dynamics
produces resonances, these must couple to the longitudinal vector bosons.
Depending on the actual masses and widths of these resonances, they might be
seen at the SSC or the LHC. However, it might also happen that they are too
heavy to be detected at the hadron colliders. In that case, the analogy to
QCD is again useful. Although the $\rho$--meson sits at 770 MeV, it leaves its
trace in the particular values of some low--energy constants which are tested
at much lower energy, say in threshold $\pi \pi$ scattering, $\sqrt{s} = 2
M_\pi \approx M_\rho / 3$. Similarly, one expects an enhancement in $V_L V_L$
pairs [209] as a signal of the strongly interacting Higgs sector. One
parametrizes the symmetry breaking sector by an effective Lagrangian and
replaces the Higgs field by an infinite tower of non--renormalizable
operators with a priori unknown coefficients [210]. A particular model of the
electroweak symmetry breaking will lead to a definite pattern of the coupling
constants. Clearly, at sufficiently low energies, $\sqrt s \sim 1$ TeV, only a
finite number of terms will contribute since they can be organized in an
energy expansion (as it is the case in QCD). For doing that, one has to set
scales. In analogy to QCD, one expects the scale of symmetry breaking  to be
$\Lambda \simeq 4 \pi v \simeq 3$ TeV and furthermore estimates the effective
theory in the one--loop approximation to be useful up to energies of
approximately $\sqrt{s} \simeq \Lambda / 2 \simeq 1.5$ TeV. If there are
lower--lying resonances with mass $M_R < \Lambda$, this window is reduced to
$\sqrt s < M_R$. However, this can only be considered as a very rough
estimate.

Since in QCD the purest reaction to test the symmetry breaking is $\pi \pi$
scattering, in the strongly coupled Higgs sector much work has been focused
on the equivalent reaction $V_L V_L \to V_L V_L$. At high energies, one can
make use of the so--called equivalence theorem [211,212]. It relates the
$S$--matrix containing the unphysical Goldstone bosons $(w^+ , w^- , z)$ to
the $S$--matrix containing the physical longitudinal vector bosons. Any
matrix--element $M$ can be decomposed as
$$M(V_L, \ldots) = M(\vec{w}, \ldots) + {\cal O}(M_V / \sqrt{s}) \eqno(7.5)$$
where the ellipsis stands for matter fields, transverse gauge bosons and so
on. The true $S$--matrix element containing the longitudinal degrees of
freedom $V_L$ and any other physical particles can be written in terms of the
Goldstone bosons $w$ as if these were real particles. Clearly, the
equivalence theorem becomes operative at high energies, $\sqrt{s} >> M_V$. It
is important to notice that it holds to all orders in the scalar coupling
$\lambda$ [212] which is essential to make it useful for a strongly coupled
theory. In fact, there are some subtleties related to eq.(7.5). Let me just
discuss one of them. While the l.h.s.of eq.(7.5)
is manifestly gauge--independent, the r.h.s. is not. One can, however,
show that all gauge--dependent terms are multplied by a factor $M_V / \sqrt
s$ (at least) and are thus suppressed. Further subtleties of the equivalence
theorem are discussed in ref.[213].

There is one more essential phenomenological ingredient which has to be
discussed before writing down the effective Lagrangian at next--to--leading
order. The MSM requires only a breaking from $G = SU(2)_L \times U(1)_Y$ down
to $H = U(1)_{\rm em}$. However, the Higgs potential (7.3) shows all features
of a Goldstone realized chiral $SU(2)$, $G = SU(2) \times SU(2) \to H =
SU(2)$, since $O(4) \sim SU(2) \times SU(2)$ and $O(3) \sim SU(2)$. To see
that in more detail, define the field $H$ as a scalar and the $\vec w$ as
pseudoscalars. Then, $V(\vec w , H)$ is invariant under parity. One can
eliminate the scalars via a field redifinition, which is nothing but choosing
the complex field $\Phi$ in the unitary gauge. The resulting effective theory
is a non--linear $\sigma$--model with $F = v$. Such a chiral $SU(2)$ is a
feature of a large class of experimentally viable models of the electroweak
symmetry breaking. It has the property that the $\rho$--parameter, which
measures
the ratio of neutral to charged current amplitudes [214]
$$\rho = {M_W^2 \over M_Z^2 \cos^2{\theta_W} }
= 1 + {\cal O} (\alpha) \eqno(7.6)$$
is protected against strong interaction corrections [215] and only subject to
small radiative corrections as denoted by the terms of order $\alpha$ (here,
$\alpha$ is a genuine symbol for the electroweak gauge
couplings)\footnote{$^*$}{Clearly, one has to define $\rho$ differently if one
chooses eq.(7.6) to define the weak mixing angle.}. In nature, $\rho = 0.998
\pm 0.0089$ [216], i.e. $\rho = 1$ is fulfilled within a few per cent. In the
MSM with $G = SU(2)_L \times U(1)_Y$, eq.(7.6) holds at tree level. However,
in that case there is no relation between the third generator proportional to
$z$  and the first two proportional to $w^\pm$ and thus loop corrections will
lead to sizeable deviations from unity. In case of the larger symmetry group
$G = SU(2)_L \times SU(2)_C$, the three Goldstone bosons belong to a triplet
and thus (7.6) is protected.  This is completely analogous to the case of
isospin symmetry in two--flavor QCD which protects the vector charges against
strong renormalization [217]. The additional $SU(2)_C$ is called the
"custodial" symmetry or one speaks of weak isospin. In most of the following
discussion, I will assume such a custodial symmetry or a larger group in
which it is embedded. So we are now in the position to write down  the
pertinent effective Lagrangian and calculate various processes.
\medskip
\goodbreak
\noindent{\bf 7.2. Effective Lagrangian at next--to--leading order}
\medskip
\goodbreak
In this section, I will discuss the effective Lagrangian to order $p^4$
assuming the existence of a custodial $SU(2)$ symmetry. The embedding into a
larger group $G$ will also be discussed briefly. In contrast to the case of
QCD, one has to quantize the external sources since they are part of a gauge
theory. I will work in the Landau
gauge ($\xi \to \infty$) since in that case the pseudoscalar Goldstone bosons
remain massless, $M_w = M_V / \xi$, and they also decouple from the ghost
sector. Furthermore, since $\rho$ is not exactly one in nature, one has to
allow for some weak isospin breaking. This can come from the quark mass
differences in the doublets (here, only the difference $m_t - m_b$ or an
eventual fourth generation is relevant), hypercharge gauge boson loops and so
on. To lowest order $p^2$, the effective Lagrangian consists of two terms
[210,218,219],
$${\cal L}^{(2)} = {v^2 \over 4} \Tr (D^\mu U^\dagger D_\mu U ) + \Delta \rho
{v^2 \over 8} \Tr [ \tau^3 (U^\dagger D_\mu U) ]^2 \eqno(7.7)$$
where $U = \exp (i \tau^i w^i / v )$ collects the Goldstone bosons and the
covariant derivative reads
$$D_\mu U = \partial_\mu U + {i \over 2} g W_\mu U - {i \over 2} g' B_\mu U
\tau^3 \eqno(7.8)$$
with $W_\mu = W_\mu^i \tau^i$. The $SU(2)_L$ and $U(1)_Y$ gauge couplings are
denoted by $g$ and $g'$, respectively. In the unitary gauge $U = 1$, the
first term of (7.7) reduces to the standard bilinear gauge boson terms
$${\cal L}^{(2)}(U=1 , \Delta \rho = 0 ) = {v^2 \over 4} \bigl( {g^2 \over 2}
W_\mu^i W_\mu^i + {{g'}^2 \over 2} B_\mu B_\mu - g g' W_\mu^3 B_\mu \bigr)
\eqno(7.9)$$
so that $M_W = g v / 2$ leads to $v = 246$ GeV as stated before.
Diagonalization of (7.9) gives the massless photon and the massive $Z$. The
second term in (7.7) brings the expected shift of the $Z$ mass so that $\rho
\ne 1$. In fact, to leading order in the quark masses, one finds $\Delta \rho
\sim (m_t - m_b)^2 / v^2 \sim m_t^2 / v^2$ [220]. Ramirez [221] has
calculated $\Delta \rho$ in two particular models, one with a heavy scalar
and the other with a techni--rho. While the effects are of opposite sign in
the two models, the contribution to $\Delta \rho$ is too small to be detected
at present. However, from this discussion it becomes clear that one has to
cleanly separate the known physics (here the $m_t^2$ contribution) from the
unknown one related to the mechanism of the symmetry breaking.

At next--to--leading order, there are much more terms. These have been
enumerated by Longhitano [218] some ten years ago. Neglecting custodial
symmetry breaking, the order $p^4$ Lagrangian reads
$$\eqalign{
\quad \quad {\cal L}^{(4)}
&= {L_1 \over 16 \pi^2} \bigl[ \Tr (D^\mu U^\dagger D_\mu U )
\bigr]^2 + {L_2 \over 16 \pi^2} \Tr (D_\mu U^\dagger D_\nu U )
\Tr (D^\mu U^\dagger D^\nu U ) \cr
&- {i g L_{9L} \over 16 \pi^2} \Tr ( W^{\mu \nu} D_\mu U D_\nu U^\dagger )
 - {i g' L_{9R} \over 16 \pi^2} \Tr ( B^{\mu \nu} D_\mu U^\dagger D_\nu U )
\cr
&+ {i g g'L_{10} \over 16 \pi^2} \Tr ( U B^{\mu \nu} U^\dagger W_{\mu \nu} )
\cr} \eqno(7.10)$$
with $2 B^{\mu \nu} = (\partial_\mu B_\nu - \partial_\nu B_\mu ) \tau^3$ and
$2 W^{\mu \nu} = \partial_\mu W_\nu - \partial_\nu W_\mu - i g [ W_\mu ,
W_\nu ] / 2$. We have used the convention of Falk et al.[222] and pulled out
a factor $(1 / 16 \pi^2 )$ so that the $L_i$ are numbers of order one. In
vectorial theories like e.g.
$SU(2)_L \times SU(2)_C \to SU(2)_{L+C}$, one has $L_{9R} = L_{9L}$. Clearly,
the terms proportional to $L_{1,2}$ are related to Goldstone boson scattering
while the last three involve external electroweak gauge bosons. There are
also ${\cal O}(p^4 )$ symmetry breaking terms which can be found in
ref.[218]. They will not be discussed in what follows.

Let me briefly consider theories with larger symmetry groups [223] with an
embedding of the custodial symmetry. In that case, there are additional
massive
scalars, the so--called pseudo--Goldstone bosons. In most models, these are
relatively light and typically carry color. So one has to enlarge the
covariant derivative (7.8) to include the embedding of $SU(3)_c \times
SU(2)_L \times U(1)_Y$ into the larger group $G$. Let $SU(3)$, $SU(2)$ and
$U(1)$ be generated by $X^\alpha = X^{\alpha A} T^A$,
$X^i = X^{i A} T^A$ and $X = X^A T^A$, respectively. For $G = G \times G$,
the pertinent covariant derivative reads
$$D_\mu U = \partial_\mu U + {i \over 2} g W_\mu^i X^i U - {i \over 2} g'
B_\mu U X + {i \over 2} g_s G_\mu^\alpha [X^\alpha , U] \eqno(7.11)$$
with $U \in G$, $G_\mu^\alpha$ the $SU(3)$ color gauge field and $g_s$ the
strong coupling constant. One also has new operators in the effective
Lagrangian. First, there is a mass term for the pseudo--Goldstone bosons,
${\cal L}^{(2)}_M = - \Tr ( U M U^\dagger M^\dagger )$, with $M$ the mass
matrix of the pseudo--Goldstone bosons. At order $p^4$, one has the following
new operators for an $SU(2N) \times SU(2N)$ global symmetry ($N > 1$) [224]
$$\eqalign{ \quad \quad
{\cal L}_{\rm new}^{(4)} &= {L_3 \over 16 \pi^2}
\Tr (D^\mu U D_\mu U^\dagger D^\nu U D_\nu U^\dagger ) +
				       {L_4 \over 16 \pi^2}
\Tr (D_\mu U D_\nu U^\dagger D^\mu U D^\nu U^\dagger ) \cr
&-  i g_s {L_{13} \over 16 \pi^2} G_{\mu \nu}^\alpha
\Tr  [X^\alpha (D^\mu U D^\nu U^\dagger + D^\mu U^\dagger D^\nu U ) ] \cr}
\eqno(7.12)$$
This extension becomes important in the context of $V_L$ pair production via
gluon fusion. Notice that Soldate and Sundrum [233] have argued that in
such models the scale of symmetry breaking is not $4 \pi v$, but rather
$4 \pi v / N $.
Let us now apply the effective Lagrangian to various physical
processes.
\bigskip
\goodbreak
\noindent{\bf 7.3. Longitudinal vector boson scattering}
\medskip
\goodbreak
In this section, I will be concerned with elastic scattering of
longitudinally polarized vector bosons which is the equivalent to $\pi \pi$
scattering in QCD. After some general remarks on the scattering amplitude, I
will discuss some physics issues like e.g. how the longitudinally polarized
gauge bosons can be produced at hadron colliders like the SSC or the LHC. I
will keep the discussion short but provide sufficient references for the
reader who wants to go deeper into this subject.

At low energies, $M_V << \sqrt s << 4 \pi v$ (or $M_R$, where $M_R$ denotes
the mass of the lowest resonance), one can derive a low--energy theorem for
the scattering process $V_L V_L \to V_L V_L$ in complete analogy with
Weinberg's prediction for $\pi \pi$ scattering [62]. It is favorable to work
in the Landau gauge so that the Goldstone bosons remain massless.
Furthermore, one can make use of the equivalence theorem. Therefore, I will
frequently interchange the symbols $'V_L '$ and $'w^\pm , z '$. To lowest
order, the scattering amplitude can be given uniquely in terms of a single
function $A(s,t,u)$ of the pertinent Mandelstam variables (with $s+t+u = 0
$),
$$\eqalign{ \quad &M(w^+ w^- \to z z ) = A(s,t,u) = {s \over v^2} \cr
\quad &M(w^+ w^- \to w^+ w^- ) = - {u \over v^2} \, \, , \, \, M(z z \to z z
) = 0 \cr} \eqno(7.13)$$
which agress with eq.(4.3)  for $F_\pi \to v$ and holds for $\rho = 1$.
The other channels, i.e. elastic $w^\pm z$, $w^+ w^+$ and $w^- w^-$
scattering, follow by crossing. If there is no custodial symmetry or if it is
broken ($\rho \ne 1$), Chanowitz et al. [225] have generalized (7.13),
$$M(w^+ w^- \to z z ) =  {s \over v^2} {1 \over \rho} \, \, , \, \,
M(w^+ w^- \to w^+ w^- ) = - {u \over v^2} \bigl[ 4 - {3 \over \rho} \bigr]
\eqno(7.14)$$
and the third prediction in (7.13) remains unaffected.	The empirical
information $\rho \approx 1$ tells us that in the low energy region, symmetry
requirements uniquely determine the $V_L V_L$ scattering amplitudes in terms
of a single scale, $v$. It is instructive to see how in the MSM this LET
emerges. Exchange of a scalar Higgs field gives
$$A(s,t,u) = - {M_H^2 \over v^2} {s \over s - M_H^2} = {s \over v^2} +
{s^2 \over v^2 M_H^2} + {\cal O}(s^3 ) \eqno(7.15)$$
The first term in the expansion for $s << v^2$ agrees with the LET, the
second one depends on the Higgs mass, which in this case is the information
about the symmetry breaking sector. Therefore, the idea is to measure
deviations from the low--energy behaviour $\sim s / v^2$  and try to relate
these to the models of the dynamics underlying the symmetry breaking.

For the order $p^4$ chiral Lagrangian (7.7,7.10) (with $\Delta \rho = 0$),
the
$V_L V_L$ scattering amplitude contains two unknown coefficients related to
the low--energy constants $L_1$ and $L_2$. This is, of course, completely
analogous to Lehmann's [65] one--loop analysis of $\pi \pi$   scattering in
the chiral limit. At next--to--leading order, the scattering amplitude is
$$\eqalign{
A(s,t,u) &= {s \over v^2}  + {1 \over 16 \pi^2 v^4} \biggl\lbrace 8 L_1
(\mu) s^2 +
4 L_2 (\mu ) t^2 \cr &- {s^2 \over 2} \ln \bigl(-{s \over \mu^2}\bigr) - {1
\over 12} ( 3 t^2 + u^2 - s^2 ) \ln \bigl(-{t \over \mu^2}\bigr) + ( t
\leftrightarrow u ) \biggr\rbrace \cr} \eqno(7.16)$$
Notice that there might be additional constant terms in the $p^4$ pieces
depending on the renormalization scheme one uses (see ref.[219] for a
detailed discussion). It is now instructive to look at the various partial
waves for given isospin ($I$) and angular momentum ($J$).
In the $I = 0$
$S$--wave,
the contribution $11 L_1 (\mu ) + 7 L_2 (\mu )$ enters whereas the $I=J=1$
amplitude is sensitive to $L_2 (\mu ) - 2 L_1 (\mu )$ (which actually is a
scale--independent quantity) and the exotic $S$--wave measures
$L_1 (\mu ) - 2 L_2 (\mu )$ [226]. The unitarity effects are entirely
determined  by the scale $v$ and the scale of dimensional regularization,
$\mu$. In the MSM, the one loop effects have been calculated in refs.[227].
Apart from the tree level Higgs exchange (7.15) one has corrections of the
type ln($M_H^2$). To get an idea in the case of strong coupling, let us
imagine models which contain nearby resonances that saturate the low--energy
constants (as already discussed in section 2). For a heavy scalar--isoscalar
Higgs boson, one finds after matching at the scale $M_H$ [13,224]
$$\eqalign{L_1 (\mu ) &= {64 \pi^3 \over 3} {\Gamma_H v^4 \over M_H^5} + {1
\over 24} \log \bigl( {M_H^2 \over \mu^2} \bigr) \cr
	   L_2 (\mu ) &=
{1 \over 12} \log \bigl( {M_H^2 \over \mu^2} \bigr) \cr} \eqno(7.17)$$
where $\Gamma_H = 3 M_H^3 / 32 \pi v^2$ is the standard model width for the
Higgs. For a Higgs with mass $M_H = 2$ TeV, one finds at the scale $\mu =
1.5$ TeV, $L_1 (\mu ) = 0.33$ and $L_2 (\mu ) = 0.01$ (these are the
renormalized values). For a model with an isovector--vector exchange, one
finds at the scale $M_\rho$ [13,224]
$$\eqalign{L_1 (\mu ) &= -192 \pi^3 {\Gamma_\rho v^4 \over M_\rho^5} + {1
\over 24} \log \bigl( {M_\rho^2 \over \mu^2} \bigr) \cr
	   L_2 (\mu ) &= - L_1 (\mu ) +
{1 \over 8} \log \bigl( {M_\rho^2 \over \mu^2} \bigr) \cr} \eqno(7.18)$$
with $\Gamma_\rho$ the pertinent width. Scaling up the QCD $\rho$--meson as
described before, one finds $L_1 (\mu ) = -0.31$ and $L_2 (\mu ) = 0.38$ at
$\mu = 1.5$ TeV. However, there are other contributions to the $L_i$. The
$\mu$--dependence is actually generated by the Goldstone boson loops and then
there is also a fermionic contribution, which can be calculated. For example,
Dawson and Valencia [228] have considered the effect of a heavy top quark
($m_t \le 200$ MeV) and found it contributes insignificantly. This again
underlines the fact that one has to subtract the known physics from the
inferred low--energy constants to get a handle on the underlying dynamics.
Let us consider the effect of the particular values of the $L_{1,2}$ as given
by eqs.(7.17,7.18) in $V_L V_L$ scattering [226,229,230,231]. In the case of
a heavy Higgs, the  $I = J = 0$ partial wave is enhanced and $T^1_1$ is
suppressed. The effect is opposite in the case of an isovector--vector
exchange, one finds suppression in $T^0_0$ and enhancement in $T^1_1$. Notice
also that the process $z z \to z z$ is very sensitive to the actual values of
$L_{1,2}$ since its tree level amplitude vanishes. The Higgs model would
predict a positive amplitude, while technicolor keeps it essentially zero
since the techni--rho does not couple to neutral gauge bosons.	For more
detailed information on these topics, including also a discussion of various
unitarization models, I refer to the recent review by Hikasa [232].

\midinsert
\vskip 12.0truecm
{\noindent\narrower \it Fig.~15:\quad
Mechanism of producing longitudinal vector gauge boson pairs. These are
a) quark--antiquark annihilation, b) gluon fusion and c) vector boson
fusion. Solid lines denote (anti)quarks, wiggly lines gluons and dashed
lines vector bosons.
\smallskip}
\vskip -1.0truecm
\endinsert
The question arises how to access $VV$ scattering at hadron colliders? In most
calculations, one assumes factorization of the production of the $V_L V_L$
pairs, their scattering and their subsequent decay. Let us focus on the first
part of this chain. In $pp$ colliders, there are essentially three mechanisms
to produce vector boson pairs, see fig.15. The annihilation of light quarks
and antiquarks (fig.15a) gives rise to mostly transversely polarized $V's$
and it is therefore an important background. Appropriate cuts have to be
chosen to separate it from the longitudinal pairs. However, as it is the case
of the $\rho$--resonance in $e^+ e^-$ annihilation, the produced longitudinal
pairs are mostly in an isospin--one state [230] making this process sensitive
to models with techni--rho's or alike. Also, if there are more general
couplings with $L_{9L} \ne L_{9R}$, this mechanism can be of importance
[222]. Second, there is the fusion of two gluons (fig.15b). Only if the
gluons couple to a loop of heavy quarks, a sizeable amount of $V_L V_L$ pairs
is produced. This mechanism is sensitive to isospin--zero  resonances and
thus a prime way of exploiting a heavy Higgs boson. Also, in models with
additional pseudo--Goldstone bosons, it is considerably enhanced due to large
color factors [223]. Third, there is the so--called vector boson fusion as
shown in fig.15c. It is, of course, only relevant if the Higgs sector is
strongly interacting [212]. This mechanism gives the cleanest signal if one
is able to isolate it. The luminosities of the initial $VV$ pair (transverse
or longitudinal) are generally calculated in the "effective W--approximation"
[234], in which the $V's$ are considered as partons, i.e. as on--shell,
physical bosons. Since they are radiated off the (anti)quarks, one also needs
the quark distribution functions in the incoming hadrons. For a detailed
study of this topic, see e.g. ref.[208].

After the scattering, one has to detect the $V's$ and separate the scattering
process from the background. The most detailed study of the reaction $pp \to
V_L V_L X$ has been performed by Bagger, Dawson and Valencia [223,224,235]
and Bagger [236] has given rate estimates for the LHC and the SSC and
discussed the various cuts, tags and vetoes which go into the detection of
the "gold--plated" decays $W^\pm \to \ell^\pm \bar{\nu}_\ell$
and $Z \to \ell^+ \ell^-$ (for
$\ell = e , \mu$) in each of the final states $W^+ W^-$, $W^+ Z$, $W^+ W^+$
and $Z Z$. The tagging and veto methods are discussed in detail in
refs.[237,238]. One finds that the events are clean, but have a low rate if
there are only high mass resonances, say $M_R \ge 2$ TeV. However, even in
the case of a standard model Higgs with $M_H = 1$ TeV, the event rates are
not large. Isospin--zero resonances are most cleanly seen in the
$W^+ W^-$ and $Z Z$ channels, while the $W^+ Z$ channel would be dominated by
isospin--one resonances. Models without low mass resonances tend to be most
visible in the $W^+ W^+$ final state. This means that there is some
complementarity  in the pertinent signals differentiating the various models
of the strongly coupled Higgs sector. For typical cuts and the nominal SSC
and LHC luminosities  and energies (these are 10$^4$ and 10$^5$ pb$^{-1}$ and
$\sqrt s
= 40$ and 16 TeV for the SSC and the LHC, respectively) and assuming $m_t =
140$ GeV, the event rates are of the order of ${\cal O}(10)$ per year and
somewhat larger than the expected background for most models. Clearly, only
high statistics experiments will be able to unravel the nature of the
mechanism triggering the electroweak symmetry breaking in the case of strong
coupling [236]. It is also instructive to see to which kind of new physics
the various final states are sensitive in terms of the low--energy constants.
If the $V_L V_L $ pairs are created via $\bar q q$ annihilation, the
$W^+_L W^-_L$ and $W_L^\pm Z_L$ final states are senstive to $L_{9R,L}$
because the anomalous three gauge boson vertex enters here.
This was first pointed out by Falk, Luke and Simmons [222] for hadron
colliders like the SSC or the LHC. They estimated from the phenomenology
of $WZ$ production that the SSC will be sensitive to $-16 \le L_{9L}
\le 7$ and the LHC sets the limits $-22 \le L_{9L} \le 12$ from
the transverse momentum distribution of the $Z's$ produced in $p p
\to W^\pm Z + X$. The channel $W^\pm \gamma $ gives limits on the
sum $\hat L = L_{10} + (L_{9L} + L_{9R})/ 2 $, but these are only of
the order $|\hat L| \le 50 \, (60)$ for the SSC (LHC). A similar study
was recently performed by  Dobado and Urdiales [239]. The trilinear
gauge boson vertex was also considered by Holdom [240] who investigated
technicolor effects on the reaction $ e^+ e^- \to W^+ W^-$.
In the case of vector boson fusion, one is
naturally sensitive to the couplings $L_{1,2}$. Gluon fusion plays a role in
the case of colored pseudo--Goldstone bosons as already stated and would
enhance the $W^+_L W^-_L$ and $Z_L Z_L$ final states considerably. Further
discussions of these topics, concerning also $e^+ e^-$ machines, can be found
in refs.[232,241].
Finally, constraints on the symmetry breaking sector from a Adler--Weisberger
type sum rule have been discussed by Weinberg [242] and Pham [243].
\bigskip
\goodbreak
\noindent{\bf 7.4. Electroweak radiative corrections}
\medskip
\goodbreak
In this section, I will be concerned with the use of EFT methods for
calculating electroweak radiative corrections. The high--precision
measurements of weak--interaction parameters at LEP have spurred much
interest in precise calculations of these radiative corrections from
both within and beyond the standard model (SM). For the SM, these
have already been calculated using conventional techniques [209,244].
For an introduction on these topics, I refer to the lectures by Peskin
[245]. In particular, much work has been focused on the so--called
"oblique" corrections, that is corrections to the gauge boson propagators
(vacuum polarization effects) [246]. There are several reasons why it
is advantagous	to consider oblique corrections. First, if there are
heavy particles which do not directly couple to the light fermions, they
can only be detected via their loop contributions to the gauge boson
propagators (for the presently operating accelerators). This is due to
the fact that in most models of the electroweak symmetry breaking the
decoupling theorem [247] is expected to be inoperative. Second, vacuum
polarization effects are universal. They do not depend on the specific
process under consideration in contrast to e.g. vertex or box corrections.
Furthermore, for reactions involving exclusively light external fermions,
one can neglect their masses as compared to the $Z$ mass and therefore
only has to take into account the transverse ($g_{\mu \nu}$) part of
the gauge boson propagator. Following Peskin and Takeuchi [248,249],
the oblique corrections can be parametrized in terms of three combinations
of the gauge boson self--energies and their derivatives,
$$\eqalign{
\alpha S &=  4 e^2 [ \Pi'_{33} (0) - \Pi'_{3Q} (0) ]   \cr
\alpha T &= {e^2 \over s^2 c^2 M^2_Z} [ \Pi_{11} (0) - \Pi_{33}
	    (0) ]				       \cr
\alpha U &=  4 e^2 [ \Pi'_{11} (0) - \Pi'_{33} (0) ]   \cr}
\eqno(7.19)$$
with $s = \sin \theta_W$, $c = \cos \theta_W$ and the indices '1,2,3,Q'
refer to the $W$, $Z$ and $\gamma$ particles. The quantity $S$ is
isospin--symmetric, whereas $T$ (which is nothing but $\Delta \rho /
\alpha$ already discussed) and $U$ are isospin--asymmetric. Since $U$
is related to isospin--breaking in the derivatives of the
self--energies, it is supposed
to be small in models with a custodial symmetry and often neglected.
Notice that Peskin and Takeuchi [249] define these parameters as deviations
from the SM ones for a given reference point $m_t = 150$ GeV and $M_H =
1$ TeV. The relation to the epsilon parameters of Altarelli et al. [250]
is $\epsilon_1 = \alpha T$, $\epsilon_2 = -\alpha U / 4 s^2$ and
$\epsilon_3 = \alpha S / 4 s^2$. The recent analysis of these authors
gives $S = 0.1 \pm 1.1$, $T = -0.06 \pm 0.69$ and $U = 1.4 \pm 1.3$
using exclusively LEP results. If one furthermore includes low--energy
data from atomic parity violation in $Cs$ [251], one finds $S = -0.34
\pm 0.68$, $T = -0.03 \pm 0.48$ and $U = 0.78 \pm 0.98$.

The use of EFT methods in the analysis of oblique corrections was
pioneered by Golden and Randall [252], Holdom and Terning [253],
Holdom [254] and Dobado et al.[255]. The most lucid discussion of these
methods can be found in the paper of Georgi [256] to which I will
return later.  First, let me give an elementary discussion of the physics
behind the parameter $S$ and its relation to the chiral Lagrangian. In
the basis of the non--physical fields $A_\mu^3$ and $B_\mu$ it is most
simple to calculate the one--loop diagram since this gives exactly the
mixing which defines $S$. The graph is divergent, it can be made finite
by a counterterm of the type
$$ L_{10} g g' B_{\mu \nu} W_a^{\mu \nu} \Tr [ U^\dagger \tau_3 U
\tau_a ]      \eqno(7.20) $$
from which it immediately follows that
$$ S = - 16 \pi L_{10}^r	 \eqno(7.21) $$
Obviously, one has to specify the renormalization prescription to make
this equation precise. As already stressed, $S$ contains a part coming
from the SM (denoted  by $S_{SM})$ and eventually a part due to physics
from some higher scale $\Lambda \gg M_Z$ (denoted by $S_{NP}$). A
particular model for the strongly coupled Higgs sector will lead to a
certain value of $L_{10}$ and thus of $S_{NP}$ [253,254]. $S_{SM}$ gets
contributions from the quarks and also the SM Higgs. For $M_H =1$ TeV,
Altarelli et al. [250] give $S_{SM} \simeq 0.65$ whereas Sint finds
$S_{SM} \simeq -0.21$ [219]. It is not yet clear what the source of this
discrepancy is. In any case, there is not much room left for non--SM
physics if one compares to $S_{\rm exp} = -0.34 \pm 0.68$. As stressed
in refs.[248,254], simple technicolor models give rise to a positive
contribution to $S_{NP}$, approximately [249,257]
$$ S_{NP} \simeq 0.3 {N_{TF} \over 2} {N_{TC} \over 3}	 \eqno(7.22)$$
for a model with $N_{TC}$ technicolors and
$N_{TF}$ technifermions. Since
the empirical value of $S$ is negative, QCD--like technicolor theories
with a large technisector seem to be ruled out. However, it is possible
to construct models which lead to negative values of $S$ [256,258]. In
any case, a fully consistent technicolor theory is not yet available and
therefore the emiprical information on $S$, $T$ and $U$ should be used
to constrain any model of the electroweak symmetry breaking.
Finally,	notice that $S$ can be written in terms of a Weinberg sum
rule [219,248,257] in complete analogy to the QCD case for $L_{10}$ or
$\bar{\ell}_5$ [13,14].

Let me now return to the paper of Georgi [256]. First, it is pointed out why
the EFT formalism is useful in the calculation of radiative corrections.
These reasons are clarity, simplicity and, most important, generality. The
EFT allows one to pick up the relevant operators and avoids most
complications due to scheme dependence, which is the case in the conventional
approach. When one integrates out the particles heavier than some scale
$\Lambda$, one produces a tower of non--renormalizable operators in the EFT.
This integrating out of the heavy particles in general leads to an effective
${\underline {action}}$
which is highly non--local. To arrive at the effective
Lagrangian, which is local, one has to perform an operator product expansion.
At the scale $\Lambda$, one has to match the physics in the two theories. In
that way, the low--energy theory keeps the knowledge about the heavy
particles no longer present as active degress of freedom. This matching leads
to a power dependence on the heavy particle masses. In addition, there is a
logarithmic dependence on these masses which originates from using the
renormalization group to evolve the EFT to some smaller scale $\mu$, $\mu <
\Lambda$. The general problem of matching in effective field theories which
are not necessarily separated by  large scales has been discussed by Georgi
[259]. The proper  matching conditions disentangle the short--distance from
the long--distance physics. While the former is incorporated in the
coefficients of the effective Lagrangian, the latter remains, of course,
explicit in the EFT. In ref.[256], a general analysis of electroweak
radiative corrections is performed under a few assumptions (custodial
symmetry, no extra Goldstone bosons and no interactions which let the light
quarks and leptons participate in the mechanism of the symmetry breaking).
First, the energy domain $M_H > \Lambda > m_t$ is considered. The leading
terms
beyond the ones from the standard model can be organized in powers of the
$U(1)$ gauge coupling. At one loop, one has a term like (7.20) contributing
to $S$ and an  operator from virtual gauge boson exchange contributing to $T$
plus a whole tower of higher order (suppressed) operators. Going down to the
regime where $m_t > \Lambda > M_Z$, there appear additional contributions to
$S$ and $T$, the most important one arises when one integrates out the
$t$--quark. At this point, it is obligatory to differentiate between the
effective action and the effective Lagrangian since the $b$--quark, which
forms a doublet with the $t$, is retained in the EFT. As an example, a
calculation is performed for large $\alpha T$, since this is the only oblique
correction that scales with $m_t^2$. It is straightforward to derive the
non--perturbative relation
$${M_W^2 \over M_Z^2} = {\rho_\star \over 2} \biggl( 1 + \sqrt{1 - {4 \pi
\alpha_\star \over \sqrt{2} G_F M_Z^2 \rho_\star }} \biggr) \eqno(7.23)$$
with $\rho_\star = 1 + \alpha T$ and $\alpha_\star = \alpha (M_Z)$. This
relation was originally conjectured in ref.[260], but it can be derived in a
much simpler fashion using the EFT approach. Further corrections to eq.(7.23)
are also discussed in ref.[256]. In any case, I strongly advice the reader to
work through that paper.

Finally, let me just point out that the use of EFT methods in the calculation
of radiative electroweak corrections is only in its beginning stage and much
more work remains to be done.
\goodbreak
\bigskip
\noindent {\bf VIII. MISCELLANEOUS OMISSIONS}
\bigskip
Here, I essentially assemble references for some of the topics not covered
in the preceeding sections. However, due to the rapid developments in the
field I do not even attempt to offer a complete list but rather refer the
reader to his/her favorite data base.
\medskip
\item{$\bullet$}{
{$\underline {More \, \, mesons:}$}
One of the central topics of CHPT
was, is and will be the dynamics and interactions of the pseudoscalar
Goldstone bosons at  low and moderate energies. Of particular interest are
the reactions without counterterm contributions at next--to--leading order
since they can serve as a clean test of the chiral sector of QCD. These are
$K_L^0 \to \pi^0 \gamma \gamma$ [99], $K_S \to \gamma \gamma$ [262] and
$\gamma \gamma \to \pi^0 \pi^0$ [43] as well as the related process leading
to the pion electromagnetic polarizabilities [263]. And, of course, there are
plentiful pion, kaon and eta semi-- and non--leptonic decays which have not
yet been measured with a high precision. A thorough discussion of this can be
found in the user's guide to the "chiral temple", known to the non--chiral
world also as the Frascati $\Phi$--factory DA$\Phi$NE [265]. For further
references, see also the lectures by Ecker [266] and the recent talk by
Leutwyler [267].}

\item{$\bullet$}{{$\underline
{Union \, \, of \, \, heavy \, \, quark \, \,
and \, \, chiral \, \, symmetries:}$} In
section 6.2, the nucleon was described as a very heavy (static) source
surrounded by a cloud of light pions. This is the physical picture underlying
the so--called "heavy quark symmetries" of mesons consisting of one very
heavy ($c, \, b$) quark and one very light ($u, \, d, \, s$) antiquark. In
the limit of the heavy mass becoming infinite, this symmetry becomes exact
and one can e.g. derive relations
between certain properties of the pseudoscalar $D, \, B$ mesons with their
vectorial excitations $D^\star , \, B^\star$.  For the emission of pions or
kaons from these mesons, e.g. $D^{\star +} \to D^0 \pi^+$ or
$B \to D^\star \pi l \nu  $, one works with an effective Lagrangian which
unites the heavy quark and the chiral symmetry of QCD. This a very new and
wide field, and some first work has been reported in refs.[268].}

\item{$\bullet$}{{$\underline
{Large \, \, N_f:}$} QCD becomes much simpler if one
considers the case of a large number of colors, $N_C \to \infty$ [26,27,28].
Similarly, one hopes to gain a deeper understanding of the chiral expansion
by considering the artificial limit of a large number of massless
(light) flavors. Some pertinent papers are given in ref.[269]. However, after
approximately ten years of experience in the context of Skyrme--type models,
which are based on a $1/N_C$ expansion, one is inclined to view this
development rather sceptically. Also, recent lattice studies seem to indicate
a qualitatively different behaviour for theories with many flavors.}

\item{$\bullet$}{{$\underline
{First \, \, reading:}$} There exist several introductory
lectures concerned with CHPT and its applications. Let me mention here the
ones by Donoghue [270], Ecker [266], Gasser [182] and Leutwyler [271]. More
specialized lectures are the ones by Truong [272], Jenkins and Manohar [149]
and the author [144]. There are also a few books on the market which deal
with EFT methods and CHPT. First, there is the classical book by Georgi [21]
and more recently, a broader exposition has been given by Donoghue, Golowich
and Holstein [273]. Finally, a state of the art summary as of the fall of
1991 is given in the workshop proceedings [274].}
\bigskip
\bigskip
\bigskip
\noindent{\bf Acknowledgements}

I express my thanks to all my collaborators and everybody who taught me about
the material discussed. I am grateful to V\'eronique Bernard for a
careful reading of the manuscript.
I also wish to thank all members of the Instiute for
Theoretical Physics at the University of Berne for creating  most
stimulating working conditions.
\vfill
\eject
\bigskip
\vfill
\eject
\noindent {\bf APPENDIX A: THE CASE OF $SU(2) \times SU(2)$}
\medskip
In the two--flavor case, it is most convenient to work with real $O(4)$
fields of
unit length, e.g. the pions are collected in $U^A (x) = (U^0 (x) , U^i (x) )$
with $U^T U = 1$. The pertinent covariant derivatives are
$$\eqalign{
 \nabla_\mu U^0&= \partial_\mu U^0 + a_\mu^i U^i    \cr
 \nabla_\mu U^i&= \partial_\mu U^i +
\epsilon^{ikl} v_\mu^k U^l - a_\mu^i U^i
\cr}
\eqno(A.1)     $$
disregarding the isoscalar vector and axial currents. Defining further
$$ \chi^A = 2 \tilde{B} (s^0 , p^i ) \, \, \, , \, \,
\tilde{\chi}^A = 2 \tilde{B} (p^0 , -s^i ) \, \, \, ,
\eqno(A.2) $$
where the value of $\tilde{B}$ is slightly different from the one of $B$ (see
ref.[14]). The next--to--leading order chiral Lagrangian reads:
$$ \leff = \sum^7_{i=1} \ell_i \, P_i \, \, + \, \,
\sum^3_{j=1} h_j \, \tilde{P}_j
\eqno(A.3) $$
with (we do not exhibit the high--energy terms)
$$\eqalign{
P_1 &= \Tr (\nabla^\mu U^ T	 \nabla_\mu U)^2  \cr
P_2 &= \Tr (\nabla_\mu U^ T	 \nabla_\nu U)
       \Tr (\nabla^\mu U^ T	 \nabla^\nu U) \cr
P_3 &= \bigl[\Tr (\chi^T       U       )	  \bigr]^2 \cr
P_4 &= \Tr (\nabla^\mu \chi^T	 \nabla_\mu U) \cr
P_5 &= \Tr (U^T F_{\mu \nu} F^{\mu \nu} U )   \cr
P_6 &= \Tr (\nabla^\mu U^T F_{\mu \nu} \nabla^\nu U )	\cr
P_7 &= \bigl[\Tr (\tilde{\chi}^T  U )		 \bigr ]^2 \cr}
\eqno(A.4) $$
One can introduce scale--independent low--energy constants $\bar{\ell}_i$ via
$$ \ell^r_i = {\gamma_i \over 32 \pi^2} \bigl( \bar{\ell}_i + \ln (M^2 /
\mu^2 ) \bigr) \, \, \, (i= 1 , \ldots , 6)
\eqno(A.5) $$
with
$$\gamma_1 = {1 \over 3} \, \, , \, \,
  \gamma_2 = {2 \over 3} \, \, , \, \,
  \gamma_3 = -{1 \over 2} \, \, , \, \,
  \gamma_4 = 2 \, \, , \, \,
  \gamma_5 = -{1 \over 6} \, \, , \, \,
  \gamma_6 = -{1 \over 3} \, \, .     \eqno(A.6)  $$
Notice that $\ell_7$ is not renormalized. The explicit relation of these
low--energy constants to the $L^r_1 , \ldots , L^r_{10}$ is spelled out in
ref.[14].
\bigskip
\vfill
\eject
\noindent{\bf APPENDIX B: $SU(3)$ MESON-BARYON LAGRANGIAN}
\medskip
\goodbreak
Here, I wish to provide the necessary definitions for the three flavor
meson--baryon system. It is most convenient to write the eight meson and
baryon fields in terms of $SU(3)$ matrices $M$ and $B$, respectively,
$$ M = {1 \over \sqrt 2} \left(
\matrix { {1\over \sqrt 2} \pi^0 + {1 \over \sqrt 6} \eta
&\pi^+ &K^+ \cr
\pi^-
	& -{1\over \sqrt 2} \pi^0 + {1 \over \sqrt 6} \eta & K^0 \cr
K^-
	&  \bar{K^0}&- {2 \over \sqrt 6} \eta  \cr} \right)
\eqno(B.1a) $$
$$ B =	\left(
\matrix  { {1\over \sqrt 2} \Sigma^0 + {1 \over \sqrt 6} \Lambda
&\Sigma^+ &  p \cr
\Sigma^-
    & -{1\over \sqrt 2} \Sigma^0 + {1 \over \sqrt 6} \Lambda & n \cr
\Xi^-
	&	\Xi^0 &- {2 \over \sqrt 6} \Lambda \cr} \right)
\eqno(B.1b) $$
with
$$ U(M) = u^2 (M) = \exp \lbrace 2 i M / F \rbrace
\eqno(B.2) $$
and the covariant derivative acting on $B$ reads
$$\eqalign{
D_\mu B &= \partial_\mu B + [ \Gamma_\mu , B ] \cr
\Gamma_\mu &= {1 \over 2} \bigl\lbrace u^\dagger [ \partial_\mu - i ( v_\mu +
a_\mu )]u + u [ \partial_\mu - i ( v_\mu - a_\mu )]u^\dagger \bigr\rbrace
\cr}
\eqno(B.3) $$
with $v_\mu$ and $a_\mu$ external vector and axial--vector sources. Under
$SU(3)_L \times SU(3)_R$, $B$ and $D_\mu B$ transform as
$$ B' = K B K^\dagger \, \, \, , \, \,\, (D_\mu B)' = K (D_\mu B) K^\dagger
\eqno(B.4)  $$
It is now straightforward to construct the lowest--order ${\cal O}(p)$
meson--baryon Lagrangian,
$${\cal L}_{\rm MB}^{(1)} = \Tr \bigl\lbrace i \bar{B} \gamma^\mu D_\mu B
- \krig{m} \bar{B} B + {1 \over 2} D \bar{B} \gamma^\mu \lbrace
u_\mu , B \rbrace
+ {1 \over 2} F \bar{B} \gamma^\mu \gamma_5 [ u_\mu , B ] \bigr\rbrace
\eqno(B.5)  $$
where $\krig{m}$ stands for the (average) octet mass in the chiral limit. The
trace in (B.5) runs over the flavor indices. Notice that in contrast to the
$SU(2)$ case, one has two possibilities of coupling the axial $u_\mu$ to
baryon bilinears. These are the conventional $F$ and $D$ couplings subject to
the constraint $F + D = g_A = 1.26$. At order ${\cal O}(p^2)$ the baryon mass
degeneracy is lifted by the terms written in eq.(6.61). However, there are
many other terms at this order. If one works in the one--loop approximation,
one also needs the terms of order ${\cal O}(p^3)$. The complete local
effective Lagrangians ${\cal L}_{\rm MB}^{(2)}$ and ${\cal L}_{\rm MB}^{(3)}$
are given by Krause [143]. The extension of this to the heavy mass formalism
is straightforward, I refer to the article by Jenkins and Manohar [149]
(which, however, contains not all terms of ${\cal L}_{\rm MB}^{(2)}$ and none
of ${\cal L}_{\rm MB}^{(3)}$).
\bigskip
\vfill
\eject
\centerline{\bf REFERENCES}
\medskip
\item{1.}E. Euler, {\it Ann. Phys.\/} (Leipzig) {\bf 26} (1936) 398;

E. Euler and  W. Heisenberg, {\it Z. Phys.\/} {\bf 98} (1936) 714;

S. Schweber, "An introduction to relativistic quantum field theory",
Harper and

Row, New York, 1964.
\smallskip
\item{2.}W. Dittrich and M. Reuter, "Effective Lagrangians in Quantum
Electrodynamics", Springer Verlag, Heidelberg, 1985.
\smallskip
\item{3.}J. Gasser and H. Leutwyler, {\it Phys. Reports\/} {\bf C87} (1982) 77.
\smallskip
\item{4.}S. L. Adler and R. F. Dashen, "Current Algebras and applications to
particle physics", Benjamin, New York, 1968.
\smallskip
\item{5.}R. D. Peccei and J. Sola, {\it Nucl. Phys.\/} {\bf
B281} (1987) 1;

C. A. Dominguez and J. Sola, {\it Z. Phys.\/} {\bf
C40} (1988) 63.
\smallskip
\item{6.}Y. Nambu and G. Jona--Lasinio, {\it Phys. Rev.\/} {\bf 122}
(1961) 345; {\bf 124} (1961) 246.
\smallskip
\item{7.}J. Goldstone, {\it Nuovo Cim.\/}
{\bf 19} (1961) 154.
\smallskip
\item{8.}H. Pagels, {\it Phys. Rep.\/} {\bf 16} (1975) 219.
\smallskip
\item{9.}R. Dashen and M. Weinstein, {\it Phys. Rev.\/} {\bf 183}
(1969) 1261.
\smallskip
\item{10.}S. Weinberg,
{\it Phys. Rev.\/} {\bf
166} (1968) 1568.
\smallskip
\item{11.}S. Coleman, J. Wess and B. Zumino,
{\it Phys. Rev.\/} {\bf 177} (1969) 2239;

C. G. Callan, S. Coleman, J. Wess and B. Zumino,
{\it Phys. Rev.\/} {\bf 177} (1969) 2247.
\smallskip
\item{12.}S. Weinberg, {\it Physica} {\bf 96A} (1979) 327.
\smallskip
\item{13.}J. Gasser and H. Leutwyler, {\it Ann. Phys. (N.Y.)\/}
 {\bf 158} (1984) 142.
\smallskip
\item{14.}J. Gasser and H. Leutwyler, {\it Nucl. Phys.\/}
 {\bf B250} (1985) 465.
\smallskip
\item{15.}C. Roiesnel and T. N. Truong,  {\it Nucl. Phys.\/} {\bf
B187} (1981) 293.
\smallskip
\item{16.}D. G. Boulware and L. S. Brown,  {\it Ann. Phys.\/} (N.Y.) {\bf
138} (1982) 392.
\smallskip
\item{17.}M. Gell-Mann, CALTECH report CTSL--20, 1961;

S. Okubo,  {\it Prog. Theor. Phys.\/} {\bf
27} (1962) 949.
\smallskip
\item{18.}M. Gell-Mann, R. J. Oakes and B. Renner,  {\it Phys. Rev.\/} {\bf
175} (1968) 2195.
\smallskip
\item{19.}M. A. Shifman, A. I. Vainshtein and V. I. Zahkarov,  {\it
Nucl. Phys.\/} {\bf B147} (1979) 385, 448, 519.
\smallskip
\item{20.}A. Manohar and H. Georgi, {\it Nucl. Phys.\/} {\bf B234} (1984) 189.
\smallskip
\item{21.}H. Georgi, ``Weak Interactions and Modern ParticlePhysics'',
Benjamin/Cummings, Reading, MA, 1984.
\smallskip
\item{22.}J. Wess and B. Zumino, {\it Phys. Lett.\/} {\bf 37B} (1971)
95.
\smallskip
\item{23.}
E. Witten, {\it Nucl. Phys.\/} {\bf B223} (1983) 422.
\smallskip
\item{24.}Ulf-G. Mei{\ss}ner, {\it Phys. Rep.\/} {\bf 161} (1988) 213.
\smallskip
\item{25.}J. Schwinger,  {\it Phys. Rev.\/} {\bf
93} (1951) 664.

B. DeWitt, "Dynamical theory of groups and fields",
Gordon and Breach, New

York, 1965.

R. Seeley, {\it Am. Math. Soc. Proc. Symp. Pure Math.\/} {\bf
10} (1967) 288.
\smallskip
\item{26.}G. 't Hooft,  {\it Nucl. Phys.\/} {\bf
B72} (1974) 461.
\smallskip
\item{27.}E. Witten,  {\it Nucl. Phys.\/} {\bf
B160} (1979) 57.
\smallskip
\item{28.}G. Veneziano, {\it Nucl. Phys.\/} {\bf
B117} (1976) 519.
\smallskip
\item{29.}C. Riggenbach, J. F. Donoghue, J. Gasser and B. Holstein,
{\it Phys. Rev.\/} {\bf D43} (1991) 127.
\smallskip
\item{30.}G. J. Gounaris and J. J. Sakurai, {\it Phys. Rev. Lett.\/} {\bf
21} (1968) 244.
\smallskip
\item{31.}S. R. Amendolia et al., {\it Nucl. Phys.\/} {\bf
B277} (1986) 168.
\smallskip
\item{32.}G. Ecker, J. Gasser, A. Pich and E. de Rafael,
{\it Nucl. Phys.\/} {\bf B321}
(1989) 311.
\smallskip
\item{33.}J. F. Donoghue, C. Ramirez and G. Valencia,
{\it Phys. Rev.\/} {\bf D39}
(1989) 1947.
\smallskip
\item{34.}
T. H. Hansson, M. Prakash and I. Zahed,
{\it Nucl. Phys.} {\bf B335} (1990) 67;

V. Bernard and Ulf--G. Mei{\ss}ner,
{\it Phys. Lett.} {\bf B266} (1991) 403;

C. Sch\"uren, E. Ruiz Arriola and K. Goeke,
{\it Nucl. Phys.} {\bf A547} (1992) 612.
\smallskip
\item{35.}
J. Balog, {\it Phys. Lett.} {\bf B149} (1984) 197;

D. Ebert and H. Reinhardt,
{\it Nucl. Phys.} {\bf B271} (1986) 188;

E. Ruiz Arriola, {\it Phys. Lett.} {\bf B253} (1991) 430;

J. Bijnens, C. Bruno and E. de Rafael, preprint CERN--TH.6521/92, 1992.
\smallskip
\item{36.}D. Espriu, E. de Rafael and J. Taron, {\it Nucl. Phys.\/} {\bf
B345} (1990) 22.
\smallskip
\item{37.}B. Holdom, J. Terning and K. Verbeck, {\it Phys. Lett.\/} {\bf
B245} (1990) 612.
\smallskip
\item{38.}M. Voloshin and V. Zakharov, {\it Phys. Rev. Lett.\/} {\bf
45} (1980) 688.
\smallskip
\item{39.}J. F. Donoghue and H. Leutwyler, {\it Z. Phys.\/} {\bf
C52} (1991) 343.
\smallskip
\item{40.}J. F. Donoghue, J. Gasser and H. Leutwyler, {\it Nucl. Phys.\/}
{\bf B343} (1990) 341.
\smallskip
\item{41.}A. Salam and J. Strathdee, {\it Phys. Rev.\/} {\bf
184} (1969) 1760;

R. J. Crewther, {\it Phys. Rev. Lett.\/} {\bf
28} (1972) 1421;

M. S. Chanowitz and J. Ellis, {\it Phys. Lett.\/} {\bf
B40} (1972) 397;

J. Schechter, {\it Phys. Rev.\/} {\bf
D21} (1980) 3393.
\smallskip
\item{42.}J. F. Donoghue and D. Wyler,
{\it Phys. Rev.\/} {\bf
D45} (1992) 892.
\smallskip
\item{43.}J. Bijnens and F. Cornet, {\it Nucl. Phys.\/}
 {\bf B296} (1988) 557.
\smallskip
\item{44.}S. L. Glashow and S. Weinberg, {\it Phys. Rev. Lett.\/} {\bf
20} (1968) 224.
\smallskip
\item{45.}R. Dashen,  {\it Phys. Rev.\/} {\bf
183} (1969) 1245.
\smallskip
\item{46.}T. Das, G. Guralnick, V. Mathur, F. Low and J. Young,
{\it Phys.
Rev. Lett.\/} {\bf 18} (1967) 759.
\smallskip
\item{47.}S. Weinberg, {\it Trans. N.Y. Acad. Sci.\/} {\bf
38} (1977) 185.
\smallskip
\item{48.}D. B. Kaplan and A. V. Manohar,  {\it Phys. Rev. Lett.\/} {\bf
56} (1986) 2004.
\smallskip
\item{49.}S. Raby and G. B. West, {\it Phys. Rev.\/} {\bf
D38} (1988) 3488.
\smallskip
\item{50.}G. t'Hooft,  {\it Phys. Rev.\/} {\bf
D14} (1976) 3432;

R. Jackiw and C. Rebbi, {\it Phys. Rev. Lett.\/} {\bf 37} (1976) 172;

C. Callan, R. Dashen and D. Gross, {\it Phys. Lett.\/} {\bf B63} (1976) 334.
\smallskip
\item{51.}K. Choi, C. W. Kim and W. K. Sze,  {\it Phys. Rev. Lett.\/} {\bf
61} (1988) 794;

K. Choi and C. W. Kim, {\it Phys. Rev.\/} {\bf D40} (1989) 890.
\smallskip
\item{52.}H. Leutwyler and M. Roos, {\it Z. Phys.\/} {\bf C25} (1984) 91.
\smallskip
\item{53.}H. Leutwyler, {\it Nucl. Phys.\/} {\bf
B337} (1990) 108.
\smallskip
\item{54.}J. F. Donoghue, B. R. Holstein and Y. C. R. Lin,  {\it Phys. Rev.
Lett.\/} {\bf 55} (1985) 2766;

F. Gilman and R. Kaufman, {\it Phys. Rev.\/}
{\bf D36} (1987) 2761;

Riazuddin and Fayyazuddin, {\it Phys. Rev.\/}
{\bf D37} (1988) 149.
\smallskip
\item{55.}J. Gasser, {\it Ann. Phys.\/} (N.Y.) {\bf
136} (1981) 62.
\smallskip
\item{56.}V. Bernard, R. L. Jaffe and Ulf--G. Mei{\ss}ner,  {\it Nucl.
Phys.\/} {\bf B308} (1988) 753.
\smallskip
\item{57.}V. Novikov et al., {\it Nucl.
Phys.\/} {\bf B165} (1980) 55.
\smallskip
\item{58.}J. F. Donoghue, B. R. Holstein and D. Wyler,
{\it Phys. Rev. Lett.\/} {\bf 69} (1992) 3444.
\smallskip
\item{59.}R. Socolow, {\it Phys.
Rev.\/} {\bf B137} (1965) 1221.
\smallskip
\item{60.}J. Gasser and H. Leutwyler, {\it Nucl. Phys.\/}
 {\bf B250} (1985) 539.
\smallskip
\item{61.}J. Schechter and A. Subbaraman, {\it Int. J. Mod. Phys.\/} {\bf
A7} (1992) 7135.
\smallskip
\item{62.}
S. Weinberg, {\it Phys. Rev. Lett.\/} {\bf 17} (1966) 616.
\smallskip
\item{63.}J. Gasser and H. Leutwyler,
{\it Phys. Lett.\/}
 {\bf 125B} (1983) 325.
\smallskip
\item{64.}J. Gasser and Ulf-G. Mei{\ss}ner,
{\it Phys. Lett.\/} {\bf B258}
(1991) 219.
\smallskip
\item{65.}H. Lehmann, {\it Phys. Lett.\/} {\bf
B41} (1972) 529; {\it Acta Phys. Austriaca Suppl.\/} {\bf 11} (1973) 139.
\smallskip
\item{66.}V. Bernard, N. Kaiser and Ulf-G. Mei{\ss}ner,
{\it Nucl. Phys.\/} {\bf
B357} (1991) 129.
\smallskip
\item{67.}J.~L.~Petersen, ``The $\pi \pi$ Interaction'',
CERN yellow report 77--04, 1977.
\smallskip
\item{68.}S. M. Roy, {\it Phys. Lett.\/} {\bf
B36} (1971) 353.
\smallskip
\item{69.}J. L. Basdevant, C. D. Froggatt, and J. L. Petersen,
{\it Nucl. Phys.\/}
{\bf B72} (1974) 413;
J. L. Basdevant, P. Chapelle, C.Lopez and M. Sigelle,
{\it Nucl. Phys.\/}
{\bf B98} (1975) 285; C. D. Froggatt and J.~L.~Petersen,
{\it Nucl. Phys.\/}
{\bf B129} (1977) 89.
\smallskip
\item{70.}J. F. Donoghue, B. R. Holstein and G. Valencia, {\it Int. J. Mod.
Phys.\/} {\bf A2} (1987) 319.
\smallskip
\item{71.}V. Bernard, N. Kaiser and Ulf--G. Mei{\ss}ner, {\it Nucl. Phys.\/}
{\bf B364} (1991) 283.
\smallskip
\item{72.}
T.T. Wu and C.N. Yang, {\it Phys. Rev. Lett.\/} {\bf 13} (1964) 380.
\smallskip
\item{73.}W. Ochs, Max--Planck--Institute preprint MPI--Ph/Ph 91--35, 1991.
\smallskip
\item{74.}J. F. Donoghue, C. Ramirez and G. Valencia,
{\it Phys. Rev.\/} {\bf D38}
(1988) 2195.
\smallskip
\item{75.}M. D. Scadron and H. F. Jones, {\it Phys. Rev.\/} {\bf
D10} (1974) 967.
\smallskip
\item{76.}H. Szadijan and J. Stern, {\it Nucl. Phys.\/} {\bf
B94} (1975) 163.
\smallskip
\item{77.}N.  H. Fuchs, H. Szadijan and J. Stern, {\it Phys. Lett.\/} {\bf
B238} (1990) 380.
\smallskip
\item{78.}C. A. Dominguez, {\it Riv. Nuovo Cim.\/} {\bf 8}
(1985) No. 6.
\smallskip
\item{79.}N.  H. Fuchs, H. Szadijan and J. Stern, {\it Phys. Lett.\/} {\bf
B269} (1991) 183.
\smallskip
\item{80.}L. Rosselet {\it et al.},
{\it Phys. Rev.\/} {\bf D15}
(1977) 574.
\smallskip
\item{81.}L. L. Nemenov, {\it Yad. Fiz.\/} {\bf
41} (1985) 980.
\smallskip
\item{82.}J. Uretsky and J. Palfrey, {\it Phys. Rev.\/} {\bf
121} (1961) 1798.
\smallskip
\item{83.}G. Czapek et al., letter of intent CERN/SPSLC 92--44, 1992.
\smallskip
\item{84.}R. J. Crewther, {\it Phys. Lett.\/} {\bf
B176} (1986) 172.
\smallskip
\item{85.}see e.g. K. M. Bitar et al., {\it Nucl. Phys.\/} (Proc. Suppl.) {\bf
26} (1992) 259.
\smallskip
\item{86.}V. Bernard, N. Kaiser and Ulf-G. Mei{\ss}ner,
{\it Phys. Rev.\/} {\bf D43} (1991) R3557.
\smallskip
\item{87.}R. W. Griffith, {\it Phys. Rev.\/} {\bf 176} (1968) 1705.
\smallskip
\item{88.}N. O. Johannesson and G. Nilson, {\it Nuovo Cim.\/} {\bf 43A}
(1978) 376.
\smallskip
\item{89.}C. B. Lang, {\it Fortschritte der Physik}
{\bf 26} (1978) 509;
\smallskip
\item{90.}C. B. Lang and W. Porod, {Phys. Rev.\/}
{\bf D21} (1980) 1295.
\smallskip
\item{91.}T. N. Truong, {\it Phys. Rev. Lett.\/} {\bf
61} (1988) 2526.
\smallskip
\item{92.}A. Dobado, M.J. Herrero and T.N. Truong,
{\it Phys. Lett.\/} {\bf B235} (1990) 134.
\smallskip
\item{93.}J. Gasser and Ulf-G. Mei{\ss}ner,
{\it Nucl.Phys.\/} {\bf B357} (1991) 90.
\smallskip
\item{94.}K.L. Au, D. Morgan and M.R. Pennington,
{\it Phys. Rev.\/} {\bf D35} (1987) 1633.
\smallskip
\item{95.}K. M. Watson, {\it Phys. Rev.\/} {\bf
95} (1955) 228.
\smallskip
\item{96.}Ulf-G. Mei{\ss}ner,
{\it Comm. Nucl. Part.
Phys.\/} {\bf 20}
(1991) 119.
\smallskip
\item{97.}A. Dobado and J. R. Pelaez, Universidad Complutense de Madrid
preprint FT/UCM/10/92, 1992.
\smallskip
\item{98.}A. Dobado and J. R. Pelaez, Universidad Complutense de Madrid
preprint FT/UCM/9/92, 1992.
\smallskip
\item{99.}G. Ecker, A. Pich and E. de Rafael, {\it Phys. Lett.\/} {\bf
B189} (1987) 363; {\it Nucl. Phys.\/} {\bf B291} (1987) 692;
{\it Nucl. Phys.\/} {\bf B303} (1988) 665.
\smallskip
\item{100.}G. Ecker, A. Pich and E. de Rafael, {\it Phys. Rep.\/}
in preparation.
\smallskip
\item{101.}A. Pich, B. Guberina and E. de Rafael,
{\it Nucl. Phys.\/} {\bf B277} (1986) 197.
\smallskip
\item{102.}J. Kambor, J. Missimer and D. Wyler, {\it Nucl. Phys.\/} {\bf
B345} (1990) 17.
\smallskip
\item{103.}C. Bernard, T. Draper, A. Soni, H. Politzer and M. Wise, {\it
Phys.Rev.\/} {\bf D32} (1985) 2343.
\smallskip
\item{104.}J. Kambor, J. Missimer and D. Wyler, {\it Phys. Lett.\/} {\bf
B261} (1991) 496.
\smallskip
\item{105.}T. N. Truong, {\it Phys. Lett.\/} {\bf
B207} (1988) 495.
\smallskip
\item{106.}Ulf--G. Mei{\ss}ner, in Proc. Fifth International Workshop on
"Perspectives in Nuclear Physics at Intermediate Energies", ed. S. Boffi,
C. Gioffi degli Atti and M. Giannini, World Scientific, Singapore, 1991.
\smallskip
\item{107.}J. F. Donoghue, E. Golowich and B. R. Holstein,
{\it Phys. Rev.\/} {\bf D30} (1984) 587.
\smallskip
\item{108.}J. Kambor, J. F. Donoghue, B. R. Holstein, J. Missimer and D.
Wyler, {\it Phys. Rev. Lett.\/} {\bf 68} (1992) 1818.
\smallskip
\item{109.}J. Gasser and H. Leutwyler, {\it Phys. Lett.\/} {\bf
B184} (1987) 83; {\it ibid} {\bf 188} (1987) 477.
\smallskip
\item{110.}H. Leutwyler, {\it Phys. Lett.\/} {\bf
B189} (1987) 197.
\smallskip
\item{111.}P. Gerber and H. Leutwyler, {\it Nucl. Phys.\/} {\bf
B321} (1989) 387.
\smallskip
\item{112.}J. Gasser and H. Leutwyler, {\it Nucl. Phys.\/} {\bf
B307} (1988) 763.
\smallskip
\item{113.}H. Leutwyler, {\it Nucl. Phys.\/} (Proc. Suppl.) {\bf
4} (1988) 248.
\smallskip
\item{114.}M. L{\"u}scher, {\it Ann. Phys.\/} (N.Y.) {\bf
142} (1982) 359.
\smallskip
\item{115.}P. Bin\'etruy and M. K. Gaillard,  {\it Phys. Rev.\/} {\bf
D32} (1985) 931.
\smallskip
\item{116.}H. Leutwyler, in "Effective field theories of the
standard model", ed. Ulf--G. Mei{\ss}ner, World Scientific, Singapore,
1992.
\smallskip
\item{117.}P. Gerber and H. Leutwyler, Bern University preprint, 1992.
\smallskip
\item{118.}V. Bernard and Ulf-G. Mei{\ss}ner, {\it Phys. Lett.\/}
{\bf B227} (1989) 465.
\smallskip
\item{119.}J. L. Goity, in "Effective field theories of the
standard model", ed. Ulf--G. Mei{\ss}ner, World Scientific, Singapore,
1992.
\smallskip
\item{120.}M. L{\"u}scher, {\it Nucl. Phys.\/} {\bf
B354} (1991) 531.
\smallskip
\item{121.}H. Neuberger, {\it Phys. Rev. Lett.\/} {\bf 60} (1988)
880;
{\it Nucl. Phys.\/} {\bf B300} [FS22] (1988) 180.
\smallskip
\item{122.}P. Hasenfratz and H. Leutwyler, {\it Nucl. Phys.\/} {\bf
B343} (1990) 241.
\smallskip
\item{123.}A. M. Polyakov, {\it Phys. Lett.\/} {\bf
B103} (1981) 207.
\smallskip
\item{124.}I. Dimitrovic, P. Hasenfratz, J. Nager and F. Niedermayer, {\it
Nucl. Phys.\/} {\bf B350} (1991) 893.
\smallskip
\item{125.}A. Hasenfratz et al., {\it Z. Phys.} {\bf C46} (1990) 257;
{\it Nucl. Phys.\/} {\bf B356} (1991) 332.
\smallskip
\item{126.}M. G{\"o}ckeler and H. Leutwyler,
{\it Nucl. Phys.\/} {\bf B350} (1991) 228.
\smallskip
\item{127.}I. Dimitrovic, J. Nager K. Jansen and T. Neuhaus, {\it
Phys. Lett.\/} {\bf B268} (1991) 408.
\smallskip
\item{128.}F. C. Hansen,
{\it Nucl. Phys.\/} {\bf B345} (1990) 685.
\smallskip
\item{129.}F. C. Hansen and H. Leutwyler,
{\it Nucl. Phys.\/} {\bf B350} (1991) 201.
\smallskip
\item{130.}S. R. Sharpe, {\it Phys. Rev.} {\bf D41} (1991) 3233;

S. R. Sharpe, R. Gupta and G. W. Kilcup,
{\it Nucl. Phys.\/} {\bf B383} (1992) 309;
\smallskip
\item{131.}D. Bernard and M. Golterman, {\it Phys. Rev.} {\bf D46} (1992)
853.
\smallskip
\item{132.}H. Leutwyler and A. V. Smilga, Bern University preprint
BUTP--92/10, 1992, {\it Phys. Rev.} {\bf D}, in print.
\smallskip
\item{133.}P. Hasenfratz and F. Niedermayer, {\it Phys. Lett.\/} {\bf
B268} (1990) 231.
\smallskip
\item{134.}N. D. Mermin and H. Wagner, {\it Phys. Rev. Lett.\/}
{\bf 17} (1966) 1133;

S. Coleman, {\it Comm. Math. Phys.} {\bf
31} (1973) 259.
\smallskip
\item{135.}P. Hasenfratz , M. Maggiore and F. Niedermayer, {\it Phys.
Lett.\/} {\bf B245} (1990) 522.
\smallskip
\item{136.}G. Aeppli et al., {\it Phys. Rev. Lett.\/} {\bf
62} (1989) 2052.
\smallskip
\item{137.}S. Chakravarty, B. I . Halperin and D. R. Nelson, {\it Phys. Rev.
Lett.\/} {\bf 60} (1988) 1057; {\it Phys. Rev.} {\bf B39} (1989) 2344.
\smallskip
\item{138.}R. J. Birgenau et al., in Proceedings of ICNS'91, 1991.
\smallskip
\item{139.}P. Hasenfratz, {\it Nucl. Phys.\/} (Proc. Suppl.) {\bf
26} (1992) 247.
\smallskip
\item{140.}U.--J. Wiese and H.--P. Ying,
Bern University preprint BUTP--92/50, 1992.

P. Hasenfratz and F. Niedermayer,
Bern University preprint BUTP--92/46, 1992.
\smallskip
\item{141.}P. Langacker and H. Pagels,
{\it Phys. Rev.\/} {\bf D8} (1971) 4595.
\smallskip
\item{142.}J. Gasser, M.E. Sainio and A. ${\check {\rm S}}$varc,
{\it Nucl. Phys.\/}
 {\bf
B 307} (1988) 779.
\smallskip
\item{143.}A. Krause, {\it Helv. Phys. Acta\/} {\bf
63} (1990) 3.
\smallskip
\item{144.}
Ulf-G. Mei{\ss}ner, in "Nucleon structure and nucleon
resonances", ed. G.A.
Miller, World Scientific, Singapore, 1992; {\it Int. J. Mod. Phys.}
{\bf E1} (1992) 561.
\smallskip
\item{145.}M. L. Goldberger and S. B. Treiman,
{\it Phys. Rev.\/} {\bf 110} (1958) 1178.
\smallskip
\item{146.}W.E. Caswell and G.P. Lepage, {\it Phys. Lett.\/} {\bf B167} (1986)
437;

M.B. Voloshin and M. Shifman, {\it Sov. J. Nucl. Phys.\/} {\bf 45} (1986) 463,
{\bf 47} (1988) 511;

N. Isgur and M.B. Wise, {\it Phys. Lett.\/} {\bf B232} (1989) 113, {\bf B237}
(1990) 527;

E. Eichten and B. Hill, {\it Phys. Lett.\/} {\bf B234} (1990) 511;

H. Georgi, {\it Phys. Lett.\/} {\bf B242} (1990) 427.
\smallskip
\item{147.}E. Jenkins and A.V. Manohar, {\it Phys. Lett.\/} {\bf B255} (1991)
558.
\smallskip
\item{148.}E. Jenkins and A.V. Manohar, {\it Phys. Lett.\/} {\bf B259} (1991)
353.
\smallskip
\item{149.}E. Jenkins and A.V. Manohar, in "Effective field theories of the
standard model", ed. Ulf--G. Mei{\ss}ner, World Scientific, Singapore,
1992.
\smallskip
\item{150.}J. G. K{\"o}rner and  G. Thompson, {\it Phys. Lett.\/} {\bf
B264} (1991) 185.
\smallskip
\item{151.}T. Mannel, W. Roberts and Z. Ryzak, {\it Nucl. Phys.\/} {\bf B368}
(1992) 264.
\smallskip
\item{152.}V. Bernard, N. Kaiser, J. Kambor
and Ulf-G. Mei{\ss}ner, {\it Nucl. Phys.\/} {\bf B388} (1992) 315.
\smallskip
\item{153.}D.G. Caldi and H. Pagels, {\it Phys. Rev.\/} {\bf D10} (1974) 3739.
\smallskip
\item{154.}T. Ericson and W. Weise, "Pions and Nuclei", Clarendon Press,
Oxford,
1988.
\smallskip
\item{155.}K.W. Rose {\it et al.}, {\it Phys. Lett.\/} {\bf B234}
(1990) 460;

F.J. Federspiel {\it et al.}, {\it Phys. Rev. Lett.\/} {\bf 67}
(1991) 1511;

J. Schmiedmayer {\it et al.}, {\it Phys. Rev. Lett.\/} {\bf 66}
(1991) 1015.
\smallskip
\item{156.}M. Damashek and F. Gilman, {\it Phys. Rev.\/} {\bf D1} (1970) 1319;

V.A. Petrunkin, {\it Sov. J. Nucl. Phys.\/} {\bf 12} (1981) 278.
\smallskip
\item{157.}V. Bernard, N. Kaiser and Ulf-G. Mei{\ss}ner,
{\it Phys. Rev. Lett.\/}
{\bf 67} (1991) 1515.
\smallskip
\item{158.}V. Bernard, N. Kaiser, and Ulf-G. Mei{\ss}ner, {\it Nucl. Phys.\/}
{\bf B373} (1992) 364.
\smallskip
\item{159.}M. N. Butler and M. J. Savage, {\it Phys. Lett.} {\bf B294} (1992)
369.
\smallskip
\item{160.}V. Bernard, N. Kaiser, J. Kambor and Ulf-G. Mei{\ss}ner,
{\it Phys. Rev.\/}
{\bf D46} (1992) 2756.
\smallskip
\item{161.}N. Kaiser, in "Baryons as Skyrme Solitons",
ed. G. Holzwarth, World Scientific, Singapore,
1993.
\smallskip
\item{162.}V. Bernard, N. Kaiser, and Ulf-G. Mei{\ss}ner, Bern University
preprint BUTP--92/51, 1992.
\smallskip
\item{163.}I.A. Vainshtein and V.I. Zakharov,
{\it Sov. J. Nucl. Phys.\/} {\bf 12} (1971) 333;
{\it Nucl. Phys.\/} {\bf B36} (1972) 589;

P. de Baenst, {\it Nucl. Phys.\/} {\bf B24} (1970) 633.
\smallskip
\item{164.}E. Mazzucato {\it et al.}, {\it Phys. Rev. Lett.\/} {\bf 57} (1986)
3144;

R. Beck {\it et al.}, {\it Phys. Rev. Lett.\/} {\bf 65} (1990) 1841.
\smallskip
\item{165.}D. Drechsel and L. Tiator,
{\it J. Phys. G: Nucl. Part. Phys.\/} {\bf 18} (1992)
449.
\smallskip
\item{166.}V. Bernard, J. Gasser, N. Kaiser and Ulf-G. Mei{\ss}ner,
{\it Phys. Lett.\/} {\bf B268} (1991) 291.
\smallskip
\item{167.}V. Bernard, N. Kaiser and Ulf-G. Mei{\ss}ner,
{\it Phys. Lett.\/} {\bf B282} (1992) 448.
\smallskip
\item{168.}S. Scherer and J. H. Koch,
{\it Nucl. Phys.\/} {\bf A534}
(1991) 461.
\smallskip
\item{169.}V. Bernard, N. Kaiser and Ulf-G. Mei{\ss}ner,
{\it Nucl. Phys.\/}
{\bf B383} (1992) 442.
\smallskip
\item{170.}M. Schumacher, private communication.
\smallskip
\item{171.}V. Bernard, N. Kaiser and Ulf-G. Mei{\ss}ner,
``Testing nuclear QCD: $\gamma p \to \pi^0 p$ at threshold'',
to appear in the $\pi N$ {\it Newsletter} No. 7 (1992).
\smallskip
\item{172.}T. P. Welch et al., {\it Phys. Rev. Lett.\/}
{\bf 69} (1992) 2761.
\smallskip
\item{173.}V. Bernard, N. Kaiser, T.--S. H. Lee and Ulf-G. Mei{\ss}ner,
{\it Phys. Rev. Lett.} {\bf 70} (1993) 387.
\smallskip
\item{174.}N.M. Kroll and M.A. Ruderman,
{\it Phys. Rev.\/} {\bf 93} (1954) 233.
\smallskip
\item{175.}V. Bernard, N. Kaiser and Ulf-G. Mei{\ss}ner,
{\it Phys. Rev. Lett.\/} {\bf 69} (1992) 1877.
\smallskip
\item{176.}Y. Nambu and D. Luri\'e, {\it Phys. Rev.\/} {\bf 125}
(1962) 1429;

Y. Nambu and E. Shrauner, {\it Phys. Rev.\/} {\bf 128}
(1962) 862.
\smallskip
\item{177.}A. del Guerra  {\it et al.}, {\it Nucl. Phys.\/}
 {\bf B107} (1976) 65;

M.G. Olsson, E.T. Osypowski and E.H. Monsay, {\it
Phys. Rev.\/} {\bf D17} (1978) 2938.
\smallskip
\item{178.}T. Kitagaki
{\it et al.}, {\it Phys. Rev.\/} {\bf D28}
(1983) 436;

L.A. Ahrens
{\it et al.}, {\it Phys. Rev.\/} {\bf D35}
(1987) 785;

L.A. Ahrens
{\it et al.}, {\it Phys. Lett.\/} {\bf B202}
(1988) 284.
\smallskip
\item{179.}E. Jenkins, {\it Nucl. Phys.\/}
{\bf B368} (1992) 190.
\smallskip
\item{180.}T.P. Cheng and R. Dashen, {\it Phys. Rev. Lett.\/}
{\bf 26} (1971) 594.
\smallskip
\item{181.}G. H\"ohler, in Landolt--B\"ornstein, vol.9 b2, ed. H. Schopper
(Springer, Berlin, 1983).
\smallskip
\item{182.}J. Gasser, in "Hadrons and Hadronic Matter", eds. D. Vautherin et
al., Plenum Press, New York, 1990.
\smallskip
\item{183.}R. Koch,
{\it Z. Phys.\/}
 {\bf C15} (1982) 161.
\smallskip
\item{184.}J. Gasser, H. Leutwyler and M.E. Sainio, {\it Phys. Lett.\/}
 {\bf 253B} (1991) 252.
\smallskip
\item{185.}L. S. Brown, W. J. Pardee and R. D. Peccei, {\it Phys. Rev.\/}
{\bf D4} (1971) 2801.
\smallskip
\item{186.}J. Gasser, H. Leutwyler and M.E. Sainio, {\it Phys. Lett.\/}
 {\bf 253B} (1991) 260.
\smallskip
\item{187.}E. Jenkins and A.V. Manohar, {\it Phys. Lett.\/} {\bf B281} (1992)
336.
\smallskip
\item{188.}M. E. Sainio, in "Effective field theories of the
standard model", ed. Ulf--G. Mei{\ss}ner, World Scientific, Singapore,
1992.
\smallskip
\item{189.}R. Marshak, Riazuddin and C.P. Ryan, "Theory of weak
interactions in particle physics", Wiley--Interscience, New York, 1969.
\smallskip
\item{190.}B. W. Lee,
{\it Phys. Rev. Lett.\/} {\bf 12} (1964) 83;

H. Sugawara,
{\it Prog. Theor. Phys.\/} {\bf 31} (1964) 213.
\smallskip
\item{191.}E. Jenkins, {\it Nucl. Phys.\/}
{\bf B375} (1992) 561.
\smallskip
\item{192.}J. Bijnens, H. Sonoda and M. B. Wise, {\it Nucl. Phys.\/}
{\bf B261} (1985) 185.
\smallskip
\item{193.} H. Neufeld, University of Vienna preprint UWThPh--1992--43, 1992;

E. Jenkins, M. Luke, A. V. Manohar and M. J. Savage, preprint

CERN--TH.6690/92, 1992.
\smallskip
\item{194.}S. Weinberg, {\it Phys. Lett.\/} {\bf
B251} (1990) 288.
\smallskip
\item{195.}S. Weinberg, {\it Nucl. Phys.\/} {\bf
B363} (1991) 3.
\smallskip
\item{196.}C. Ordonez and U. van Kolck, {\it Phys. Lett.\/} {\bf
B291} (1992) 459.
\smallskip
\item{197.} M. Rho, {\it Phys. Rev. Lett.\/} {\bf 66} (1991) 1275.
\smallskip
\item{198.} T.--S. Park, D.--P. Min and M. Rho, preprint SNUTP 92--82, 1992.
\smallskip
\item{199.} H. Yukawa, {\it Proc. Phys. Math. Soc. Japan\/} {\bf 17} (1935)
48.
\smallskip
\item{200.}H. Witala, W. Gl\"ockle and Th. Cornelius,
{\it Few Body Systems\/}
{\bf 6} (1989) 79; {\it Nucl. Phys.} {\bf A496} (1989) 446;
{\it Phys. Rev.} {\bf C39} (1989) 384;

I. Slaus, R. Machleidt, W. Tornow, W. Gl\"ockle and H. Witala,

{\it Comments Nucl. Part. Phys.} {\bf 20} (1991) 85;

W. Gl\"ockle, private communication.
\smallskip
\item{201.}
B. Frois and J.--F. Mathiot,
{\it Comments Part. Nucl. Phys.\/} {\bf 18} (1989)
291 (and references therein).
\smallskip
\item{202.} K. Kubodera, J. Delorme and M. Rho, {\it Phys. Rev. Lett.\/} {\bf
40} (1978) 755

M. Rho and G. E. Brown, {\it Comments Part. Nucl. Phys.\/} {\bf 10} (1981)
201.
\smallskip
\item{203.}S. Weinberg, {\it Phys. Lett.\/} {\bf
B295} (1992) 114.
\smallskip
\item{204.}J. Gunion, H. Haber, G. Kane and S. Dawson, "The Higgs Hunter's
Guide", Addison--Wesley, Menlo Park, 1990.
\smallskip
\item{205.}D. Dicus and V. Mathur, {\it Phys. Rev.} {\bf D7} (1973) 3111.
\smallskip
\item{206.}B. W. Lee, C. Quigg and H. B. Thacker, {\it Phys. Rev.} {\bf D16}
(1977) 1519.
\smallskip
\item{207.}E. Farhi and L. Susskind, {\it Phys. Reports} {\bf 74}
(1981) 277.
\smallskip
\item{208.}E. Eichten, I. Hinchcliffe, K. Lane and C. Quigg,
{\it Rev. Mod. Phys.} {\bf 56} (1984) 579.
\smallskip
\item{209.}M. Veltman, {\it Acta Phys. Pol.} {\bf B8} (1977) 475.
\smallskip
\item{210.}T. Appelquist and C. Bernard, {\it Phys. Rev.} {\bf D22} (1980)
200.
\smallskip
\item{211.}J. M. Cornwall, D. N. Levin and G. Tiktopoulos, {\it Phys. Rev.}
{\bf D10} (1974) 1145.
\smallskip
\item{212.}M. S. Chanowitz and M. K. Gaillard, {\it Nucl. Phys.} {\bf B261}
(1985) 379.
\smallskip
\item{213.}J. Bagger and C. Schmidt, {\it Phys. Rev.} {\bf D41} (1990) 264;

H.-J. He, Y.-P. Kuang and X. Li,  {\it Phys. Rev. Lett.}
{\bf 69} (1992) 2619.
\smallskip
\item{214.}D. A. Roos and M. Veltman, {\it Nucl. Phys.} {\bf B95}
(1975) 135.
\smallskip
\item{215.}P. Sikvie, L. Susskind, M. Voloshin and V. Zakharov, {\it Nucl.
Phys.} {\bf B173} (1980) 189.
\smallskip
\item{216.}U. Amaldi et al., {\it Phys. Rev.} {\bf D36} (1987) 1385.
\smallskip
\item{217.}M. Ademello and R. Gatto,  {\it Phys. Rev. Lett.}
{\bf 13} (1964) 264.
\smallskip
\item{218.}A. Longhitano, {\it Nucl. Phys.} {\bf B188} (1981) 118.
\smallskip
\item{219.}S. Sint, Diploma thesis, Universit{\"a}t Hamburg, 1991.
\smallskip
\item{220.}M. Veltman, {\it Nucl. Phys.} {\bf B123} (1977) 89;

M. S. Chanowitz, M. A. Furman and I. Hinchcliffe, {\it Phys. Lett.} {\bf
B78} (1978) 285.
\smallskip
\item{221.}C. Ramirez, {\it Phys. Rev.} {\bf D42} (1990) 1726.
\smallskip
\item{222.}A. Falk, M. Luke and E. Simmons, {\it Nucl.
Phys.} {\bf B365} (1991) 523.
\smallskip
\item{223.}J. Bagger, S. Dawson and G. Valencia, {\it Phys. Rev. Lett.}
{\bf 67} (1991) 2256.
\smallskip
\item{224.}J. Bagger, S. Dawson and G. Valencia, preprint
FERMILAB--PUB--92/75--T, to appear in {\it Nucl. Phys.} {\bf B}, 1992.
\smallskip
\item{225.}M. Chanowitz, M. Golden and H. Georgi, {\it Phys. Rev.}
{\bf D36} (1987) 1490.
\smallskip
\item{226.}J. F. Donoghue and C. Ramirez, {\it Phys. Lett.}
{\bf B234} (1990) 361.
\smallskip
\item{227.}S. Dawson and S. Willenbrock, {\it Phys. Rev.}
{\bf D40} (1989) 2280;

M. Veltman and F. J. Yndurain, {\it Nucl. Phys.} {\bf B325} (1989) 1.
\smallskip
\item{228.}S. Dawson and G. Valencia, {\it Nucl. Phys.} {\bf B348} (1991) 23.
\smallskip
\item{229.}A. Dobado and M. J. Herrero, {\it Phys. Lett.} {\bf B228} (1989)
495;

A. Dobado, M. J. Herrero and T. N. Truong, {\it Phys. Lett.} {\bf B235}
(1990) 129.
\smallskip
\item{230.}S. Dawson and G. Valencia, {\it Nucl. Phys.} {\bf B352} (1991) 27.
\smallskip
\item{231.}D. A. Dicus and W. W. Repko, {\it Phys. Lett.} {\bf B228} (1989)
503.
\smallskip
\item{232.}K.--I. Hikasa, KEK preprint KEK--TH--319, 1992.
\smallskip
\item{233.}M. Soldate and R. Sundrum, {\it Nucl. Phys.} {\bf B340} (1990) 1.
\smallskip
\item{234.}G. L. Kane, W. W. Repko and W. B. Rolnick {\it Phys. Lett.} {\bf
B148} (1984) 367;

S. Dawson, {\it Nucl. Phys.} {\bf B249} (1985) 42.
\smallskip
\item{235.}J. Bagger, S. Dawson and G. Valencia, {\it Phys. Lett.}
{\bf B292} (1992) 137.
\smallskip
\item{236.}J. Bagger, talk given at the DPF meeting, Fermilab, 1992.
\smallskip
\item{237.}R. Cahn, S. Ellis, R. Kleiss and W. Stirling, {\it Phys. Rev.}
{\bf D35} (1987) 1626;

U. Baur and E. Glover, {\it Nucl. Phys.} {\bf B347} (1990) 12;

V. Barger, K. Cheung, T. Han and D. Zeppenfeld, {\it Phys. Rev.}
{\bf D44} (1991) 2701.
\smallskip
\item{238.}
V. Barger, K. Cheung, T. Han and R. Phillips, {\it Phys. Rev.}
{\bf D42} (1990) 3052;

D. Dicus, J. Gunion, L. Orr and R. Vega, {\it Nucl. Phys.} {\bf B377} (1992)
31.
\smallskip
\item{239.}A. Dobado and M. Urdiales, {\it Phys. Lett.} {\bf B292} (1992)
128.
\smallskip
\item{240.}B. Holdom, {\it Phys. Lett.} {\bf B258} (1991) 156.
\smallskip
\item{241.}M. Chanowitz, {\it Ann. Rev. Nucl. Part. Sci.} {\bf 38} (1988)
323.
\smallskip
\item{242.}S. Weinberg, {\it Phys. Rev. Lett.\/} {\bf
65} (1990) 1177.
\smallskip
\item{243.}T. N. Pham, {\it Phys. Lett.\/} {\bf
B255} (1991) 451.
\smallskip
\item{244.}
A. Sirlin, {\it Nucl. Phys.} {\bf B71} (1974) 29;
{\it Nucl. Phys.} {\bf B100} (1975) 291;
{\it Phys. Rev.} {\bf D22} (1980) 971;

M. Veltman, {\it Nucl. Phys.} {\bf B123} (1977) 89;

W. Marciano and A. Sirlin,
{\it Phys. Rev.} {\bf D22} (1980) 2695;

M. Peskin and R. G. Stuart, SLAC--PUB--3725, 1985.
\smallskip
\item{245.}M. Peskin, in "Physics at the 100 GeV mass scale", ed. E. C.
Brennan, SLAC--Report--361, 1989.
\smallskip
\smallskip
\item{246.}D. Kennedy and B. W. Lynn, {\it Nucl. Phys.} {\bf B322} (1989) 1;

M. Kuroda, G. Moultaka and D. Schildknecht, {\it Nucl. Phys.} {\bf B350}
(1991) 25.
\smallskip
\item{247.}T. Appelquist and J. Carrazone, {\it Phys. Rev.}
{\bf D11} (1975) 2856.
\smallskip
\item{248.}M. Peskin and T. Takeuchi,  {\it Phys. Rev. Lett.}
{\bf 65} (1990) 964.
\smallskip
\item{249.}M. Peskin and T. Takeuchi,  {\it Phys. Rev.}
{\bf D46} (1992) 381.
\smallskip
\item{250.}G. Altarelli, R. Barbieri and S. Jadach,
{\it Nucl. Phys.} {\bf B369} (1992) 3.
\smallskip
\item{251.}W. Marciano and J. Rosner,  {\it Phys. Rev. Lett.}
{\bf 65} (1990) 2963.
\smallskip
\item{252.}M. Golden and L. Randall, {\it Nucl. Phys.} {\bf B361} (1991) 3.
\smallskip
\item{253.}B. Holdom and J. Terning, {\it Phys. Lett.} {\bf B247} (1990) 88.
\smallskip
\item{254.}B. Holdom, {\it Phys. Lett.} {\bf B259} (1991) 329.
\smallskip
\item{255.}A. Dobado, D. Espriu and M. J. Herrero, {\it Phys. Lett.} {\bf
B255} (1991) 405.
\smallskip
\item{256.}H. Georgi, {\it Nucl. Phys.} {\bf B363} (1991) 301.
\smallskip
\item{257.}
C. Roiesnel and T. N. Truong, {\it Phys. Lett.} {\bf B253} (1991) 439.
\smallskip
\item{258.}
S. Bertolini and A. Sirlin, {\it Phys. Lett.} {\bf B257} (1991) 179;
{\it Phys. Rev.} {\bf D22} (1980) 971;

E. Gates and J. Terning, {\it Phys. Rev. Lett.} {\bf 67} (1991) 1840;

M. Dugan and L. Randall,
{\it Phys. Lett.} {\bf B264} (1981) 154.
\smallskip
\item{259.}H. Georgi, {\it Nucl. Phys.} {\bf B361} (1991) 339.
\smallskip
\item{260.}
M. Consoli, W. Hollik and F. Jegerlehner,
{\it Phys. Lett.} {\bf B227} (1989) 167.
\smallskip
\item{261.}
J. F. Donoghue and B. R. Holstein, {\it Phys. Rev.} {\bf D46} (1992) 4076.
\smallskip
\item{262.}
G. D'Ambrosio and D. Espriu, {\it Phys. Lett.} {\bf B175} (1986) 237;

J. L. Goity, {\it Z. Phys.} {\bf C34} (1987) 341.
\smallskip
\item{263.}
B. R. Holstein, {\it Comments Nucl. Part. Phys.} {\bf 19} (1990) 221.
\smallskip
\item{264.}
J. Bijnens, A. Bramon and F. Cornet, {\it Phys. Rev. Lett.} {\bf 61} (1988)
1453;

J. F. Donoghue and D. Wyler, {\it Nucl. Phys.} {\bf B316} (1989) 289;

T. N. Pham,
{\it Phys. Lett.} {\bf B246} (1990) 175;

D. Issler, Preprint SLAC--PUB--4943 and 5200, 1990;

J. Bijnens, {\it Nucl. Phys.} {\bf B367} (1991) 709;

for an overview, see J. Bijnens, in
"Effective field theories of the
standard model",

ed. Ulf--G. Mei{\ss}ner, World Scientific, Singapore,
1992.
\smallskip
\item{265.}
The DAFNE Physics Handbook, eds. L. Maiani, G. Pancheri and N. Paver, INFN
Frascati, to appear.
\smallskip
\item{266.}
G. Ecker, preprint CERN--TH.6600/92 and UWThPh--1992--44, 1992.
\smallskip
\item{267.}
H. Leutwyler, preprint BUTP/92--42, 1992.
\smallskip
\item{268.}
C. J. C. Im, preprint SLAC--PUB--5627, 1991;

A. Dobado and J. R. Pelaez,
{\it Phys. Lett.} {\bf B286} (1992) 136.
\smallskip
\item{269.}
M. Wise, {\it Phys. Rev.} {\bf D45} (1992) 2188;

G. Burdman and J. F. Donoghue,
{\it Phys. Lett.} {\bf B280} (1992) 287;

T. M. Yan et al., {\it Phys. Rev.} {\bf D46} (1992) 1148;

J. L. Goity, {\it Phys. Rev.} {\bf D46} (1992) 3929;

U. Kilian, J. G. K\"orner and D. Pirjol, {\it Phys. Lett.} {\bf B288}
(1992) 360;

A. F. Falk and M. Luke, {\it Phys. Lett.} {\bf B292}
(1992) 119;

J. F. Amundson et al., preprint CERN--TH.6650/92, 1992.

M. A. Novak, M. Rho and I. Zahed, preprint SUNY--NTG--92/27, 1992.
\smallskip
\item{270.}
J. F. Donoghue, in "Medium energy antiprotons and the quark--gluon structure
of hadrons", eds. R. Landua, J. M. Richard and R. Klapish, Plenum Press, New
York, 1992.
\smallskip
\item{271.}
H. Leutwyler, in ``Recent Aspects of Quantum Fields'', eds. H. Mitter
and M. Gausterer, Springer Verlag, Berlin, 1991.
\smallskip
\item{272.}
T. N. Truong, in "Medium energy antiprotons and the quark--gluon structure
of hadrons", eds. R. Landua, J. M. Richard and R. Klapish, Plenum Press, New
York, 1992.
\smallskip
\item{273.}
J. F. Donoghue, E. Golowich and B. R. Holstein, " Dynamics of the  Standard
Model", Cambridge Univ. Press, Cambridge, 1992.
\smallskip
\item{274.}"Effective field theories of the
standard model", ed. Ulf--G. Mei{\ss}ner, World Scientific, Singapore,
1992.

\par
\end